\documentclass{emulateapj}

\usepackage{graphicx,psfig,fancyhdr,natbib,subfigure}
\usepackage{epsfig, psfig, epsf}
\usepackage{amsmath, cancel}
\usepackage{amssymb}
\usepackage{lscape}

 \def\apjl{ApJ~Lett.} 
\def\apjs{ApJS}

 \def\aap{Astron.~\&~Astrophys.}

\def \Omb {\Omega_{\rm b}}

\def \Omlam {\Omega_{\Lambda}}
\def \Omm {\Omega_{\rm m}}

\def \kms {{\rm ~km~s}^{-1}}

\def \hmpc{~h^{-1}~{\rm Mpc}} 
\def \hkpc{\;h^{-1}{\rm kpc}}

\def \xis{\xi(s)}
\def \xisp{\xi(r_{p}, \pi)}

\def \xir{\xi(r)}

\def \gsim { \lower .75ex \hbox{$\sim$} \llap{\raise .27ex \hbox{$>$}} }
\def \lsim { \lower .75ex \hbox{$\sim$} \llap{\raise .27ex \hbox{$<$}} }

\def \msun {M_{\odot}}
\def \Msol {M_{\odot}}

\def \wp {w_{p}(r_{p})}

\def \BEST   {{\tt best} }
\def \TARGET {{\tt target} }

\def \QSO    {{\tt TARGET\_QSO}}
\def \HIZ    {{\tt TARGET\_HIZ}}

\def \FIRST  {{\tt TARGET\_FIRST}}

\shorttitle{SDSS Quasar Clustering at redshift $z\leq2.2$}
\shortauthors{N.~P.~Ross et al.}

\begin{document}
\title{Clustering of Low-Redshift ($z\leq2.2$) Quasars 
  from the Sloan Digital Sky Survey}

\author{Nicholas P. Ross\altaffilmark{1}, Yue Shen\altaffilmark{2}, Michael A. Strauss\altaffilmark{2}, 
Daniel E. Vanden Berk\altaffilmark{1}, Andrew J. Connolly\altaffilmark{3}, Gordon T. Richards\altaffilmark{4},  Donald P. Schneider\altaffilmark{1}, David H. Weinberg\altaffilmark{5}, Patrick B. Hall\altaffilmark{6}, Neta A. Bahcall\altaffilmark{2}, Robert J. Brunner\altaffilmark{7}}
\altaffiltext{1}{Department of Astronomy and Astrophysics, The Pennsylvania State University, 525 Davey Laboratory, University Park, PA 16802, U.S.A.; npr@astro.psu.edu}
\altaffiltext{2}{Princeton University Observatory, Princeton, NJ 08544, U.S.A.}
\altaffiltext{3}{Department of Astronomy, University of Washington, Box 351580, Seattle, WA 98195, U.S.A.}
\altaffiltext{4}{Department of Physics, Drexel University, 3141 Chestnut Street, Philadelphia, PA 19104, U.S.A}
\altaffiltext{5}{Astronomy Department and Center for Cosmology and AstroParticle Physics, Ohio State University, Columbus, OH 43210, U.S.A.}
\altaffiltext{6}{Department of Physics and Astronomy, York University, Toronto, ON M3J 1P3, Canada}
\altaffiltext{7}{Department of Astronomy, MC-221, University of Illinois, 1002 West Green Street, Urbana, IL 61801, U.S.A.}

\begin{abstract}
We present measurements of the quasar two-point correlation function,
$\xi_{Q}$, over the redshift range $0.3 \leq z \leq 2.2$ based upon
data from the Sloan Digital Sky Survey (SDSS). Using a homogeneous
sample of \hbox{30,239} quasars with spectroscopic redshifts from the
Data Release 5 Quasar Catalogue, our study represents the largest
sample  used for this type of investigation to date. With this
redshift range and an areal coverage of $\approx$4,000 deg$^{2}$, we
sample over \hbox{25 $h^{-3}$ Gpc$^{3}$} (comoving) of the Universe in
volume, assuming the current $\Lambda$CDM cosmology. 
Over this redshift range, we find that the redshift-space correlation
function, $\xis$, is adequately fit by a single power-law, with
$s_{0}=5.95\pm0.45\hmpc$ and \hbox{$\gamma_{s}=1.16^{+0.11}_{-0.16}$}
when fit over \hbox{$1.0 \leq s \leq 25.0 \hmpc$}.  We find no evidence for
deviation from $\xis=0$ at scales of $s>100 \hmpc$, but do observe
redshift-space distortions in the 2-D $\xi(r_{p},\pi)$ measurement.
Using the projected correlation function, $\wp$, we calculate the
real-space correlation length, \hbox{$r_{0}=5.45^{+0.35}_{-0.45}
\hmpc$} and $\gamma=1.90^{+0.04}_{-0.03}$, over scales of \hbox{$1.0 \leq
r_{p} \leq 130.0 \hmpc$}.
Dividing the sample into redshift slices, we find very little, if
any, evidence for the evolution of quasar clustering, with the
redshift-space correlation length staying roughly constant at
\hbox{$s_{0}\sim 6-7 \hmpc$} at $z\lesssim2.2$ (and only increasing at
redshifts greater than this). We do, however, see tentative evidence
for evolution in the real-space correlation length, $r_{0}$, at
$z>1.7$. Our results are consistent with those from the 2QZ survey
and previous SDSS quasar measurements using photometric
redshifts. Comparing our clustering measurements to those reported for
X-ray selected AGN at $z\sim0.5-1$, we find reasonable agreement in
some cases but significantly lower correlation lengths in others. 
Assuming a standard $\Lambda$CDM cosmology, we find that the linear
bias evolves from $b\sim1.4$ at $z=0.5$ to $b\sim3$ at $z=2.2$, with
$b(z=1.27)=2.06\pm0.03$ for the full sample. We compare our data to
analytical models and infer that quasars inhabit dark matter haloes of
constant mass $M_{\rm halo} \sim 2\times10^{12} h^{-1}\msun$ from
redshifts $z\sim2.5$ (the peak of quasar activity) to $z \sim 0$;
therefore the ratio of the halo mass for a typical quasar to the mean
halo mass at the same epoch drops with decreasing redshift. The
measured evolution of the clustering amplitude is in reasonable
agreement with recent theoretical models, although measurements to
fainter limits will be needed to distinguish different scenarios for
quasar feeding and black hole growth. 
\end{abstract}

\keywords{clustering -- quasars: general -- cosmology: observations --
large-scale structure of Universe. general -- surveys}

\section{Introduction}

Understanding how and when the structures in the local Universe formed
from the initial conditions present in the early Universe is one of
the fundamental goals of modern observational cosmology. Tracing the
evolution of clustering with cosmic epoch offers the potential to
understand the growth of structure and its relation to the energy and
matter content of the Universe, including the relationship between the
dark matter and the luminous galaxies and quasars that we observe. 

As such, one of the primary science goals of the Sloan Digital Sky
Survey \citep[SDSS;][]{York00} is to measure the large-scale
distribution  of galaxies and quasars, and in particular, to determine
the spatial clustering of quasars as a function of
redshift. \citet{Shen07} report on the clustering of high redshift
($z\geq2.9$)  quasars from the SDSS; in this paper, we investigate the
spatial clustering from redshift $z=2.2$ to the present day, i.e. the
evolution of quasar clustering over nearly 80\% of the age of the
Universe (the gap in redshift being a consequence of the optical
selection techniques used in the SDSS). 

Due to their high intrinsic luminosities, quasars are seen to large
cosmological distances, and are thus good probes of large-scale
structure (LSS) and its evolution. However, until recently, quasar
studies were plagued by low-number statistics, leading to shot noise,
and samples covered only small areas of sky, leading to sample
variance. With the advent of large solid angle ($\gtrsim 1000$
deg$^2$) surveys with efficient selection techniques, these
limitations have been overcome, and the number of known quasars has
increased by more than an order of magnitude in the last decade, thanks
mainly to the 2dF QSO Redshift Survey \citep[2QZ;][]{Boyle00, Croom04}
and the SDSS. The latest SDSS quasar catalogue \citep{Schneider07}
contains nearly \hbox{80 000} objects. Using the data from these large
surveys, we are now in a position to make high-precision measurements
of quasar clustering properties. 

The two-point correlation function (2PCF), $\xi$, is a simple but
powerful statistic commonly employed to quantify the clustering
properties of a given class of object \citep{Peebles80}. The observed
value of $\xi$ for quasars can be related to the underlying (dark)
matter density distribution via
\begin{equation}
  \xi(r)_{\rm quasar} = b_{Q}^{2} \,\, \xi(r)_{\rm matter}
  \label{eq:bias}
\end{equation}
where $\xi(r)_{\rm matter}$ is the mass correlation function and
$b_{Q}$ is the linear bias parameter for quasars. 
Although equation~(\ref{eq:bias}) defines $b_{Q}$, and there are
theoretical arguments suggesting that $b_{Q}$ is scale-independent on
large scales, e.g. \citet[][]{Scherrer98}, we do not know {\it a priori}
if this is the case.  

With certain reasonable assumptions, the measurement and
interpretation of the bias can lead to determination of the dark
matter halo properties of quasars and to quasar lifetimes
\citep[$t_{q}$, ][]{Martini01, Haiman01}. In the standard scenario,
quasar activity is triggered by accretion onto a central, supermassive
black hole \citep[SMBH, e.g.][]{Salpeter64, LyndenBell69,
Rees84}. Given the possible connection between the SMBH and host halo,
and the fact that halo properties are correlated with the local
density contrast, clustering measurements can be used to constrain
this potential halo-SMBH connection and provide an insight into quasar
and black hole physics \citep[e.g.][]{Baes03, WyitheLoeb05a, Wyithe06,
Adelberger05, Fine06, daAngela08}.  This information, combined with
the  quasar luminosity function (QLF), constrains $\eta$, the fraction
of  the Eddington luminosity at which quasars shine, and their duty
cycle \citep{WyitheLoeb05a, Shankar07}.

Early measurements of the quasar 2PCF \citep[e.g.][]{Arp70, Hawkins75,
Osmer81, Shanks83c, Shanks87} measured statistically significant
clustering on scales of a few$\hmpc$, for both the quasar
auto-correlation function and cross-correlation with galaxies. This
result has been confirmed with data from more recent surveys,
\citep[e.g.][]{Croom05, PMN04}. The Quasar 2PCF is typically fit to a
single power law of the form, 
\begin{equation} 
  \xi(r) = ( r / r_{0})^{-\gamma} 
  \label{eq:xi_PL_model}
\end{equation} 
over the range $1 \hmpc \leq r \leq 100 \hmpc$. Here, $r_{0}$ is the
correlation length quoted in comoving coordinates and $\gamma$ is the
power-law slope. Typical measured correlation lengths and slopes for
quasars at redshift $z\sim1.5$ are $r_{0}= 5 - 6 \hmpc$ and  $\gamma
\sim 1.5$, respectively. 

The evolution of the quasar correlation function has been disputed for
a long time, with some authors reporting that $r_{0}$ either decreased
or only weakly evolved with redshift \citep[e.g.][]{Iovino88,
Croom96}, while others reported an increase with redshift
\citep[e.g.][]{Kundic97, LaFranca98}. However, with the advent of the
2QZ Survey, $r_{0}$ has been shown to evolve at the $\sim90 - 99\%$
confidence level, in the sense that quasar clustering increases with
redshift, although the actual degree of evolution is weak
\citep{Croom01, PMN04, Croom05}. In particular, \citet{Croom05} used
over \hbox{20 000}  objects from the final 2QZ dataset to measure the
redshift-space two-point correlation function, $\xi(s)$, over the
redshift range $0.3<z<2.2$ and found  a significant increase in the
clustering amplitude at high redshift. The quasar bias, where the bias
depends on the underlying CDM model such that a constant $r_{0}$ can
imply a strongly varying $b$, was found to be a strong function of
redshift, with an empirical dependence of
\begin{equation} 
   b_{Q}(z)=(0.53\pm0.19)+(0.289\pm0.035)(1+z)^{2}. 
\label{eq:Croom_bias_model}
\end{equation} 
These values were used to derive the mean dark matter halo (DMH) mass
occupied by quasars, giving a redshift-independent value of
$M_{\rm DMH}=(3.0\pm1.6)\times10^{12} h^{-1} \Msol$. Independent analysis
of the 2QZ data by \citet{PMN04} confirmed these findings. 

Using the SDSS, \citet{Shen07} found that redshift $2.9\leq z\leq5.4$
quasars are significantly more clustered than their $z\sim1.5$
counterparts, having a real-space correlation length and power-law
slope of $r_{0}=15.2\pm2.7\hmpc$ and $\gamma=2.0\pm0.3$, respectively,
over the scales $4\hmpc \leq r_{p} \leq 150 \hmpc$ (where $r_{p}$ is
the separation from the projected correlation function,  $\wp$).
\citet{Shen07} also found that bias increases with redshift, with,
$b_{Q} \sim 8$ at $z=3.0$ and $b_{Q} \sim 16$ at $z=4.5$. 

\citet{Myers06, Myers07a}, also using SDSS data, examined the
clustering of photometrically-selected quasar candidates over  $\sim50
\hkpc$ to $\sim20\hmpc$ scales. In this sample, quasar redshifts were
assigned from photometric rather than spectral information
\citep{Richards01}. They found that the linear bias, $b_{Q}$,
increases with redshift, from $b_{Q}=1.93$ at redshifts $0.4 \leq z
<1.0$ to $b_{Q}=2.84$ at $2.1 \leq z < 2.8$, consistent with
equation~(\ref{eq:Croom_bias_model}) \citep[Fig. 4 of][]{Myers07a}. 

\citet{Padmanabhan08} measured the clustering of
photometrically-selected luminous red galaxies (LRGs) around a low
redshift, $0.2<z<0.6$, sample of quasars, with both LRG and quasar
samples coming from the SDSS. They determined a large-scale quasar
bias $b_{Q} = 1.09\pm0.15$ at a median redshift of $z=0.43$. After
taking into account measurement and interpretation subtleties, the
results from \citet{Padmanabhan08}, are in qualitative agreement with
those from \citet{Serber06}, who find that $M_{i}\leq-22$, $z\leq0.4$
quasars are located in higher local galaxy overdensities than typical
$L^{*}$ galaxies. \citet{Serber06} suggested that quasars typically
reside in $L^{*}$ galaxies, but have a local excess of neighbours
within $\sim0.15-0.7 \hmpc$, which contributes  to the triggering of
quasar activity through mergers and other
interactions. \citet{Strand08} using photometric redshift cuts,
confirm the basic overdensity values measured by
\citet{Serber06}. \citet{Hennawi06}, \citet{Myers07b} and
\citet{Myers08} reached similar conclusions by examining pairs of
quasars on $<1\hmpc$ scales. The quasar correlation function shows a
small scale excess over a power law, and \citet{Hennawi06} suggested
that the small-scale excess can be attributed to dissipative
interaction events that trigger quasar activity in rich environments.

Due to the evolution of the quasar luminosity function and the
flux-limited nature of most quasar samples, there is a strong
correlation between redshift and luminosity in these samples, making
it difficult to isolate luminosity dependence of clustering from
redshift dependence.  Recently, \citet{daAngela08} combined data from
the 2QZ and the 2SLAQ Survey \citep[2dF-SDSS LRG And QSO
Survey;][]{Croom08}, to investigate quasar clustering and break this
degeneracy. \citet{daAngela08} estimate the mass of the dark matter haloes
which quasars inhabit to be $\sim 3 \times 10^{12} h^{-1} \Msol$, in
agreement with \citet{Croom05}, a value that does not evolve strongly
with redshift or depend on QSO luminosity. Their results also suggest
that quasars of different luminosities may contain black holes of
similar mass.

There have also been recent advances in theoretical predictions of the
quasar correlation function and its evolution with redshift 
\citep{Lidz06, Hopkins07, Shankar07, Hopkins08, Basilakos08} 
and we discuss these models in more detail in Sections 4 and 5. 

In this paper, we shall measure the quasar 2PCF for redshifts
$z\leq2.2$,  using the largest sample of spectroscopically identified
quasars to date. We will investigate the dependence of quasar
clustering strength with redshift and luminosity, 
allowing tests of current quasar formation and evolution models.

This paper is organised as follows. In Section 2 we present our data
sample, mentioning several effects that could give rise to systematics
in the measurements. In Section 3 we briefly describe the
techniques involved in measuring the two-point correlation function
and in Section 4 we present our results.  In Section 5 we compare and
contrast our evolutionary results with recent observational
results in the literature, and we conclude in Section 6. Appendix A 
gives technical details for the SDSS, Appendix B describes our error
analysis and Appendix C carries out a series of systematic checks. 

In our companion paper \citep[][]{Shen09}, we expand our
investigations on the clustering of SDSS quasars. Using the same data
as we examine here, Shen et al. study the  dependence of quasar
clustering on luminosity, virial black hole mass, quasar colour and
radio loudness. 

We assume the currently preferred flat, ``Lambda Cold Dark Matter''
($\Lambda$CDM) cosmology where $\Omb=0.042$, $\Omm=0.237$,
$\Omlam=0.763$ \citep{Sanchez06, Spergel07}  and quote distances in
units of $\hmpc$ to aid in ease of comparisons with previous results
in the literature. Since we are measuring objects with redshifts
resulting from the Hubble flow, all distances herein are given in
comoving coordinates.  Where a value of Hubble's Constant is assumed
e.g. for absolute magnitudes, this will be quoted explicitly. Our
magnitudes are based on the AB zero-point system \citep{Oke83}.

\section{Data}\label{section_data} 
Much care must be taken when constructing a dataset that is valid for
a statistical analysis. In this section and Appendix A we describe the
various samples we use to investigate potential systematic effects in
our clustering measurements. Appendix A provides some of the relevant
technical details of the SDSS, discussing the Catalogue Archive Server
(CAS) and the SDSS Survey geometry. 

    \subsection{The Sloan Digital Sky Survey}

    The SDSS uses a dedicated 2.5m wide-field telescope \citep{Gunn06}
    to collect light for 30 2k$\times$2k CCDs \citep{Gunn98} over five
    broad bands - {\it ugriz} \citep{Fukugita96} - in order to image
    $\sim\pi$ steradians of the sky. The imaging data are taken on dark
    photometric nights of good seeing \citep{Hogg01} and are calibrated
    photometrically \citep{Smith02, Ivezic04, Tucker06, Padmanabhan08a},
    and astrometrically \citep{Pier03}, and object parameters are measured
    \citep{Lupton01, Stoughton02}.
    
    Using the imaging data, quasar target candidates are selected for
    spectroscopic observation based on their colours, magnitudes and
    detection in the FIRST radio survey \citep{Becker95}, as described by
    \citet{Richards02}. Unless stated otherwise, all quoted SDSS
    photometry has been corrected for Galactic extinction following
    \citet[][]{Schlegel98}. Here we are concerned with only those quasars
    selected from the main quasar selection
    \citep{Richards02}. Low-redshift, $z\lesssim3$, quasar targets are
    selected based on their location in $ugri$-colour space and the
    high-redshift, $z\gtrsim3$, objects in $griz$-colour space.  Quasar
    candidates passing the $ugri$-colour selection  are selected to a flux
    limit of $i=19.1$, but since high-redshift quasars are rare, objects
    lying in regions of colour-space corresponding to quasars at $z>3$ are
    targetted to $i=20.2$. Furthermore, if an unresolved, $i\leq19.1$ SDSS
    object is matched to within $2''$ of a source in the FIRST catalogue, it
    is included in the quasar selection.
    
    A tiling algorithm then assigns these candidates to specific
    spectroscopic plates, in order to maximise target completeness
    \citep{Blanton03}. Each 3$^{\circ}$ diameter spectroscopic plate holds
    640 fibres and quasar candidates are allocated at a density of
    approximately 18 fibers deg$^{-2}$. No two fibres can be placed closer
    than $55''$, corresponding to $\sim 0.7 \hmpc$ at $\langle z
    \rangle=1.27$, the mean redshift of our sample
    (Fig.~\ref{fig:fibre_collisions}). In the case of conflicts because of
    this $55''$ constraint, the main quasar selection candidates were
    given targetting priority over the MAIN galaxy and LRG survey targets
    \citep[][respectively]{Strauss02, Eisenstein01}. Therefore, excluding
    subtle effects due to gravitational lensing \citep{Scranton05,
      Mountrichas07}, the LSS `footprint' of these foreground galaxies
    should not affect our LSS quasar measurements. Some targets, including
    brown dwarf and hot subdwarf calibration star candidates, were given
    higher priority than the main quasar candidates. However, since the
    surface density  of these Galactic objects is very low ($\ll$ 1
    deg$^{-2}$), this should not have any significant impact on our
    results. We investigate the effects of quasar-quasar fibre collisions
    in Appendix~\ref{sec:fibre_collisions}.

    \subsection{Quasar Samples}
    For our analysis, we use the SDSS Data Release Five
    \citep[DR5;][]{Adelman-McCarthy07} and select quasars from the latest
    version of the quasar catalogue \citep[DR5Q;][]{Schneider07}. This
    catalogue consists of spectroscopically identified quasars that have
    luminosities larger than $M_{i} = -22.0$ (measured in the rest frame)
    and at least one emission line with FWHM larger than 1000
    $\kms$. Every object in the DR5Q had its spectrum manually inspected.
    There are \hbox{77 429} confirmed quasars over the \hbox{5 740}
    deg$^2$ spectroscopic DR5 footprint; the \hbox{65 660} DR5Q quasars
    with redshifts $z\leq2.2$ will be the parent sample we use in this
    investigation. 
    
    At $z\geq2.2$ the ``ultra-violet excess'' (UVX) method of
    selecting quasars begins to fail due to the Ly$\alpha$-forest
    suppressing flux as it moves through the SDSS $u$-band, and quasars
    have colours similar to those of F-stars \citep{Fan99}. Thus, for
    $2.2<z\leq2.9$, the completeness of the survey is dramatically lowered
    as is discussed in detail by \citet{Richards06}. A lower redshift
    limit of $z=0.30$ is chosen to match that of the 2QZ. Therefore,
    although we will present results in the redshift ranges $z<0.30$ and
    $2.2 < z \leq 2.9$, we will not place strong significance on these
    data. The number of quasars used in this study is twice that of the
    previous largest quasar survey, the 2QZ \citep[][]{Boyle00, Croom05}
    and allows division of our sample in luminosity and redshift bins
    while retaining statistical power. As shown in Sections 4 and 5, these
    new data complement the existing 2QZ and 2SLAQ quasar survey results,
    and together improve constraints on theoretical models.

    \begin{deluxetable*}{lrccrrr}
      \tablecolumns{7}
      \tablewidth{10cm}
      \tablecaption{The SDSS Spectroscopic Quasar Samples\label{tab:The_SDSS_minmax_z_MIs}}
      \tablehead{Sample      &  & Area            & Number    & $z_{\rm min}$ & $z_{\rm max}$ & $z_{\rm med}$ \\
        Description &  & /deg$^{2}$      & in sample &             &              &               }
      \startdata
          DR5Q                &              & $\approx5740$ &  77 429    &  0.078  & 5.414 & 1.538  \\ 
          \;\;\;\; " \;\;\;\; & $z \leq2.9$  &               &  71 375    &  0.078  & 2.900 & 1.372  \\ 
          \;\;\;\; " \;\;\;\; & $0.3\leq z \leq2.9$ &        &  69 692    &  0.300  & 2.900 & 1.400  \\ 
          \;\;\;\; " \;\;\;\; & $z \leq2.2$  &               &  65 660    &  0.078  & 2.200 & 1.278  \\
          \;\;\;\; " \;\;\;\; & $0.3\leq z \leq2.2$ &        &  63 977    &  0.300  & 2.200 & 1.306  \\
          \hline
          PRIMARY             &              & 5713         &  55 577     &  0.080  & 5.414 & 1.543  \\ 
          \;\;\;\; " \;\;\;\; & $z \leq2.9$  &              &  50 062     &  0.080  & 2.900 & 1.326  \\ 
          \;\;\;\; " \;\;\;\; & $0.3\leq z \leq2.9$  &      &  48 526     &  0.300  & 2.900 & 1.360  \\ 
          \;\;\;\; " \;\;\;\; & $z \leq2.2$  &              &  46 272     &  0.080  & 2.200 & 1.234 \\ 
          \;\;\;\; " \;\;\;\; & $0.3\leq z \leq2.2$  &      &  44 736     &  0.300  & 2.200 & 1.268 \\
          \hline 
          UNIFORM             &               & 4013        &  38 208     &  0.084  & 5.338 & 1.575 \\ 
          \;\;\;\; " \;\;\;\; &  $z \leq2.9$  &             &  33 699     &  0.084  & 2.900 & 1.319 \\ 
          \;\;\;\; " \;\;\;\; & $0.3\leq z \leq2.9$  &      &  32 648     &  0.300  & 2.900 & 1.234 \\ 
          \;\;\;\; " \;\;\;\; & $z \leq2.2$   &             &  31 290     &  0.084  & 2.200 & 1.354 \\ 
          \;\;\;\; " \;\;\;\; & ${\bf 0.3\leq {\it z} \leq2.2}$ && {\bf 30 239} &  {\bf 0.300}  & {\bf 2.200} & {\bf 1.269} \\  
          \enddata
          \tablecomments{\footnotesize The SDSS Spectroscopic Quasar Samples used in our analysis, with
          minimum, maximum and median redshifts. 
         The DR5Q is the catalogue presented in \citet{Schneider07}, while
         the PRIMARY and UNIFORM samples are described in Section 2.
         The results for the UNIFORM sample indicated in boldface are given in Section 4.}
    \end{deluxetable*}

    \begin{figure}
      \includegraphics[height=8.0cm,width=8.0cm]
      {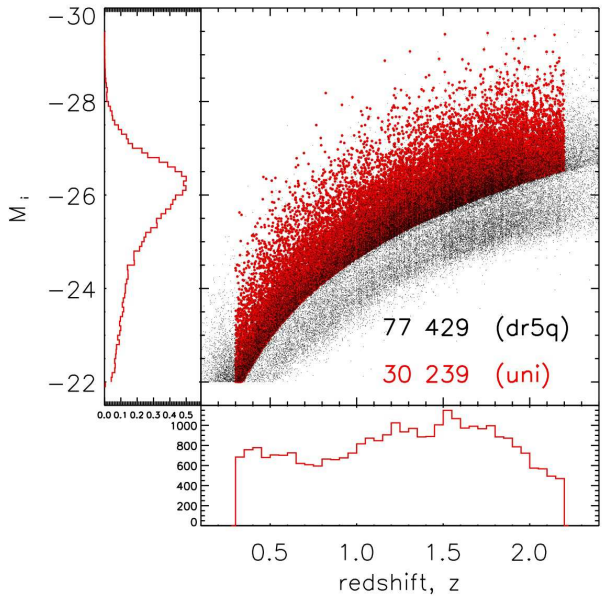}
      \centering
      \caption[The SDSS DR5 Quasar $L-z$ plane.]
              {The SDSS DR5 Quasar $L-z$ plane for the DR5Q (black points)
               and the UNIFORM sample (red points). 
               The affect of the $i=19.1$ magnitude limit
               can clearly be seen. $M_{i}$ is the $i$-band absolute 
               magnitude at the plotted redshift where we use the 
               $K$-correction given by Table 4 of \citet{Richards06}.
               } 
      \label{fig:SDSS_Lz_PRIMARY}
    \end{figure}
        
    We construct two subsamples from DR5Q. The first is designated as
    the ``PRIMARY'' Sample, which will include those objects in the DR5Q
    which were targetted as primary quasar candidates \citep{Richards02},
    having satisfied one, or more, of the \QSO$\,$, \HIZ$\,$ or \FIRST$\,$
    selections \cite[see][Section 4.8, for more details on these
    flags]{Stoughton02}. The SDSS quasar survey was designed to be
    complete in the primary sample, and no attempt was made at
    completeness for the quasars selected by other means. In total there
    are \hbox{55 577} quasars in the DR5Q that had their target flags set
    to one (or more) of these primary flags, with \hbox{46 272} quasars
    satisfying our high redshift limit (Table 1). 
    
    The SDSS quasar selection algorithm was in flux in the early part
    of the survey, and was only finalised after DR1. We define the
    ``UNIFORM'' sample to be those primary objects selected with this
    final version. The UNIFORM sample is flux limited to $i=19.1$ at
    $z\leq2.9$  and contains \hbox{38 208} objects, dropping to \hbox{31
      290} when a redshift cut of $z\leq2.2$ is applied. We show the
    distribution of objects in the redshift-luminosity plane for the full
    DR5Q and $0.30\leq z \leq 2.2$ UNIFORM sample in
    Figure~\ref{fig:SDSS_Lz_PRIMARY}. We will use both the PRIMARY and
    the UNIFORM samples in what follows, but will find inconsistent
    results between the two samples at scales $\gtrsim 60\hmpc$. This is
    investigated further in Appendix C.
    
    The quasar correlation function is sensitive to a number of
    potential systematic effects, including bad photometry and improperly
    corrected dust reddening. Since quasars are selected by their optical
    colors, we shall perform checks on both our PRIMARY and UNIFORM
    samples in Appendix C to see what effect regions with poor photometry
    \citep[as defined by ][]{Richards06, Shen07} has on our clustering
    measurements. 

    While all selection for the quasar sample is undertaken using
    dereddened colors (Richards et al 2001), if there remain systematic
    errors in the reddening model they can induce excess power into the
    clustering in a number of different ways. Appendix C describes how
    these effects affect our $\xis$ measurements and the interpretations
    based thereon. Briefly, we find that: the UNIFORM sample is the most
    stable sample for our studies; reddening and bad fields produce
    insignificant effects to our measurements; our results are insensitive
    to the choice of the upper bound of the integral in
    equation~(\ref{eq:wp_with_pi_max}) ($\pi_{\rm max}$, see Section 3.2)
    and the comoving $z_{\rm max}$ and fibre collisions are not a concern
    on the scales we investigate.

\section{Techniques}   
In this section we describe the techniques we shall use to calculate
the Quasar $z\leq2.2$ 2PCF. The interested reader is referred to the
comprehensive texts of \citet{Peebles80, Peebles93book,
Peacock99, Coles02} and \citet{Martinez02book} for full details on the 2PCF. 
 
    \subsection{Estimating the 2-Point Quasar Correlation Function}
    In practice, $\xi$ is measured by comparing the actual quasar
    distribution to a catalogue of ``random'' points, which have the same
    selection function, angular mask and radial distribution as the data,
    but are spatially distributed in a ``random'' manner  - i.e. are not
    clustered. The construction of this random sample shall be described
    in  Section~\ref{sec:randoms}.    
    
    We use the estimator of \citet{LS93} to calculate $\xi$, as this
    has been found to be the most reliable estimator for 2PCF studies
    \citep{Kerscher00}. Comparing our results to those using the
    estimators of \citet[DP,][]{Davis83} and \citet{Hamilton92},  we find the
    DP estimator causes systematic errors on large
    scales with too much power at $s \geq 40 \hmpc$, as this estimator  is
    less robust to errors in  the estimation of mean density. The LS
    estimator is given by, 

    \begin{eqnarray}
      \xi_{LS}(s) &=& 1 + \left(\frac{N_{rd}}{N} \right)^{2} \frac{DD(s)}{RR(s)} -
                      2   \left( \frac{N_{rd}}{N} \right) \frac{DR(s)}{RR(s)} \\
               &\equiv&   \frac{\langle DD \rangle - \langle 2DR \rangle + 
                          \langle RR \rangle}
                         {\langle RR \rangle}
      \label{eqn:lseq} 
    \end{eqnarray}
    Here $N$ and $N_{\rm rd}$ are the number of data and random points
    in the sample, DD$(s)$ is the number of data-data pairs with
    separation between $s$ and $s+\Delta s$ in the given catalogue,
    DR$(s)$ is the number of data-random pairs, and RR$(s)$ the number of
    random-random pairs. The angled brackets denote the suitably
    normalised pair counts, since we employ at least twenty times more
    random points than data in order to reduce Poisson noise. We choose
    our bins to be logarithmically spaced, with widths of
    $\Delta\log(s/\hmpc) =0.1$.

    The measurement of a quasar redshift will not only have a (large)
    component due to the Hubble expansion, but also components due to the
    intrinsic peculiar velocities and redshift errors associated with the
    individual quasar. The peculiar velocities can been seen in the
    redshift-space correlation function, both at small and large scales
    (see Section 4). However, as noted in \citet{Schneider07} and
    discussed in detail in \citet[][Appendix A]{Shen07}, quasar redshift
    determination can have uncertainties of $\sigma_{v}=500-1450 \kms$ and
    hence $\sigma_{z}=0.003-0.01$, and these redshift errors will
    dominate any determination of the peculiar velocity signal. 

    The real-space correlation function, $\xir$, is what would be
    measured in the absence of any redshift-space distortions. We can 
    measure $\xir$ by projecting out the effects of peculiar velocities
    and redshift errors along the line of sight.
    
    One can resolve the redshift-space separation, $s$, between two
    quasars into two components, $r_{p}$ and $\pi$, where $r_{p}$ is the
    separation between two objects {\it perpendicular} to the
    line-of-sight and $\pi$ is the separation {\it parallel} to the
    line-of-sight. Thus,  
    \begin{equation}
      s^{2} = r_{p}^{2} + \pi^{2} 
    \end{equation}
    (where $r_{p}\equiv \sigma$ is also found in the literature).
    The `2-D' redshift-space correlation function, $\xisp$, can be
    calculated as before, 
    \begin{equation}
      \xi_{\rm LS}(r_{p}, \pi) = \frac{   \langle  DD(r_{p},\pi) \rangle -    
        \langle 2DR(r_{p},\pi) \rangle +
        \langle  RR(r_{p},\pi) \rangle}
      {\langle  RR(r_{p},\pi) \rangle}
    \end{equation}
    where the bin sizes are now chosen to be 
    $\Delta\log (r_{p} / \hmpc) = \Delta\log (\pi / \hmpc) = 0.2$.
    
    Redshift-space distortions affect only the radial component  of
    $\xisp$; thus by integrating along the line-of-sight direction,
    $\pi$, we obtain the projected correlation function, 
    \begin{equation}
      w_{\rm p}(r_{p}) = 2 \int^{\infty}_{0} \xisp \, d\pi.
      \label{eq:wp_sigma_def}
    \end{equation}
    In practice we set the upper limit on the integral to be
    $\pi_{\rm max}=10^{1.8}=63.1 \hmpc$ and show that although varying
    this limit does cause some difference to the deduced $\wp$, it does
    not cause significant changes to the 2PCF over the  scales of interest
    for our studies (Appendix~\ref{sec:pi_max}).

    The integral in equation~\ref{eq:wp_sigma_def} can be rewritten in terms
    of $\xi(r)$ \citep{Davis83}, 
    \begin{equation}
      w_{\rm p}(r_{p}) = 2 \int^{\pi \rm{max}}_{0} \frac{r \, \xi(r) }
                                              {\sqrt{(r^{2} - r_{p}^{2})}} dr.
      \label{eq:wp_with_pi_max}
    \end{equation}
    If we assume that $\xir$ is a power law of the form, 
    $\xi(r) = (r / r_{0})^{-\gamma}$ (which, as we shall find later, is a 
    fair assumption), then equation~\ref{eq:wp_with_pi_max} can be 
    integrated analytically, such that with $\pi_{\rm max} = \infty$,  
    \begin{equation}
     w_{p}(r_{p})       = r_{0}^{\gamma} r_{p}^{1-\gamma}
                           \left[\frac 
                                 {\Gamma(\frac{1}{2}) \,  
                                  \Gamma(\frac{\gamma-1}{2})}
                                {\Gamma(\frac{\gamma}{2})}
                           \right]
                        \equiv r_{0}^{\gamma}  r_{p}^{1-\gamma}
                           A(\gamma),
       \label{eq:wp_div_sigma}
    \end{equation}
    where $\Gamma(x)$ is the Gamma function.

    In linear theory and in the absence of small-scale velocities and
    redshift errors, the redshift-space and real-space correlation
    function can be  related via
    \begin{equation}
      \xi(s)= \xi(r) \left({ 1+\frac{2}{3}\beta(z) + \frac{1}{5}\beta^{2}(z)} \right),
      \label{eq:xis_div_xir}
    \end{equation}
    where 
    \begin{equation}
      \beta(z) = \frac{\Omm(z)^{0.55}} {b(z)}
      \label{eq:beta_Omm_b}
    \end{equation}
    parametrizes the `flattening' at large scales of the correlation
    function due to the infall of matter from underdense to overdense
    regions. The value of $\beta(z)$ has traditionally been measured
    via fits to observed data \citep[e.g.][]{Kaiser87, Fisher94, Peacock01,
      Hawkins03, Ross07, Guzzo08}.

    \begin{figure}
      \includegraphics[height=8.0cm,width=8.0cm]
      {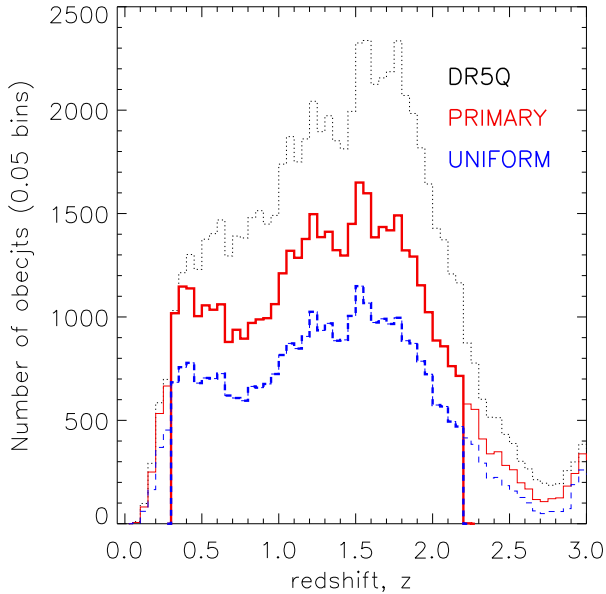}
      \centering
      \caption[The SDSS DR5 Quasar $N(z)$.]
      {The SDSS DR5 Quasar $N(z)$. The solid (red) histogram
        shows the quasar redshift distribution for the PRIMARY sample, while
        the dashed (blue) histogram shows the redshift distribution for the
        UNIFORM sample. The thin lines for both PRIMARY and UNIFORM 
        do not include the $0.3 \leq z \leq 2.2$ cuts. 
        As a comparison, the full DR5Q sample is given by the dotted 
        (black) histogram. }
      \label{fig:SDSS_Nofz}
    \end{figure}
    \subsection{Construction of the Random Catalogue}\label{sec:randoms} 
    As mentioned above, to calculate $\xi$ in practice, one needs to
    construct a random catalogue of points that mimics the data in every
    way, bar its clustering signal. The angular mask and completeness for the 
    PRIMARY and UNIFORM sample is described in detail in \hbox{Appendix A}.
    
    The radial distribution of the sample is measured from the data
    themselves. Figure~\ref{fig:SDSS_Nofz} shows the $N(z)$ distribution
    of the DR5Q quasars from our samples. We fit a tenth-order polynomial
    to both the PRIMARY and UNIFORM samples, which we use to generate the
    random sample redshift distribution. This method has proved reliable
    in previous quasar clustering studies \citep[e.g.][]{Croom05, daAngela08}.

    \subsection{Errors and Covariances}
    Recent studies \citep[e.g.][]{Scranton02, Zehavi02, Myers06,
      Ross07} have employed three main methods, {\it Poisson,
      Field-to-Field} and {\it Jackknife}  to estimate errors in correlation
    function measurements. The `simplest' of these is the Poisson error
    described by \citet{Peebles73}; this is the Poisson noise due to the
    number of pairs in the sample, 
    \begin{equation}
    	\sigma_{\rm Poi} = \frac{1+\xi(s)} {\sqrt{\rm{DD}(s)}}.
    \end{equation}
    This expression should be valid at smaller scales where the number
    of pairs is small and most pairs are independent \citep[i.e. few
    quasars are involved in more than one pair; ][]{Shanks94,
      Croom96}. However, as reported in \citet[][]{Myers05} and
    \citet{Ross07}, the Poisson error under-estimates measurement error
    when compared to e.g. the field-to-field or Jackknife errors at larger
    scales, where quasar pairs are not independent. For this work, we will
    not report any field-to-field errors, but instead concentrate on a
    jackknife resampling procedure in order to calculate the full
    covariance matrix, from which we will use just the diagonal elements.
    Full details of the jackknife procedure, including the geometry of the
    subsamples used and the justification for using only the diagonal
    elements are given in Section 4 and Appendix B.

\section{Results}

    \begin{figure}
      \includegraphics[width=8.5cm, height=8.0cm]
      {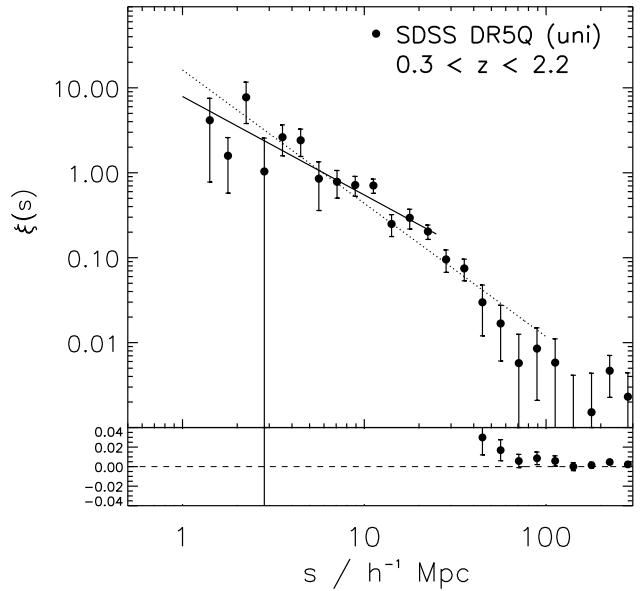}
      \centering
      \caption[The SDSS Quasar redshift-space 2PCF, $\xis$]
      {The SDSS Quasar redshift-space 2PCF, $\xis$, 
        from the UNIFORM sample (filled circles). 
        The solid line shows the best fit single power-law model over
        $1 \leq s \leq 25.0 \hmpc$, while the dotted line shows the
        best fit single power-law model over $1 \leq s \leq 100.0 \hmpc$.
        The lower panel shows the $\xis$ behaviour near zero on a linear 
        scale. The quoted errorbars are jackknife errors from the diagonal
        elements of the covariance matrix. }
      \label{fig:xis_DR5_UNI22}
    \end{figure}

    \begin{figure}
      \includegraphics[width=8.5cm, height=8.0cm]
      {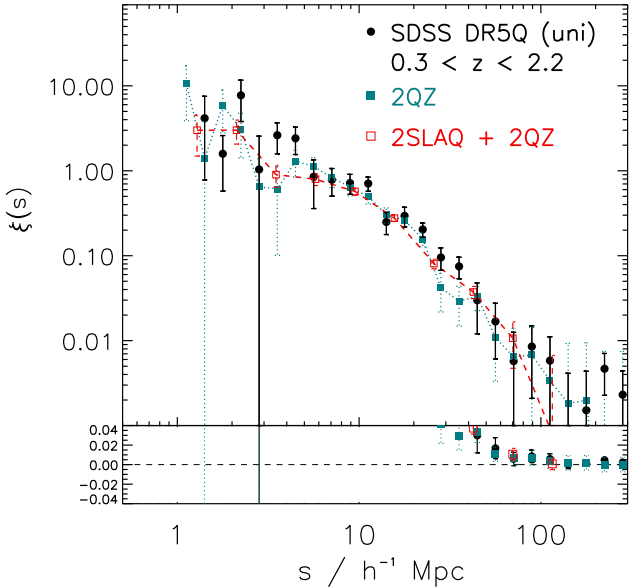}
      \centering
      \caption[The SDSS Quasar redshift-space 2PCF, $\xis$]
      {The Quasar redshift-space 2PCF, $\xis$, from the 
        UNIFORM sample as in Fig.~\ref{fig:xis_DR5_UNI22}.
        Also shown are the redshift-space correlation functions from 
        the 2QZ \citep[][]{Croom05}, shown as cerulean filled squares
        connected with a dotted line, and the 2SLAQ QSO survey 
        \citep[][]{daAngela08}, shown as the red filled squares connected
        by the dashed line. There is excellent agreement
        between the three surveys.}
      \label{fig:xis_DR5_quasars_UNI22_2QZ_2SLAQ}
    \end{figure}
    \subsection{SDSS Quasar Redshift-Space Two-Point Correlation  Function,
      $\xi(s)$ ($0.30 \leq z \leq 2.2$)}
    
    The two-point redshift-space correlation function for the SDSS
    DR5Q UNIFORM sample over the redshift interval $0.3<z<2.2$ is given in
    Figure~\ref{fig:xis_DR5_UNI22}. As described in Appendix B, the
    errorbars are jackknife errors from the diagonal elements of the
    covariance matrix, i.e. $\sigma_{i}^{2} = C_{ii}$. We justify this
    approach by considering that the covariance matrix is close to
    diagonal (Fig.~\ref{fig:regression_matrix}) and using just the
    diagonal elements of the covariance matrix produces results very close
    to that using the whole matrix, when fitting out to 25 $\hmpc$.  The
    off-diagonal elements of the covariance matrix are too noisy to be
    useful at large scales, and we therefore only use the diagonal elements 
    in all the fits and plots that follow.     
    
    We start by fitting a simple, single power-law model of the form
    in Equation~\ref{eq:xi_PL_model}. We find that a single power law with
    a redshift-space correlation length of $s_{0}=5.95\pm0.45 \hmpc$ and
    power-law slope of $\gamma_{s}=1.16^{+0.11}_{-0.08}$ provides an
    adequate description of the data over the scales $1.0 \leq s \leq 25.0
    \hmpc$ (solid line, Fig.~\ref{fig:xis_DR5_UNI22}). Here a value of
    $\chi^{2} = 11.5$ is obtained with 11 degrees of freedom (dof)
    giving $P$, the probability of acceptance (of our power law model to
    the data) of 0.402. A less suitable fit is found at larger scales due
    to the data falling below the power law. Over the range $1.0 \leq s
    \leq 100.0 \hmpc$, the best fit model has a similar correlation
    length, $s_{0}=5.90\pm0.30 \hmpc$ but a significantly steeper
    power-law slope, $\gamma_{s}=1.57^{+0.04}_{-0.05}$ (dotted line,
    Fig.~\ref{fig:xis_DR5_UNI22}). The $\chi^{2}$ for this model is 32.8
    with 15 dof and $P=5\times10^{-3}$. The data systematically deviate
    from the power-law fit, possibly due to the effects of
    redshift-distortions (on small scales), with a ``flattening'' of the
    data compared to the model at small, $s\lesssim5\hmpc$, scales and a
    steepening at large, $s\gtrsim40\hmpc$, scales - though a decline
    below a power-law at large scales is also expected from linear theory
    via the CDM real-space $\xi(r)$. 

    In Figure~\ref{fig:xis_DR5_quasars_UNI22_2QZ_2SLAQ}, we compare
    our results with the redshift-space correlation function $\xis$ from
    two other recent studies, the 2QZ \citep{Croom05} and the 2SLAQ QSO
    \citep{daAngela08} surveys. The analysis by \citet{daAngela08} uses
    data from both the 2QZ and 2SLAQ QSO surveys and thus the samples are
    not completely independent.
    
    The 2QZ and 2SLAQ QSO surveys both cover very similar redshift
    ranges to our $z<2.2$ sample. The 2QZ covers a much smaller area,
    $\approx750$deg$^{2}$, than the SDSS but has 2/3 as many quasars as
    our sample, since it reaches to a deeper limiting magnitude of $b_{\rm
      J}=20.85$ (corresponding to $g \approx 20.80$ and $i\approx
    20.42$). The 2SLAQ QSO survey has a smaller area yet, $\approx
    180$ deg$^{2}$, and reaches a magnitude deeper than the 2QZ to $g=21.85$
    $(i \approx 21.45)$ resulting in \hbox{8 500} quasars with $0.3 < z <
    2.2$.
    
    The agreement in the correlation function between surveys over $1
    \hmpc \leq s \leq 100 \hmpc$ scales is impressive but not necessarily
    unexpected, since we are essentially sampling the same type of objects
    i.e. luminous AGN, powered by supermassive black holes accreting at or
    near their Eddington limits \citep[][]{Kollmeier06, Shen08b}, quite
    possibly in similar mass environments (see Section 5). However, the
    samples have different luminosities, with mean $L_{\rm Bol, SDSS} =
    3.4 \times 10^{46}$ erg s$^{-1}$ (Table~\ref{tab:s_nought_comp_QSOs})
    compared with mean $L_{\rm Bol, 2QZ} \approx 1.3 \times 10^{46}$ erg
    s$^{-1}$  \citep[assuming $M_{b_{\rm J}}=-24.6$ and eqn. 27 from][for
    the 2QZ QSOs]{Croom05}, suggesting that variation in quasar luminosity
    is due to a variation in SMBH fueling, rather than a variation in SMBH
    mass (which maybe correlated to halo mass). We explore this luminosity
    dependence on clustering further in the companion paper
    \citep[][]{Shen09}. 
    
    \begin{figure*}
      \includegraphics[height=7.5cm,width=16.0cm]
      {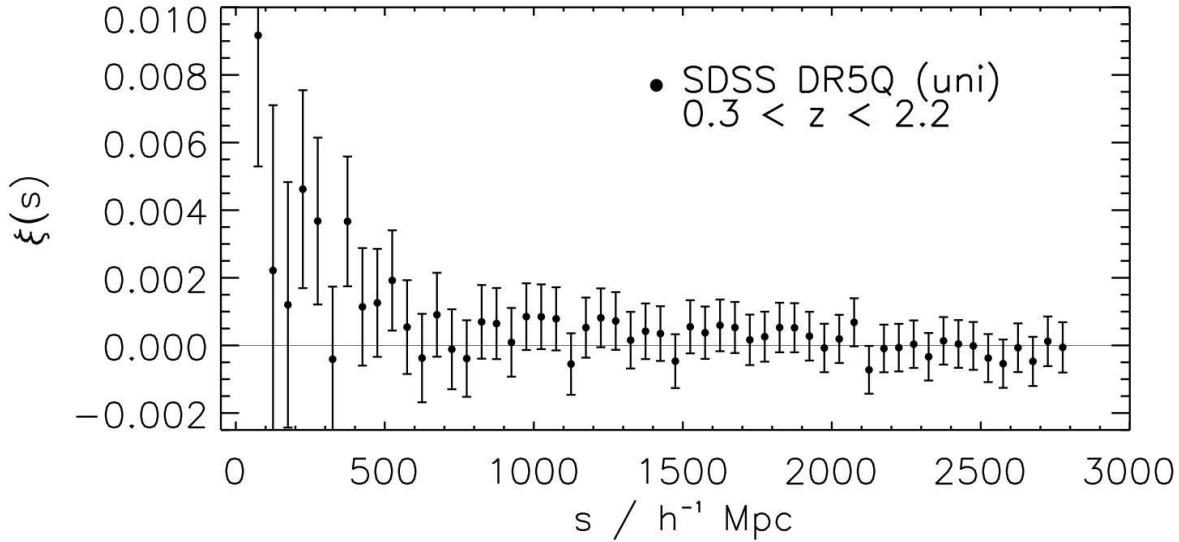}
      \centering
      \caption[The SDSS Quasar redshift-space 2PCF, $\xis$, large scales]
      {The SDSS Quasar redshift-space 2PCF, $\xis$, for our UNIFORM sample
        over the redshift range $0.3 \leq z \leq2.2$ at very large scales.
        Jackknife errors are plotted. The data are consistent 
        with $\xis=0$ out to scales of $s\sim3000 \hmpc$, which is
        the largest scales well-sampled by SDSS.} 
      \label{fig:xis_SDSS_UNI22_VERY_LARGE_SCALES}
    \end{figure*}  

    Figure~\ref{fig:xis_SDSS_UNI22_VERY_LARGE_SCALES} displays the
    very large scale $\xi(s)$ using the LS estimator.  We see that apart
    from one data point at $s\approx 400 \hmpc$, the redshift-space
    correlation function is within 1$\sigma$ of $\xis=0$ at scales greater
    than $\sim 300 \hmpc$. A $\chi^{2}$ test
    comparing the data to $\xis=0$ over the range of $100 \leq s <
    1000\hmpc$ and $100 \leq s < 3000\hmpc$ gives $\chi^{2}=8.2$ (18 dof,
    $P=0.975$) and $\chi^{2}=25.3$ (54 dof, $P=0.999$), respectively.  Our
    rms scatter is $\pm0.001$, which compares well to  the 2QZ value of
    $\pm0.002$; with a sample $\sim 50\%$ larger, we have roughly doubled
    the pair counts at these very large scales.  The dimensions of our
    sample do not allow us to probe separations beyond $3000 \hmpc$.

    \begin{figure}
      \includegraphics[width=8.4cm,height=6.0cm]
      {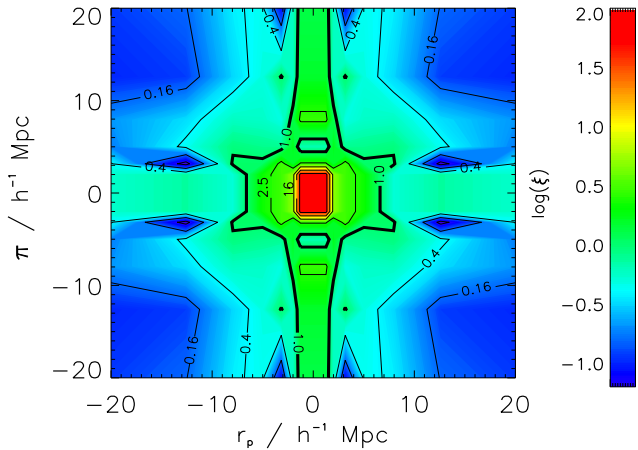}
      \centering
      \caption[The SDSS DR5 Quasar $\xisp$]
      {The SDSS DR5 Quasar $\xisp$. 
        The contours give lines of constant $\xi$ 
        having $\delta \log \xi=0.4$ between contours, with
        $\log \xi=1.6$ the highest value at the centre of 
        the plot. The thick contour is $\xi=1.0$. The actual
        $\xisp$ measurement is repeated and mirrored over 
        four quadrants to show the deviations from circular
        symmetry.}
      \label{fig:xi_si_pi_plot_810}
    \end{figure}
    \subsection{SDSS Quasar 2-D 2-Point Correlation Function, 
                            $\xisp$ ($0.30 \leq z \leq 2.2$)}
                          
    Figure~\ref{fig:xi_si_pi_plot_810} shows the SDSS DR5 Quasar 2-D
    redshift-space correlation function $\xisp$ for the UNIFORM sample,
    over $0.3\leq z \leq 2.2$. The redshift-space distortions in the
    clustering signal - seen as deviations from isotropy - are immediately
    apparent. At small $r_{p}$, the random peculiar motions and redshift
    errors of quasars cause an elongation of the clustering signal along
    the line-of-sight direction, $\pi$. This is the well-known
    ``Fingers-of-God'' effect \citep{Jackson72}. Cosmological information
    can be extracted from the Quasar 2D $\xi(r_{p},\pi)$ measurement
    \citep[e.g.][]{Hoyle02, daAngela05, daAngela08}. However, full
    treatment of the separation of the effects of large-scale `squashing'
    in $r_{p}$ (used to determine $\beta(z)$ in
    equation~\ref{eq:beta_Omm_b}) and the substantial contribution from
    the Fingers-of-God at small scales is left to a future paper.

    \begin{figure}
      \includegraphics[height=8.0cm,width=8.5cm]
      {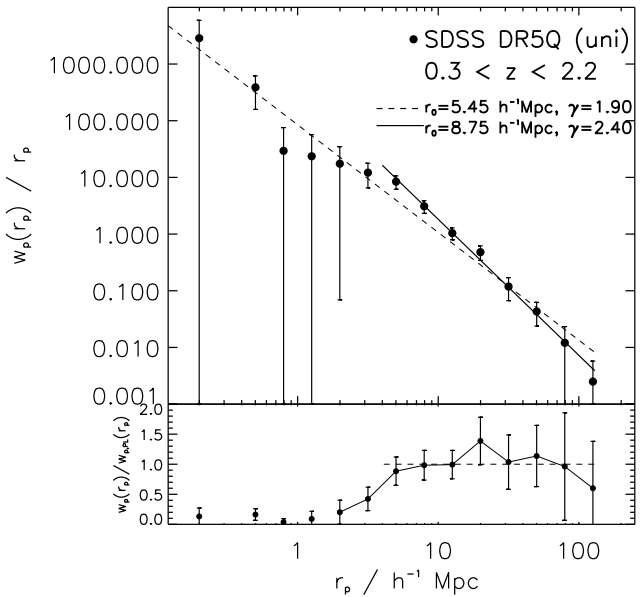}
      \centering
      \caption[The SDSS Quasar redshift-space 2PCF, $\wp$, large scales] 
      {The SDSS Quasar redshift-space 2PCF, $\wp$. The dashed line
        shows the best fit single power law to the data over our full range of
        scales, $0.1 < r_{p} < 130.0 \hmpc$. Here, the real-space correlation
        length is $r_{0}=5.45^{+0.35}_{-0.45} \hmpc$ with a slope
        $\gamma=1.90^{+0.04}_{-0.03}$. Restricting the range to $4.0 < r_{p} <
        130.0 \hmpc$, the best-fit values become $r_{0}=8.75^{+0.35}_{-0.50}
        \hmpc$ and $\gamma={2.40}^{+0.07}_{-0.10}$. The lower panel shows the
        ratio of the data divided by the power-law model over $4.0 < r_{p} <
        130.0 \hmpc$. }
      \label{fig:wp_SDSS_zlt2pnt2}
    \end{figure}
    \begin{figure*}
      \includegraphics[height=17.5cm,width=18.0cm]
      {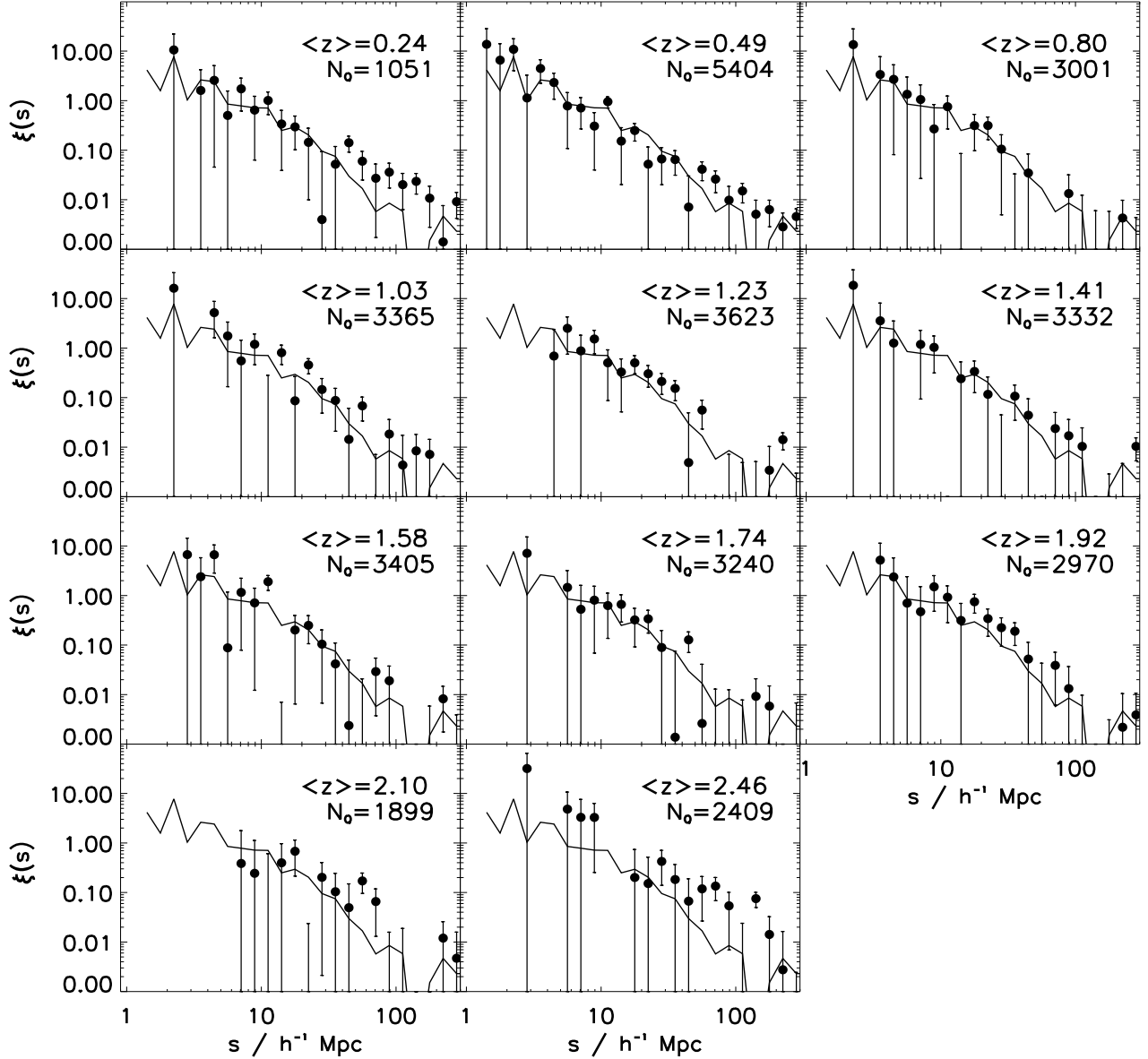}
      \centering
      \caption[The SDSS DR5 Quasar redshift-space 2PCF, $\xis$, 
      evolution with redshift, $z$]
      {The SDSS DR5 Quasar redshift-space 2PCF, $\xis$, and its
        evolution with redshift. All panels have the same scaling
        with the respective number of quasars, $N_{\rm Q}$, in each redshift 
        range given.  The thin (black) line in each 
        panel is $\xis$ for the full DR5Q UNIFORM sample, over $0.30<z<2.20$.
        The quoted errorbars are Poisson (see text for justification).}
      \label{fig:xis_DR5_quasars_3by3_for_evol}
    \end{figure*}
    \begin{figure*}
      \includegraphics[height=17.5cm,width=18.0cm]
      {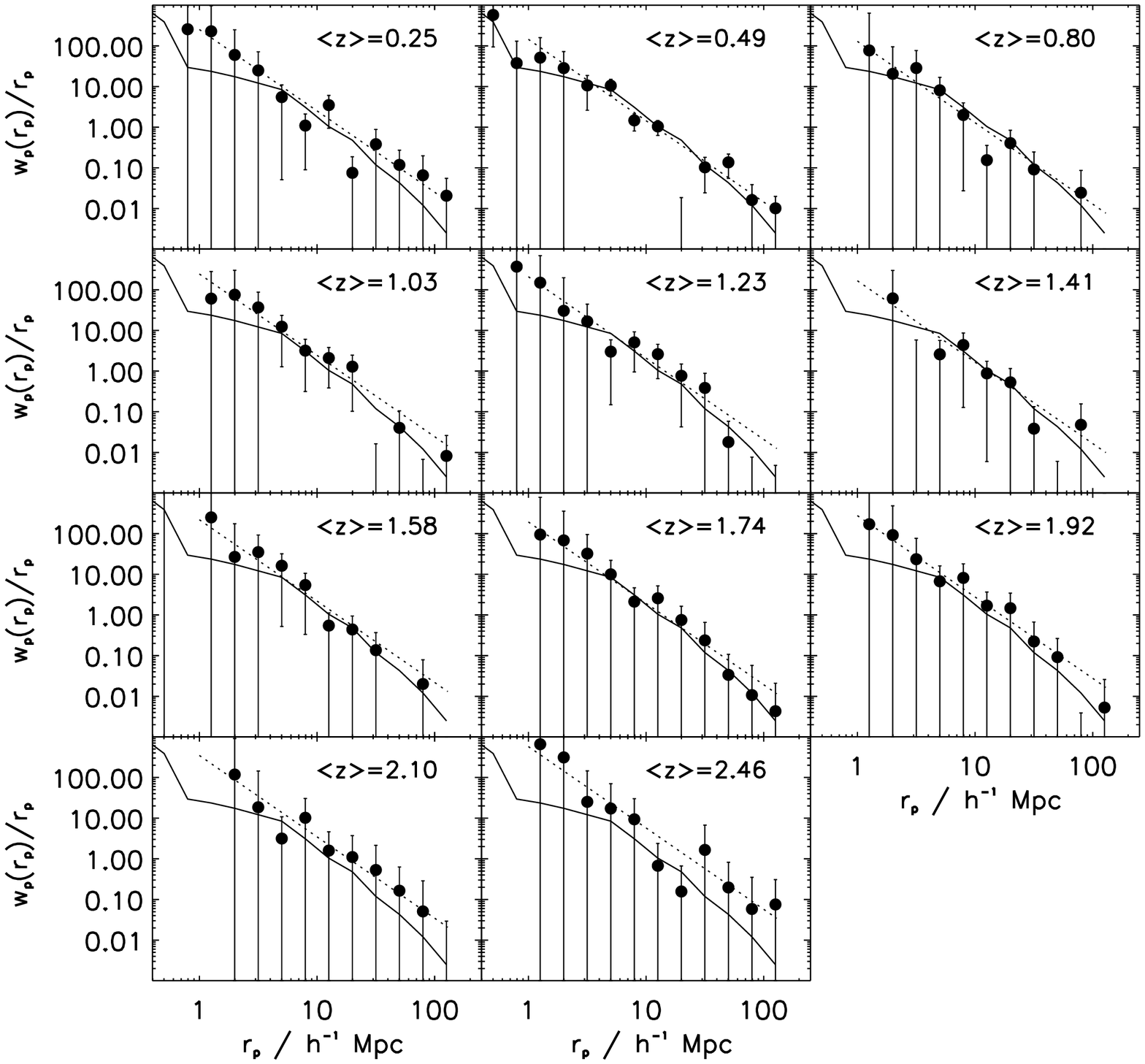}
      \centering
      \caption[The SDSS Quasar redshift-space 2PCF, $\wp$]
      {The SDSS DR5 Quasar projected 2PCF, $\wp$, and its
        evolution with redshift. 
        The solid line in each panel is $\wp/r_{p}$ 
        for the full DR5Q UNIFORM sample, over $0.30<z<2.20$.
        The quoted errorbars are scaled jackknifes (see text for details). 
        The relevant power-law fits as given in Table~\ref{tab:s_nought_comp_QSOs} 
        are shown by the dotted lines.}
      \label{fig:wp_sigma_DR5_quasars_3by3_for_evol}
    \end{figure*}
    \subsection{SDSS Quasar Projected 2-Point Correlation  Function}

    In Figure~\ref{fig:wp_SDSS_zlt2pnt2}, we show the projected
    2-point correlation function, $\wp$, calculated using
    equation~\ref{eq:wp_with_pi_max}. The reported error bars are
    jackknife errors, using the same jackknife area subsamples as for the
    $\xis$ calculation (Appendix B). Since we are fitting power laws of
    the form $\xi(r) = (r / r_{0})^{\gamma}$
    (equation~\ref{eq:wp_div_sigma}), we plot $\wp/r_{p}$ on the ordinate.
    We find the best fitting single power-law to the SDSS Quasar
    $\wp/r_{p}$ data to be $r_{0}=5.45^{+0.35}_{-0.45}\hmpc$ and
    $\gamma=1.90^{+0.04}_{-0.03}$ over our full range of scales, $0.1 <
    r_{p} < 130.0 \hmpc$. This provides a somewhat poor fit, giving a value
    of  $\chi^{2}=22.02$ with 12 degrees of freedom ($P=0.038$).  We
    remind the reader that due to fibre collisions, measurements at scales
    of $r_{p} \lesssim 1 \hmpc$ are biased low
    (Sec.~\ref{sec:fibre_collisions}). Restricting the range to $4.0 <
    r_{p} < 130.0 \hmpc$, we find the best fit power-law has an increased
    real-space correlation length of $r_{0}=8.75^{+0.35}_{-0.50} \hmpc$
    and a steeper slope of $\gamma={2.40}^{+0.07}_{-0.10}$. This power-law
    is a more acceptable fit, having $\chi^{2}=3.47$ with 6 dof
    ($P=0.748$).  We further suggest that the difference between the
    fitted results and their dependence on scale is due to a ``break'' in
    the $\wp/r_{p}$ measurements at $r_{p}\sim 2 - 5 \hmpc$. However, we
    are hesitant to offer an explanation of this behaviour of our
    measurements in terms of, e.g. the transition from the 1 to 2-halo
    regime \citep[cf. ][]{PMN04}. 
    
    Comparisons of our $\wp/r_{p}$ results to those of \citet{Shen07}
    for the $z>2.9$ redshift quasar measurements shows that the high
    redshift SDSS quasars have a much larger clustering amplitude than the
    lower redshift sample. The consequences of this are discussed in
    detail in \citet{Shen07}.

    \begin{deluxetable*}{cccccccccc}
      \tablecolumns{10} 
      \tablewidth{16cm} 
      \tablecaption{Evolution of the real-space correlation length \label{tab:s_nought_comp_QSOs}} 
      \tablehead{$z$-interval & $\bar{z}$ & $N_{\rm q}$ & $L_{\rm Bol}$  & $s_{0}/\hmpc$  &  $\gamma_{s}$  & $\chi^{2}$ & $\nu$ & $s_{0}/\hmpc$ &  $r_{0}/\hmpc$    \\
        &           &               & ($10^{46}$ erg s$^{-1}$) &  &     &           &      & $(\gamma_{s}=1.16)$ & $(\gamma=2.0)$}
      \startdata
      0.30,2.20  & 1.269 & 30 239 & 3.43 & $5.95\pm0.45$ & $1.16^{+0.11}_{-0.16}$ & 11.5 & 11 &  $5.95\pm0.45$ & $5.45^{+0.35}_{-0.45}$, $\gamma=1.90^{+0.04}_{-0.03}$ \\ 
      &       &        &      & $^{a}$$5.90\pm0.30$  & $1.57^{+0.04}_{-0.05}$ & 32.8 & 17 &  & $8.75^{+0.35}_{-0.50}$, $\gamma=2.40^{+0.07}_{-0.10}$  \\ 
      \hline
      0.08,0.30    & 0.235 &  1 051 &  0.16 & $ 6.90^{+1.35}_{-1.50}$ & $1.37^{+0.41}_{-0.31}$ &  2.6 &  9 & $6.20\pm1.55$        & 8.95$\pm$0.92    \\
      0.30,0.68    & 0.488 &  5 404 &  0.50 & $ 6.05^{+0.45}_{-0.65}$ & $1.67^{+0.23}_{-0.24}$ & 12.2 & 11 & $4.60^{+0.80}_{-0.75}$ & 6.78$\pm$0.56    \\
      0.68,0.92    & 0.801 &  3 001 &  1.39 & $ 7.05^{+1.15}_{-1.45}$ & $1.90^{+0.60}_{-0.60}$ &  6.2 &  8 & $5.40^{+1.60}_{-1.70}$ & 6.40$\pm$0.64    \\
      0.92,1.13    & 1.029 &  3 365 &  2.07 & $ 2.68^{+1.42}_{-1.28}$ & $0.57^{+0.14}_{-0.15}$ &  9.4 &  7 & $6.30^{+1.60}_{-1.65}$ & 8.80$\pm$0.84   \\
      1.13,1.32    & 1.228 &  3 623 &  2.83 & $ 7.10^{+1.45}_{-1.65}$ & $1.00^{+0.30}_{-0.25}$ &  2.5 &  6 & $7.75^{+1.50}_{-1.60}$ & 8.14$\pm$0.92   \\
      1.32,1.50    & 1.412 &  3 332 &  3.60 & $ 6.05^{+1.35}_{-1.85}$ & $2.13^{+0.87}_{-0.78}$ &  6.1 &  7 & $3.65^{+1.70}_{-1.80}$ & 7.26$\pm$0.93     \\
      1.50,1.66    & 1.577 &  3 405 &  4.40 & $ 6.10^{+1.40}_{-1.60}$ & $1.67^{+0.81}_{-0.50}$ & 13.5 &  8 & $4.65^{+1.55}_{-1.70}$ & 8.34$\pm$0.84   \\
      1.66,1.83    & 1.744 &  3 240 &  5.29 & $ 7.70^{+1.70}_{-1.90}$ & $1.11^{+0.39}_{-0.31}$ &  1.0 &  6 & $7.90^{+1.80}_{-1.85}$ & 7.83$\pm$0.71   \\
      1.83,2.02    & 1.917 &  2 970 &  6.63 & $ 7.43^{+2.37}_{-2.43}$ & $0.84^{+0.41}_{-0.30}$ &  2.5 &  7 & $8.70^{+2.05}_{-2.15}$ & 9.38$\pm$0.79    \\
      2.02,2.20    & 2.104 &  1 899 &  8.69 & $^{a}$$3.65^{+1.60}_{-1.85}$ & $1.10^{+0.29}_{-0.15}$ & 8.7 & 10 & $4.10^{+1.75}_{-1.90}$ & 10.50$\pm$0.96  \\
      2.20,2.90    & 2.462 &  2 409 & 11.64 & $10.75^{+2.15}_{-3.42}$ & $2.60^{+0.60}_{-1.10}$ &  0.2 &  4 & $7.15^{+5.50}_{-6.45}$ & 13.51$\pm$1.81      
      \enddata
      \tablecomments{\footnotesize Evolution of the redshift-space, $s_{0}$, and real-space,
          $r_{0}$, correlation lengths.  For $s_{0}$, both the correlation length
          and power-law slope were allowed to vary.  All redshift-space subsamples were fitted
          over the range $1.0 \leq s \leq 25.0 \hmpc$,  unless otherwise noted
          with $^{a}$, where the range was  $1.0 \leq s \leq 100.0 \hmpc$. 
          For $s_{0}$ we quote values both with floating and  fixed ($\gamma_{s}=1.16$) power-laws.
          For the full sample, $r_{0}$ and $\gamma$ are allowed to vary and fits 
          were performed over the scales $1.0 \leq r_{p} \leq 130.0 \hmpc$.
          While for the real-space subsamples, the calculation of $r_{0}$ was made by
          by fitting our wp(rp)/rp measurements using equation~\ref{eq:wp_div_sigma},
          over the  $1.0 \leq r_{p} \leq 130.0 \hmpc$, while keeping the power-law 
          index fixed at $\gamma=2.0$. 
          The bolometric luminosities are from the catalogue of \citet{Shen08b}.}
    \end{deluxetable*}

    \begin{figure*}
      \includegraphics[width=16.5cm, height=14.0cm]
      {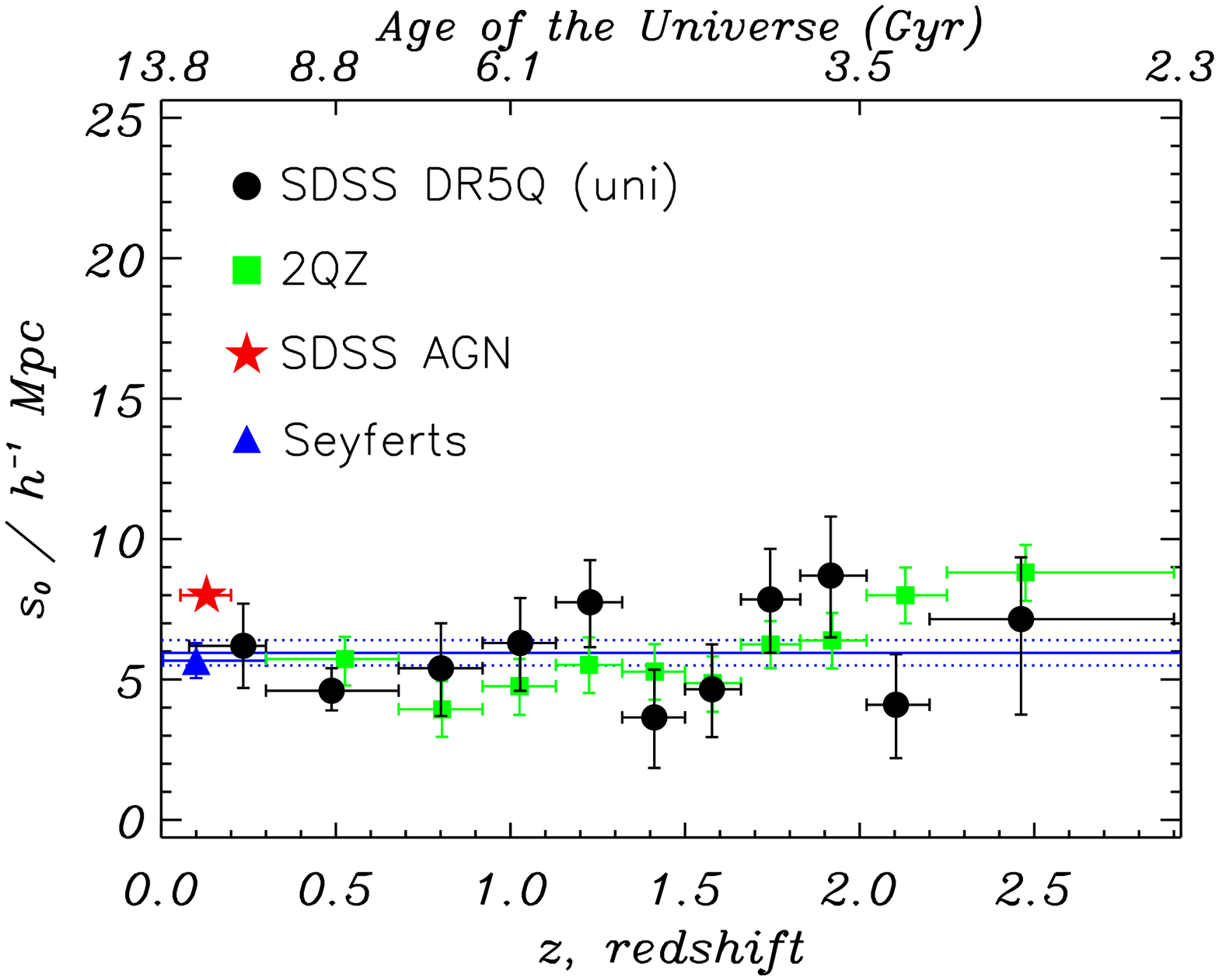}
      \centering
      \caption[Evolution of the redshift-space correlation length,$s_{0}$.]
      {Evolution of the redshift-space correlation length,
        $s_{0}$, up to redshift $z=2.9$. The (black) filled circles are from
        the DR5Q UNIFORM sample and the (blue) line gives the best-fit value
        for the whole sample with associated 1$\sigma$ errors. 
        The (green) filled squares are from the 2QZ \citep{Croom05}, while
        the (red) filled star is from a measurement of AGN clustering at 
        $z<0.2$ by \citet{Wake04}. The (blue) filled triangle is 
        the clustering measurement of Seyfert galaxies from \citet{Constantin06}.}
      \label{fig:s0_with_z}
    \end{figure*}
    \subsection{Evolution of the SDSS Quasar Correlation Function}
    
    Figures~\ref{fig:xis_DR5_quasars_3by3_for_evol}
    and~\ref{fig:wp_sigma_DR5_quasars_3by3_for_evol} present the evolution of
    the redshift-space, $\xis$, and the projected, $\wp$, 2PCF, using the
    SDSS DR5 UNIFORM Quasar sample. 
    
    We plot both $\xis$ and $\wp$ for sub-samples of the UNIFORM data,
    with the relevant redshift limits given in
    Table~\ref{tab:s_nought_comp_QSOs}. Here we choose the redshift slices
    so that we match those of the 2QZ Survey given by \citet{Croom05}. Our
    survey generally has 50\% more data in each redshift bin. However,
    since the 2QZ selects QSO candidates on the basis of their stellar
    appearance on photographic plates, low-redshift quasars with
    detectable host galaxies on the plate are preferentially rejected from
    the final 2QZ catalogue, and the SDSS Quasar UNIFORM sample has a
    larger proportion of low, $z\lesssim0.5$, redshift quasars\footnote{The 
    larger number of low redshift quasars in the SDSS sample is also at least in 
    part due to the contribution of the H$\alpha$ emission line in the $i$-band, 
    as well as host galaxy contribution at low redshift.}. We fit
    power-law models of the form given by equation~(\ref{eq:xi_PL_model}),
    over the ranges $1.0 \leq s \leq 25.0 \hmpc$ (except for our $2.02
    \leq z < 2.20$ bin, where to get finite constraints, we fit to $s_{\rm
      max}=100 \hmpc$). The best fit parameters and corresponding 1$\sigma$
    errors are given in Table~\ref{tab:s_nought_comp_QSOs}.
    
    In Fig.~\ref{fig:xis_DR5_quasars_3by3_for_evol}, we show
    measurements for $\xis$ for the redshift slices. The measurement of
    $\xis$ for the full redshift range measurement is given by the thin
    line in each panel. We show Poisson errors as these are approximately
    equal to jackknife errors on scales where the number of $DD$ pairs is
    less than the number of quasars in the (sub)sample (see
    Fig.~\ref{fig:error_ratio} and
    Appendix~\ref{sec:jackknife_appendix}). This scale is $s \sim 40-80
    \hmpc$ for the sub-samples given here. As such, the errorbars on
      scales $\gtrsim 80 \hmpc$ are most likely under-representative. The
    $\xis$ data show a trend to `lose' quasar-quasar $DD$ pairs at the
    smallest separation, as the redshift increases. Keep in mind that the
    length scale suppressed due to the 55'' fibre collision limitation
    increases from $s \sim 0.2 \hmpc$ at $z=0.5$ to $s \sim 1 \hmpc$ at
    $z=2$ (Fig.~\ref{fig:fibre_collisions}), giving rise to the apparent
    depression in the correlation function on small scales.     
    
    Fig.~\ref{fig:wp_sigma_DR5_quasars_3by3_for_evol} ($\wp/r_{p}$)
    has the same format as
    Fig.~\ref{fig:xis_DR5_quasars_3by3_for_evol}. However, here we show
    scaled jackknife errors, scaled using the $\wp/r_{p}$ Poisson and
    Jackknife error measurements from the full sample.  As can be seen
    from inspection, the errorbars plotted here have  generally larger
    magnitudes than the spread of the data alone. As such,  this leads to
    questioning whether this is due to the Poisson errors being  general
    under-estimates or the jackknifes being over-estimates of the  true
    error\footnote{The interested reader is pointed towards recent work
      by \citet{Norberg08}, who use large $N$-body simulations to
      investigate  different error estimators and the 2PCF for galaxy
      clustering.}.
    
    As a check, we calculate the ``summed variance'' Poisson errors,
    that is, we sum the variances of each bin included in the integral for
    $\wp$. This method returns smaller errors  than those shown in
    Fig.~\ref{fig:wp_sigma_DR5_quasars_3by3_for_evol},  especially at the
    smaller, $r_{p} < 10 \hmpc$, scales.      Re-assuringly, when we come
    to fit single power-law models for $\xi(r)$ (Sec. 5.2) in order to
    find values for the  real-space correlation length, $r_{0}$, the
    best-fit values we find from using the ``summed variance'' errors 
    are in good  agreement with those found using our $\wp/r_{p}$ ``averaging'' 
    method quoted in Table~\ref{tab:s_nought_comp_QSOs}. We explicitly note
    though that there still could be an issue with the jackknife errors
    being too large (for $\wp/r_{p}$) currently for reasons unknown. 
    
    Figure~\ref{fig:s0_with_z} shows the evolution of the
    redshift-space correlation length, $s_{0}$, with both redshift and the
    age of the Universe (adopting the cosmology given at the end of
    Section 1). Since there is a covariance between the best-fit $s_{0}$
    and $\gamma_{s}$, here we fix $\gamma_{s}$ to the best-fit value of
    the whole sample ($\gamma_{s}=1.16$)  and then measure the best-fit
    $s_{0}$.  We find the clustering strength remains reasonably constant
    with redshift out to $z\sim3$, which is equivalent to approximately
    80\% of the history of the Universe. This trend was also seen in
    Fig.~\ref{fig:xis_DR5_quasars_3by3_for_evol}.  The correlation length
    is measured to be $s_{0} = 5-7 \hmpc$ for bright, optically identified
    quasars in the SDSS, up to $z \sim 3$.

\section{Evolution of Galaxy, AGN and Quasar Clustering}

 \begin{figure*}
      \includegraphics[width=18.0cm, height=14.0cm]
      {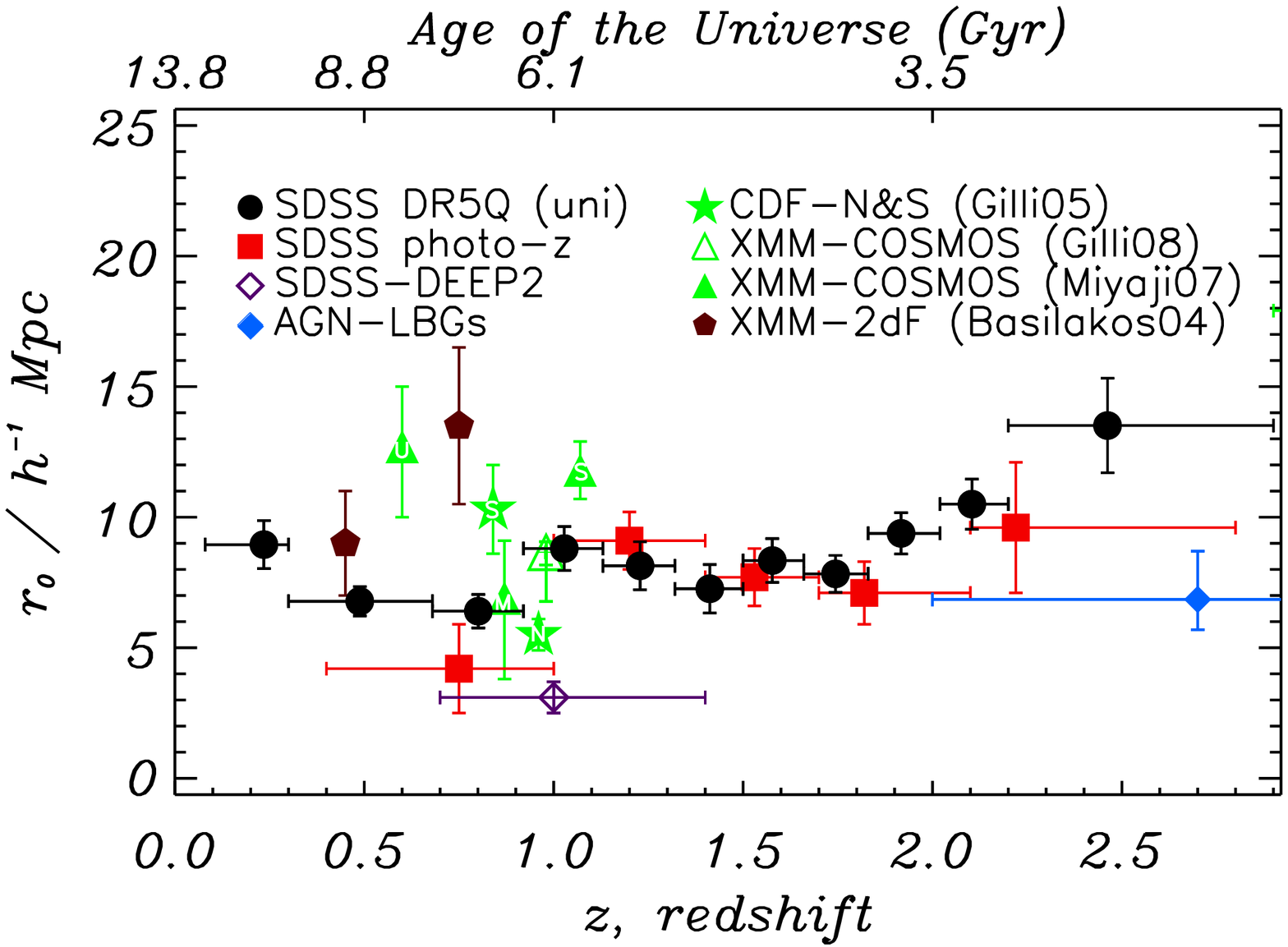}
      \centering
      \caption[Evolution of the real-space correlation length, $r_{0}$.] 
              {Evolution of the real-space correlation length, $r_{0}$,
                up to redshift $z=2.9$. The filled (black) circles are from 
                the DR5Q UNIFORM sample (this work); the filled (red) squares 
                are from the photometric sample of SDSS quasars from 
                \citet{Myers06}; The open (purple) diamond is 
                the quasar-galaxy cross-correlation from the DEEP2 Survey
                \citep{Coil07} and the solid (blue) diamond is the
                AGN-galaxy cross-correlation using Lyman Break galaxies
                \citep[LBGs, ][]{Adelberger05}.
                X-ray data from the {\it Chandra} Deep Fields 
                \citep{Gilli05} are shown by filled
                (green) 5-pointed stars (with the North and South fields 
                denoted `N' and `S' respectively); the {\it XMM-Newton}-2dF 
                survey \citep{Basilakos04} are indicated by filled (dark red) 
                pentagons; the {\it XMM-Newton}
                COSMOS survey \citep{Miyaji07}, is shown by filled
                (green) triangles,  with the `SFT' (0.5-2 keV), 'MED' (2-4.5 keV)
                and `UHD' (4.5-10 keV) band measurements denoted as S, M and U
                respectively. 
                Measurements from \citet{Gilli08}, 
                also using the {\it XMM-Newton} COSMOS survey, are given by the
                open (green) triangle.}
      \label{fig:r0_with_z}
    \end{figure*}
    \subsection{The Redshift-Space Evolution}

    In Figure~\ref{fig:s0_with_z}, we compare our measurements of the
    evolution of the redshift-space correlation length, $s_{0}$, to those
    recently published in the literature. We calculate our values for
    $s_{0}$ by fitting our $\xis$ measurements using
    equation~\ref{eq:xi_PL_model}. Motivated by the fits in
    Fig.~\ref{fig:xis_DR5_UNI22}, we hold the power-law index fixed at
    $\gamma_{s}=1.16$. The study of quasar clustering most comparable to
    our own is that presented by \citet{Croom05} for the 2QZ survey.  Our
    study using the SDSS DR5Q UNIFORM quasar sample and the 2QZ are in
    very good agreement over the full redshift range, given the associated
    uncertainties. However, in the SDSS DR5Q sample, we see very little,
    if any, evolution in the redshift-space correlation length even to
    $z\sim3$, whereas the 2QZ does show marginal evolution in $s_{0}$.
    The similarity of these results again suggests that quasar clustering
    only weakly depends on luminosity for the dynamical ranges probed in
    these samples, a topic discussed further in \citet{Shen09}. 

    The filled (red) star in Fig.~\ref{fig:s0_with_z} is from the
    study by \citet{Wake04} who use a sample of \hbox{13 605} narrow-line
    AGNs in the redshift range $0.055<z<0.2$ from the first Data Release
    of the SDSS \citep{Abazajian03}. They find that the AGN
    autocorrelation function is consistent with the observed galaxy
    autocorrelation function over $s=0.2-100 \hmpc$ scales. Furthermore,
    they show that the AGN 2PCF is dependent on the luminosity of the
    narrow [O III] emission line ($L_{\rm [O III]}$), with low $L_{\rm [O
      III]}$ AGNs having a higher clustering amplitude than high $L_{\rm [O
      III]}$ AGNs. This measurement suggests that lower activity AGNs reside
    in more massive DM haloes than do higher activity AGNs, as $L_{\rm [O
      III]}$ provides a good indicator of AGN fueling rate
    \citep[e.g.][]{Miller03, Kauffmann03}.  As such, it is
    interesting to note that our lowest redshift quasar clustering data
    point is, within the uncertainties, consistent with the measurement
    from \citet{Wake04}. We use the term `quasar' here loosely, as for our
    lowest redshift bin, the mean bolometric luminosity is
    $1.6\times10^{45}$ ergs s$^{-1}$, a factor of 20 lower than our full
    sample (Table~\ref{tab:s_nought_comp_QSOs}).
    
    \citet{Constantin06} study the clustering of specific classes of
    AGN, namely Seyert galaxies and LINERs (low-ionization nuclear
    emission-line regions) with the classes being separated on the basis
    of emission-line diagnostic diagrams \citep[e.g.][]{BPT, Kewley01}.
    They find that LINERs, which show the lowest luminosities and
    obscuration levels, exhibit strong clustering
    ($s_{0}=7.82\pm0.64\hmpc$), suggesting that these objects  reside in
    massive haloes and thus presumably have relatively massive  black
    holes that are weakly active or inefficient in their accretion,
    potentially due to the insufficiency of their fuel supply. Seyfert
    galaxies, however, have lower clustering, $s_{0}=5.67\pm0.62\hmpc$
    (Fig.~\ref{fig:s0_with_z}, blue triangle), are very luminous and show
    large emitting gas densities, suggesting that their black holes are
    less massive but accrete quickly and efficiently enough to dominate
    the ionization. Therefore, based on our lowest redshift clustering
    results, the stronger link for our low-luminosity `quasars' is to
    Seyfert galaxies rather than LINERs.

    \subsection{The Real-Space Evolution}

    In Figure~\ref{fig:r0_with_z}, we compare our measurements (black
    circles) of the evolution of the real-space correlation length,
    $r_{0}$, to those recently published in the literature. We calculate
    our values for $r_{0}$ by fitting our $\wp/r_{p}$ measurements using
    equation~\ref{eq:wp_div_sigma}, calculating an $r_{0}$ value at each
    separation where $\wp/r_{p}$ is non-zero, and reporting the standard
    error on the mean for these values in
    Table~\ref{tab:s_nought_comp_QSOs}.  Motivated by the fits in
    Fig.~\ref{fig:wp_SDSS_zlt2pnt2}, we hold the  power-law index fixed at
    $\gamma=2.0$, thus setting $A(\gamma=2)=\pi$
    (eqn.~\ref{eq:wp_div_sigma}).  We caution again however, that as can be
    seen from inspecting Fig.~\ref{fig:wp_sigma_DR5_quasars_3by3_for_evol}, 
    the scatter in the
    points is small compared to the quoted errorbars, and thus, this
    method may well under-estimate the errors associated with the 
    real-space correlation length.

    \citet{Myers06} reported a measurement of the clustering of quasars
    using $\sim\hbox{80 000}$ SDSS quasars photometrically classified from
    the catalogue of \citet{Richards04}. The $r_{0}$ measurements from
    \citet{Myers06} are given by the filled (red) squares in
    Fig.~\ref{fig:r0_with_z},  and are in very good agreement with our own
    data (we plot the data from their Table 1, from the `Deprojected
    $r_{0}$' section and the $0.75 \leq r < 89 \hmpc$ row).
    
    \citet{Coil07} calculate the cross-correlation between $\sim
    \hbox{30 000}$ redshift $0.7<z<1.4$ galaxies observed as part of the
    DEEP2 galaxy redshift survey \citep{Davis01, Davis03}, and quasars
    over the same redshift range. In total there are 36 SDSS quasars and
    16 quasars identified from the DEEP2 survey itself over the 3
    deg$^{2}$ covered by the DEEP2. \citet{Coil07} find that
    $r_{0}\sim3.4\pm0.7 \hmpc$ for the quasar-galaxy cross-correlation
    ($\xi_{QG}$). These authors measure $r_{0}\sim3.1\pm0.6 \hmpc$ for
    the inferred quasar clustering scale length, assuming that $\gamma$ is
    the same for the galaxy and the quasar samples and  the two samples
    trace each other perfectly, giving  $\xi_{QG} = \sqrt{\xi_{QQ}
      \times\xi_{GG}}$.  We show this measurement as an open (purple) diamond in
    Fig.~\ref{fig:r0_with_z}. Although still consistent with the
    low-redshift measurement of \citet{Myers06}, it is at odds with our
    measurements.  Determination of $\xi_{QQ}$ from the cross-correlation
    measurement assumes that the density fields traced by the galaxies and
    quasars, $\delta_{G}$ and $\delta_{Q}$ respectively,  are perfectly
    correlated spatially,      i.e. the correlation coefficient between
    the two is $r=+1$ \citep[e.g.][]{Blanton99, Swanson08}\footnote{The
      simplest and frequently assumed relationship between $\delta_{1}$ and
      $\delta_{2}$ is ``deterministic linear bias'', $\delta_{1} = b_{\rm
        lin} \delta_{2}$  where $b_{\rm lin}$ is a constant parameter,
      $\delta_{1}=\rho_{1}({\bf x}) / \bar{\rho}_{1} -1$ and
      $\delta_{2}=\rho_{2}({\bf x}) / \bar{\rho}_{2} -1$,
      e.g. \citet{Peebles80, Dekel99, Swanson08}.}.   Thus, as is quite
    plausible, if $z\sim 1$ quasars and galaxies sample the underlying
    mass density field differently, then one can reconcile the difference
    in correlation lengths by invoking a correlation coefficient that is
    modestly different from unity. 

    \citet{Adelberger05b, Adelberger05} studied the clustering of Lyman
    Break galaxies (LBGs) around $2 \lesssim z \lesssim 3$ AGN. The
    dynamic range in luminosity for this sample is nearly 10 magnitudes
    \citep[$16 \lesssim G_{\rm AB} \lesssim 26$, ][]{Adelberger05} and is
    thus much greater than for our SDSS DR5 UNIFORM sample. These authors
    report a value of $r_{0}=5.27^{+1.59}_{-1.36} \hmpc $ for a sample of
    38 AGN with central SMBH masses of $10^{5.8} < M_{\rm BH} / M_{\odot}
    < 10^{8}$ and $r_{0}=5.20^{+1.85}_{-1.16} \hmpc $ for a sample of 41
    AGN with $10^{8} < M_{\rm BH} / M_{\odot} < 10^{10.5}$. If we assume
    the correlation coefficient is $r=1$ and the power-law slopes are
    constant between samples, we find \citep[with  $r_{0, {\rm
        LBG-LBG}}=4.0\pm0.6\hmpc$ at $z=2.9$, ][]{Adelberger05c}  that $r_{0,
      \rm{AGN-AGN}}\approx 6.9 \hmpc$. This result is very broadly
    consistent with \citet{Myers06} but in  tension with our SDSS DR5
    UNIFORM results. \citet{Adelberger05} sample vastly different
    luminosity ranges than we do and find the clustering does not vary
    significantly with luminosity, immediately ruling out luminosity
    dependence  as an explanation of the different clustering
    amplitudes. Again, the assumption of perfect correlation is called
    into question, with a non-unity correlation coefficient $r$
    potentially reconciling both these and the \citet{Coil07} DEEP2
    results. 
    
    We also compare our results with clustering measurements of recent
    deep X-ray surveys, which are particularly well suited to finding
    intrinsically less luminous, potentially obscured objects at high
    redshift \citep{Brandt05}. An immediate caveat we place in the
    following comparison is that the SDSS DR5Q surveys $\sim4000$ deg$^2$,
    while the largest solid angle of the current deep X-ray surveys is of
    order 1 deg$^{2}$ and therefore the X-ray results are much more
    susceptible to cosmic variance. 
    
    \citet{Basilakos04} estimate $r_{0}$  using the angular
    autocorrelation function, $w(\theta)$, of hard, (2-8 keV) X-ray
    selected sources detected in a $\approx2$ deg$^{2}$ field using a
    shallow ($f_{\rm X}$[2-8 keV] $\approx 10^{-14}$ ergs cm$^{-2}$
    s$^{-1}$)  and contiguous {\it XMM-Newton} survey. The area surveyed
    consisted of 13 usable pointings, overlapping that of the 2QZ survey,
    and resulted in the detection of 171 sources.  Various models for the
    redshift distribution are given in \citet[][see their Table
    1]{Basilakos04}; for our comparison, we adopt the $r_{0}$ values
    calculated using $(\Omm,\Omlam)=(0.3,0.7)$, which either assume
    ``pure luminosity evolution'' \citep[PLE, ][]{Boyle98} or
    ``luminosity-dependent density evolution'' \citep[LDDE,][]{Ueda03}.
    As such, the PLE and LDDE models produce different mean redshifts of
    $\bar{z}=0.45$  and $\bar{z}=0.75$, respectively, for the AGN sample.
    \citet{Basilakos04}  find $r_{0}=9.0\pm2.0\hmpc$ for the PLE model and
    $r_{0}=13.5\pm3\hmpc$ for the LDDE model, fixing the power-law slope
    at $\gamma=2.2$. These observations are given by filled (dark red)
    pentagons in Fig.~\ref{fig:r0_with_z}.
    
    \citet{Gilli05} obtained a sample of nearly $260$ AGN in  the {\it
      Chandra} Deep Field North \citep[CDF-N, ][]{Alexander03, Barger03} and
    South \citep[CDF-S, ][]{Rosati02} with spectroscopic redshifts.  They
    report that in both fields the AGN have $\bar{z}\sim0.9$ and a median
    0.5-10 keV luminosity of  $\bar{L}_{X} \sim 10^{43}$ erg s$^{-1}$,
    i.e. in the local Seyfert galaxy luminosity regime. Correlation
    lengths and slopes of $r_{0}=5.5\pm0.6 \hmpc$, $\gamma=1.50\pm0.12$
    and $r_{0}=10.3\pm1.7 \hmpc$, $\gamma=5.5\pm0.6$  are found for the
    CDF-N and CDF-S respectively \citep[][their Table 2]{Gilli05}, shown
    as filled (green) stars in Fig.~\ref{fig:r0_with_z}. 

    \citet{Miyaji07} measured the angular  autocorrelation function of
    X-ray point sources detected by {\it XMM-Newton} in the $\sim 2$
    deg$^{2}$ COSMOS field \citep{Scoville07}. The measurements for the
    0.5-2 (SFT), 2-4.5 (MED) and 4.5-10 (UHD) keV bands are given by
    filled  (green) triangles in Fig.~\ref{fig:r0_with_z}.
    \citet{Gilli08} also report on the spatial clustering of AGN in the
    COSMOS field using  $\sim550$ spectroscopically identified AGN at a
    median redshift and 0.5-10 keV luminosity of $z =0.98$ and ${L}_{X}
    =6.3 \times 10^{43}$ erg s$^{-1}$ respectively. They find a value of
    $r_{0} = 8.65\pm0.5 \hmpc$  (Fig.~\ref{fig:r0_with_z}, open green
    triangle) and a power-law slope of $\gamma=1.88\pm0.07$. However, this
    result is affected by a coherent structure of 40 AGN at
    $z\sim0.36$. Removing this structure causes $r_{0}$ to drop to $\sim 6
    \hmpc$, similar to that of the previous deep X-ray AGN measurements.

    We find that our clustering measurements are in good agreement
    with the lower correlation lengths found by some of the deep X-ray
    surveys, e.g. \citet{Gilli05} for the CDF-N, \citet{Miyaji07}  for
    their MED (2-4.5 keV) band and XMM-COSMOS \citep{Gilli08} However,
    there remains much scatter in the deep X-ray data, potentially due to
    cosmic variance and the small samples used for these analyses. Thus,
    we use the method given by \citet{Somerville04} to estimate the
    ``relative cosmic variance'', $\sigma_{v}^{2} \equiv \frac{\langle
      N^{2} \rangle -  \langle N \rangle^{2}}{\langle N \rangle^{2}} -
    \frac{1}{\langle N \rangle}$, where $\langle N \rangle$ is the mean
    and $\langle N^{2} \rangle$ the variance of the probability
    distribution  function $P_{N}(V)$, which represents the probability of
    counting $N$ objects in volume $V$. The second term here is the
    correction for Poisson shot noise.  For the XMM-COSMOS study by
    \citet{Gilli08} we assume that the COSMOS area is 2 deg$^{2}$ and
    therefore the volume is  $\sim$ a few $\times 10^7 \hmpc$ \citep[from
    Fig. 1 in][]{Scoville07}.  We also assume a number density of
    $1.8\times10^{-4} h^{3} {\rm Mpc}^{-3}$ \citet[][Fig. 9]{Gilli08} for
    the COSMOS-XMM AGN, and a redshift of $z=1$, even though the range is
    known to be much wider.  This gives a bias of $b\sim2.2$ according to
    Fig. 3 of \citet{Somerville04}.  (Interestingly, this is very much in
    line with with what we find in the next section for the SDSS quasars
    at this redshift). Thus, the $\sigma_{v, \rm COSMOS}^{2}\approx 4
    \times 10^{-3}$,  which suggests that the COSMOS survey shouldn't be
    dominated by cosmic variance. However, we note that this value does
    not take into account the Poisson shot noise term, $1/\langle N
    \rangle$, which is likely to be significant considering  the
    relatively small number ($\sim$500) of objects in the XMM-COSMOS
    sample. For the Chandra Deep Fields, the cosmic variance is much
    greater due to the fact that the areas are (at least) an order of
    magnitude smaller.  (CDF-N is 0.13 deg$^2$; CDF-S is 0.04
    deg$^2$). This could well explain the difference between the two CDF
    clustering measurements (as has been discussed in the relevant
    studies).  We also note that the AGN-LBG study comes from an area of
    $\sim0.5$ deg$^{2}$ \citep{Steidel04} and so cosmic variance could
    potentially be an explanation for the difference seen in
    Fig.~\ref{fig:r0_with_z}.

    \begin{figure*}
      \includegraphics[width=16.5cm, height=14.0cm]
      {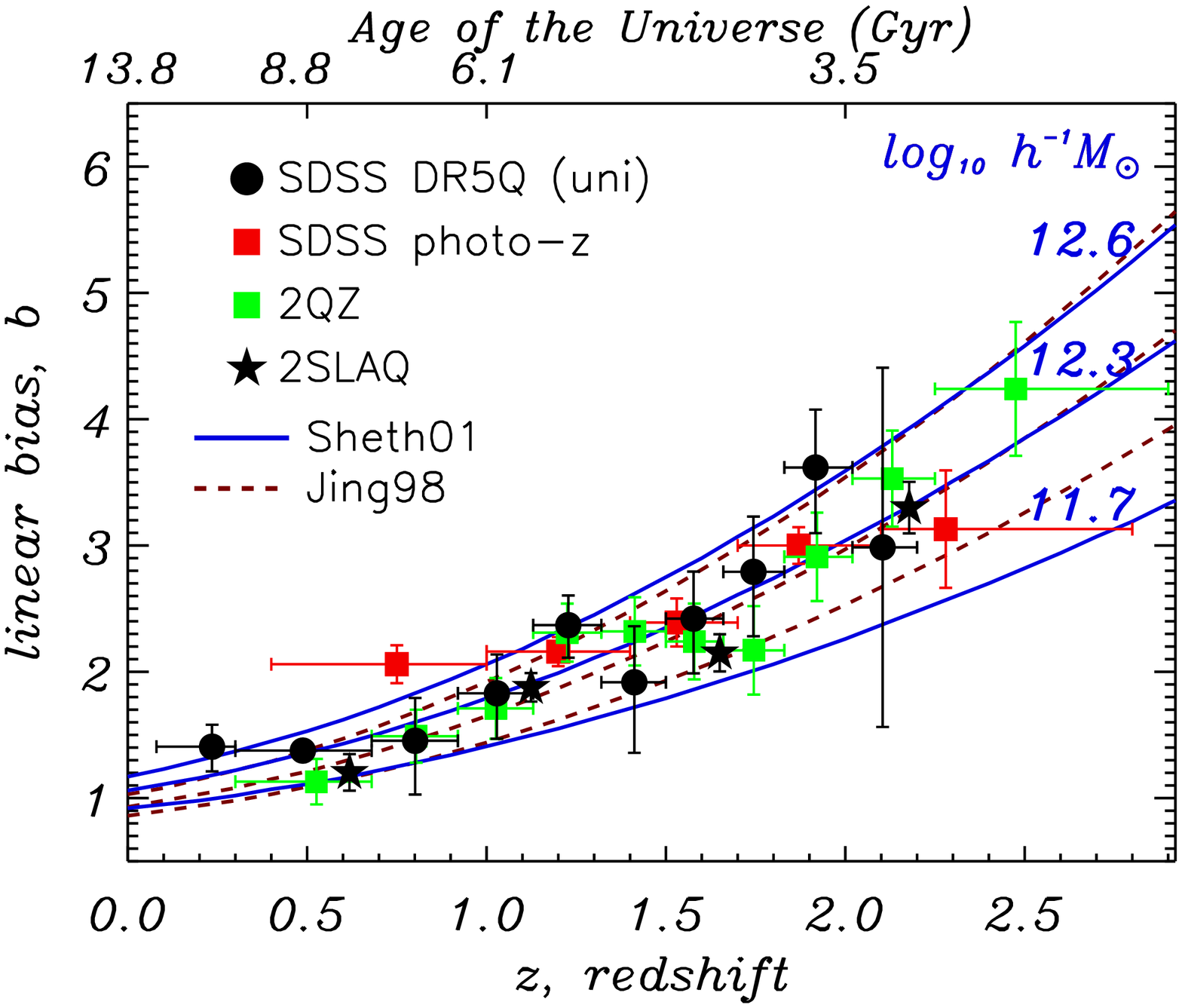}
      \centering
      \caption[Evolution of the linear bias, $b$ with redshift.]
      {Evolution of the linear bias of quasars, $b_{Q}$, with redshift, 
        to $z=3$.
        The (black) circles, are from the SDSS DR5Q UNIFORM sample
        (this work); 
        the (red) squares, from the photometric SDSS quasar measurements 
        \citep{Myers06}; 
        the (green) squares from the 2QZ survey \citep{Croom05}; 
        the (black) stars are from the 2SLAQ QSO survey \citep{daAngela08};
        The solid lines give dark halo masses from the models of \citet{Sheth01}
        with $\log h^{-1} M_{\odot}$ = 12.6, 12.3 and 11.7 from top to bottom.
        The dotted lines give dark halo masses from the models of \citet{Jing98}
        with $\log h^{-1} M_{\odot}$ = 12.3, 12.0 and 11.7 from top to bottom.
      }
      \label{fig:bias_with_z3}
    \end{figure*}
    
    \begin{deluxetable}{cccc}
      \tablecolumns{4}
      \tablewidth{8cm}
      \tablecaption{}
      \tablehead{$<z>$ &  $\bar{\xi}_{Q}(s,z)$ & $\bar{\xi}_{\rho}(r,z)$ &  $b$}
      \startdata
      \hline
      1.27     &  $0.391\pm0.011$  &  0.069  &  $2.06\pm0.03$ \\
      \hline
      0.24     &  $0.462\pm0.104$  &  0.176  &  $1.41\pm0.18$ \\
      0.49     &  $0.363\pm0.028$  &  0.138  &  $1.38\pm0.06$ \\
      0.80     &  $0.311\pm0.133$  &  0.104  &  $1.45\pm0.38$ \\
      1.03     &  $0.383\pm0.118$  &  0.085  &  $1.83\pm0.33$ \\
      1.23     &  $0.524\pm0.095$  &  0.072  &  $2.37\pm0.25$ \\	
      1.41     &  $0.309\pm0.134$  &  0.062  &  $1.92\pm0.50$ \\		
      1.58     &  $0.411\pm0.119$  &  0.054  &  $2.42\pm0.40$ \\	
      1.74     &  $0.472\pm0.141$  &  0.049  &  $2.79\pm0.47$ \\	
      1.92     &  $0.674\pm0.166$  &  0.043  &  $3.62\pm0.49$ \\
      2.10     &  $0.425\pm0.442$  &  0.039  &  $2.99\pm1.42$ \\	
      \hline
      \hline
      \enddata
      \tablecomments{\footnotesize The evolution of the linear bias for the SDSS Quasar UNIFORM sample.}
      \label{tab:bias_evol}
    \end{deluxetable}
    \subsection{Evolution of Bias}
    One key reason for measuring the correlation function as a
    function of redshift, $\xi(s,z)$, is to determine the linear bias,
    $b$, defined by the model of equation~(\ref{eq:bias}).  We shall assume
    that $b$ is independent of scale on the scales and redshift range
    under investigation here.\footnote{The precise way in which
      galaxies/luminous AGN trace the underlying matter distribution is
      still poorly understood. \citet{Blanton06}, \citet{Schulz06},
      \citet{Smith07} and \citet{Coles07} all suggest that bias is
      potentially scale dependent. We do not take this into account in the
      current analysis.} We follow the method in \citet{Croom05} and
    \citet{daAngela08} to determine $b$  using our redshift-space
    correlation function $\xi(s,z)$ measurements from Section 4. 
    
    In order to minimize non-linear effects e.g. redshift-space distortions, 
    we shall use the volume-averaged correlation function, $\bar{\xi}$, defined as
    \begin{eqnarray}
      \bar{\xi} &=& \frac{\int_{s_{\rm min}}^{s_{\rm max}} 4\pi s'^{2} \xi(s') ds'} 
                         {\int_{s_{\rm min}}^{s_{\rm max}} 4\pi s'^{2} ds'} \\
                &=& \frac{3}{ (s^{3}_{\rm max} - s^{3}_{\rm min})}  
                    \int_{s_{\rm min}}^{s_{\rm max}}  \xi(s') s'^2 ds'.
      \label{eq:xi_bar_jaca}
    \end{eqnarray} 
    where $s_{\rm min}=1.0 \hmpc$ is set in practice. Unless
    explicitly stated otherwise, $s_{\rm max}$ is always chosen to be $20
    \hmpc$, so that non-linear effects in the sample should be small due
    to the $s^{2}$ weighting and for ease of comparison with
    \citet{Croom05} and \citet{daAngela08}. In the linear regime, the
    $z$-space and real-space correlation functions can be given by
    equation~(\ref{eq:xis_div_xir}). Thus we combine
    equations~(\ref{eq:bias}) and~(\ref{eq:xis_div_xir}), and recognise
    that $\beta = \Omm^{0.55}/b$ leaves us with a quadratic equation
    in $b$.  We are assuming a flat, cosmological-constant model and hence
    the effective exponent of $\Omm$ is 0.55 \citep{Linder05,
      Guzzo08} rather than 0.6, suggested by \citet{Peebles80}, although we
    find this makes virtually no difference to our bias
    measurements. Solving the quadratic in $b$ leads to
    \begin{equation}
      b(z)  =  \sqrt{ \frac{\bar{\xi}_{Q}(s,z)} {\bar{\xi}_{\rho}(r,z)}
                     -\frac{4\Omm^{1.1}(z)}{45} 
                    }
                     -\frac{\Omm^{0.55}(z)}{3}.
      \label{eq:bias_xi_20}
    \end{equation}
    We now use our measured $\bar{\xi}_{Q}(s,z)$ together with a
    theoretical estimate of $\bar{\xi}_{\rho}(r,z)$ and $\Omm(z)$ to
    determine the bias.
    
    To estimate $\bar{\xi}_{\rho}(r,z)$, we follow \citet{Myers07a}
    and \citet{daAngela08}, and use the non-linear estimate of $P(k)$
    given  by \citet{Smith03}. The models of \citet{Smith03} predict the
    non-linear power spectrum of dark matter for a range of CDM
    cosmologies over a wide range of scale. We thus Fourier transform
    these $P(k)$ models and integrate over $s= 1 - 20 \hmpc$ to compute
    $\bar{\xi}_{\rho}(r,z)$. The cosmological parameters used in our
    chosen model are  $\Omm(z=0)=0.3$, $\Omlam(z=0)=0.7$, $\Gamma = 0.17$
    and $\sigma_{8} = 0.84$.  We find the simple form, 
    \begin{equation}
      \bar{\xi}_{\rho}(r,z) =  [A \exp(Bz) + C]\bar{\xi}_{\rho}(r, z=0)
    \end{equation}
    where $A= 0.2041$, $B=-1.082$, and $C=0.018$ models the
    evolution of $\bar{\xi}_{\rho}(r,z)$ extremely well, for $1\hmpc \leq
    s \leq 20 \hmpc$.
    
    At the mean redshift of our survey, $\Omm(z=1.27)=0.81$,  we find
    $b_{Q}(z=1.27)=2.06\pm0.03$ from the full SDSS DR5Q UNIFORM sample.
    The values for our redshift sub-samples are shown as filled circles in
    Fig.~\ref{fig:bias_with_z3} and are given in
    Table~\ref{tab:bias_evol}.  We estimate our errors by using the
    variations in $\bar{\xi}(s)$ from our 21 jackknife estimates, scaled
    using the number of $DD$ pairs in each redshift slice subsample.
    Previous measurements from the 2QZ Survey \citep[filled green circles,
    ][]{Croom05}, the 2SLAQ QSO Survey \citep[open black stars,
    ][]{daAngela08} and photometrically selected SDSS quasars
    \citep[filled red squares, ][]{Myers07a} are again in excellent
    agreement with our data. We compare these bias estimates with various
    models in Section 5.4. 
    
    Having measured $b(z)$ and assuming a cosmological model, 
    we can infer the parameter $\beta(z)$ using equation~\ref{eq:beta_Omm_b}.
    The space density of quasars is much smaller than that of
    galaxies, so the errors on the clustering measurement (e.g. $\xisp$) are
    much larger than for galaxy surveys \citep[cf.][]{Hawkins03,
      Zehavi05b, Ross07, Guzzo08}. Furthermore, as discussed in Section
    4.2, we have not included the effects from the ``Fingers-of-God''  in
    the present calculation of $\beta(z)$  but the
    peculiar velocities at small (transverse $r_{p}$) scales will very
    strongly affect the measured redshift distortion value of 
    $\beta$ \citep[][]{Fisher94, daAngela05}. 
    With $b(z=1.27)=2.06\pm0.03$ and $\Omm(z=1.27)=0.81$
    we find $\beta(z=1.27)=0.43$, but for the reasons given above 
    we present no formal error bar. 
    This result is consistent with the values of $\beta(z)$, 
    measured from redshift-space distortions in the 2QZ survey,
    $\beta(z = 1.4) = 0.45^{+0.09}_{-0.11}$ \citep{Outram04} and
    $\beta(z = 1.4) = 0.50^{+0.13}_{-0.15}$ \citep{daAngela05}.

    \subsection{Models of bias and dark matter halo mass estimation}
    We now compare our bias measurements with those of recent models for
    the relationship of quasars to their host haloes. 
    
    The fitting formula of \citet[][]{Jing98}, which is  derived from
    $N$-body simulations and assumes spherical collapse for the formation
    of haloes, is plotted in Fig.~\ref{fig:bias_with_z3} (dashed lines)
    with the assumed halo masses (top to bottom) $M_{\rm DMH}=
    2.0\times10^{12} h^{-1} M_{\odot}$,  $1.0\times10^{12} h^{-1}
    M_{\odot}$ and $5.0\times10^{11}h^{-1} M_{\odot}$, respectively.
    With the \citet{Jing98} model, we find the halo mass at which a
    `typical SDSS quasar' inhabits remains {\it constant} (given
    associated errors) with redshift, at a value of a  $M_{\rm DMH}\sim
    1\times10^{12} h^{-1} M_{\odot} $. 
    
    By incorporating the  effects of non-spherical collapse for the
    formation of dark matter haloes, \citet{Sheth01} provide fitting
    functions for the halo bias, which are also shown in
    Fig.~\ref{fig:bias_with_z3} (solid lines). Here, the three assumed
    halo masses of (top to bottom) $M_{\rm DMH}= 4.0\times10^{12} h^{-1}
    M_{\odot}$, $2.0\times10^{12} h^{-1} M_{\odot}$ and
    $5.0\times10^{11}h^{-1} M_{\odot} $, respectively are plotted.
    Comparing our results to the \citet{Sheth01} models, we again find the
    host dark matter halo mass is constant with redshift, at a value of a
    $M_{\rm DMH}\sim 2\times10^{12} h^{-1} M_{\odot}$; this mass does not
    significantly change from $z\sim2.5$ to the present day, i.e. over
    80\% the assumed age of the Universe. Therefore, as dark matter halo
    masses generally grow with time, the ratio of the halo mass for a
    typical quasar to the mean halo mass at the same epoch drops as one
    approaches redshift $z=0$. Since the ``non-spherical collapse'' model
    is likely to be more realistic, and for ease of comparison with
    previous results, we quote the \citet{Sheth01} halo mass value from
    here on.     

    Our values of halo masses of $M_{\rm DMH}\sim 2 \times10^{12}
    h^{-1} M_{\odot}$ found for the SDSS quasars compare very well to
    those of \citet{Padmanabhan08}, who find a similar value for low
    ($z<0.6$) SDSS quasars.  \citet{Croom05} also find a constant, but
    slightly higher value of $M_{\rm DMH}= 3.0\pm1.6\times10^{12} h^{-1}
    M_{\odot}$, by using the \citet{Sheth01} prescription,  over the
    redshift range $0.3<z<2.9$ for the 2QZ. \citet[][]{daAngela08} also
    find $M_{\rm DMH} \sim 3.0 \times10^{12} h^{-1} M_{\odot}$ but recall
    this analysis uses data from both the 2QZ and 2SLAQ QSO surveys.
    \citet{Myers07a} provide halo masses (also using the  \citet{Sheth01}
    prescription) for two cosmologies and we take their $\Gamma = 0.15$,
    $\sigma_{8} = 0.8$ model as this is closer to our own assumed
    cosmology. Again no evolution in the halo mass is found from
    $z\sim2.5$, but the \citet{Myers07a} value of $M_{\rm DMH} = (5.2 \pm
    0.6) \times 10^{12} h^{-1} M_{\odot}$ is appreciably higher than our
    results. \citet[][]{PMN04} applying a halo occupation distribution
    (HOD) model to the 2QZ data, find $M_{\rm DMH} \sim 1 \times 10^{13}
    h^{-1} M_{\odot}$. This is roughly an order of magnitude higher than
    the values we report and indeed at least double that of the other
    values found in the literature for luminous quasars. The
    \citet[][]{PMN04} value is in line with $M_{\rm DMH}\sim 1 - 2 \times
    10^{13} h^{-1} M_{\odot}$ which is the halo mass found for both the
    most luminous quasars or those that are FIRST-detected (i.e. radio
    loud) in the SDSS DR5Q at $z< 2.5$ \citep{Shen09}.  Thus we suggest
    some caution should be taken in the \citet[][]{PMN04} result but note
    that these authors use the halo bias formula from
    \citet{Sheth99} which is likely to contribute to  some of the
    discrepancy.  \citet{Shen07} find a {\it minimum} halo mass of $M_{\rm
      DMH} = 2-3\times10^{12} h^{-1} M_{\odot}$, and  $M_{\rm DMH} =
    4-6\times10^{12} h^{-1} M_{\odot}$, for the very luminous, higher
    clustered, high redshift SDSS quasars at $2.9\leq z\leq3.5$ and
    $z\geq3.5$ respectively. 
    
    Using semi-analytic models for BH accretion and quasar emission
    developed  on top of the Millennium Simulation \citep{Springel05},
    \citet{Bonoli08} provide a direct theoretical companion work to our
    observational study and that of Shen et al. (2009). These authors
    reproduce our findings that luminous AGNs i.e. the SDSS $z<2.2$
    quasars (with $L_{\rm Bol}\sim L^{*}$), are hosted by dark matter
    haloes with a narrow mass range centred around a few $10^{12} h^{-1}
    M_{\odot}$. The results of \citet[][e.g. their Fig. 13]{Bonoli08}
    might however suggest a slightly stronger redshift evolution for the
    host halo mass at $z<2$ than is given by our observational data, but
    this is hard to confirm given the associated errors on both the
    observational data and theoretical models.

    \begin{figure}
      \includegraphics[width=8.5cm, height=7.5cm]
      {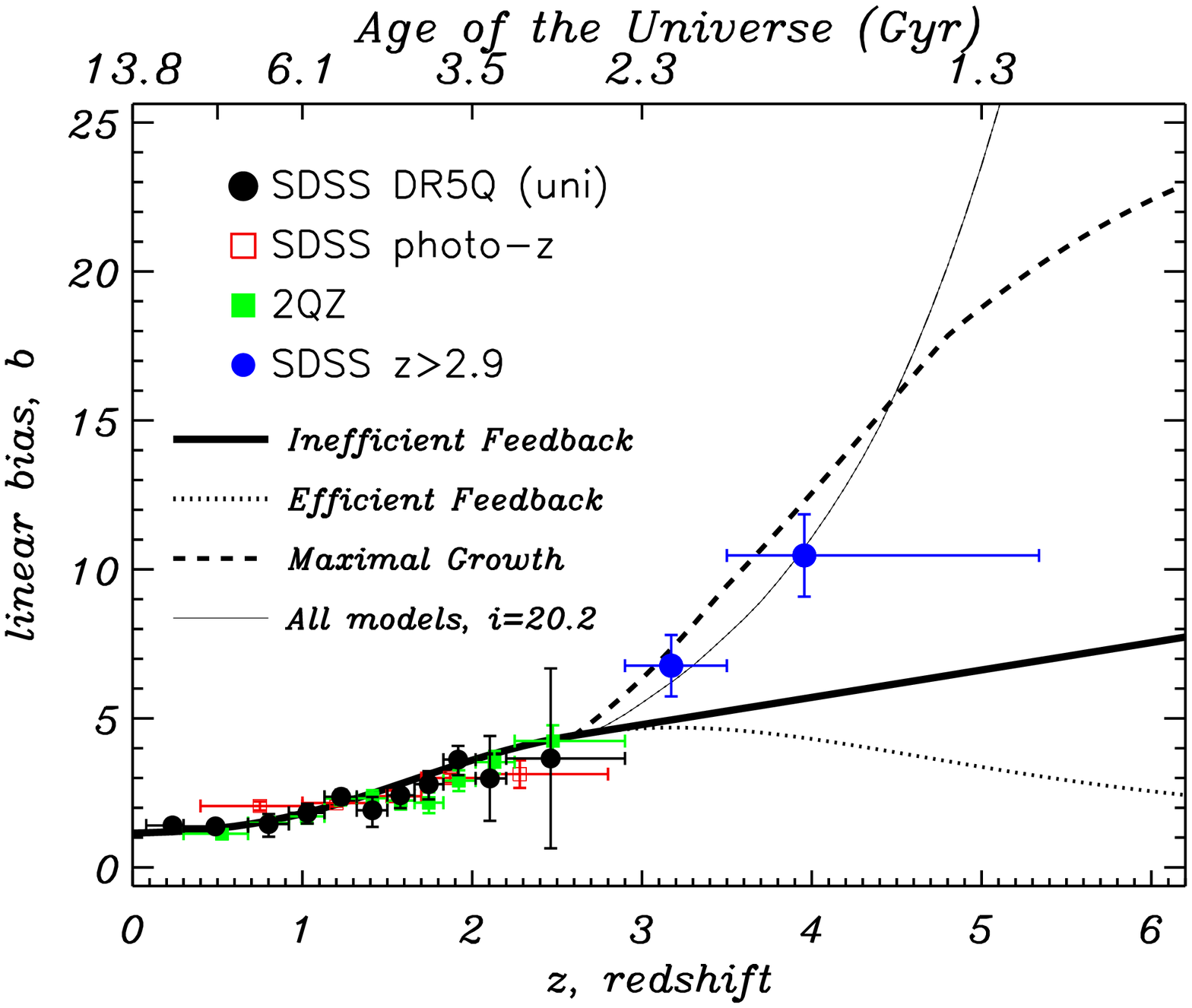}
      \centering
      \caption[Evolution of the linear bias, $b$ with redshift.]
      {Evolution of the linear bias of quasars, $b_{Q}$, with redshift to $z=6$.
        Filled (black) circles, this work; 
        filled (green) squares, \citet{Croom05}; 
        open (red) squares, \citet{Myers07a}; 
        filled (blue) circles, \citet{Shen07};
        The {\it thin} solid line shows the behaviour
        for all of the three \citet{Hopkins07} models at $i=20.2$, with
        these models having identical behaviour for $b(z)$.
        The {\it thick} solid line shows the ``Inefficient feedback'' 
        model for a magnitude limited survey of $i=30$ i.e., a 
        truly ``complete'' survey. 
        The dotted line is for the ``Efficient feedback'' model (at $i=30$) and
        the ``Maximal growth'' model is given by the dashed line (also for $i=30$).
      }
      \label{fig:bias_with_z6}
    \end{figure}
    We next compare with the models of \citet[][e.g. their
    Fig. 13]{Hopkins07}. Here three models are described for the feeding
    of quasars. All the models have the same $z<2$ behaviour, as in each
    case quasars are said to ``shut down'', i.e. there is no accretion
    onto the central SMBH at $z\lesssim2$. 
    
    The first of the \citet{Hopkins07} models is the ``Inefficient
    Feedback''  model, whose predictions are given by the solid lines in
    Fig.~\ref{fig:bias_with_z6}.  Here $z\sim6$ quasars grow either
    continuously or episodically with their host systems until the epoch
    where ``downsizing'' begins (i.e. $z\sim2$), Thus, at redshifts $z >
    2$ feedback from quasar activity is insufficient  to completely shut down
    the quasar, hence the term ``inefficient feedback''.  
    
    The second of the \citet{Hopkins07} models is the ``Extreme
    Feedback'' model, represented by the dotted line in
    Fig.~\ref{fig:bias_with_z6}. Here each SMBH only experiences one
    episode of quasar activity, after which the quasar completely shuts
    down, even if this occurs at high ($z>2$) redshifts. BH growth will
    cease  after this one-off quasar phase. If objects cannot grow after
    their quasar epoch even at high redshifts, then the subsequent decline
    of the break in the QLF at $L=L^{*}$ traces a decline in
    characteristic active masses, and the linear bias of active systems
    ``turns over''. 
    
    The third of the \citet{Hopkins07} models is the ``Maximal
    Growth'' model, represented by the dashed line in
    Fig.~\ref{fig:bias_with_z6}. In this model the BHs grow mass at their
    Eddington rate until $z\sim2$.  For example, a $\sim10^{8} M_{\odot}$
    BH at $z\sim6$ will grow to $\sim5\times10^{9} M_{\odot}$ at $z\sim2$
    at which point the growth ceases and the BH mass remains constant.
    
    The limiting factor in our ability to discriminate between models
    is the dynamic range in luminosity and redshift. We thus 
    extend our redshift baseline up to
    $z=6$ in Figure~\ref{fig:bias_with_z6} and now also plot the bias
    estimates for the $z>2.9$ SDSS quasar clustering measurements of
    \citet{Shen07}, given by the filled blue circles, where we use their
    measured values of $\bar{\xi}_{Q}(s,\bar{z}=3.2)=1.23\pm0.35$  and
    $\bar{\xi}_{Q}(s,\bar{z}=4.0)=2.41\pm0.59$ with our
    equation~(\ref{eq:bias_xi_20}) to estimate the bias. 

    A magnitude limit of $m_{i} < 20.2$ is chosen for the models to
    match the SDSS high-redshift quasar selection. As can be seen in
    Fig.~\ref{fig:bias_with_z6}, all models match the observational
    clustering data well at $z<2$.  However, at this magnitude limit the
    QLF break luminosity $L^{*}$ is only marginally resolved at $z\sim
    2-3$ \citep[e.g.][]{Richards06}; above this redshift surveys are
    systematically biased to more massive $L > L^{*}$ BHs with higher
    clustering and larger linear biases. Subsequently, the models with
    the $m_{i} < 20.2$ limit have no discriminating power at $z>2$, and
    the predicted behaviour for the linear bias from the ``Inefficient'',
    ``Efficient'' and ``Maximal Growth'' models is identical. To break
    this degeneracy,  deeper observational data at high redshift will be
    needed. Fortunately,  these data should be in hand within the next few
    years, which will be able to discriminate and test these models, such
    as those with an effectively infinitely deep flux limit of $i=30$ that
    are also plotted  in Fig.~\ref{fig:bias_with_z6}. Therefore, further
    investigations into the link between AGN/quasar activity, the build-up
    of SMBH mass and the formation and evolution of quasars and galaxies
    using clustering measurements are left to future investigations.

\section{Conclusions}
We have calculated the two-point correlation function using a
homogeneous sample of \hbox{30 239} quasars from the Fifth Data release
of the SDSS Quasar Survey, covering a solid angle of $\approx4000$
deg$^{2}$, a redshift range of $0.3 \leq z \leq 2.2$ and thus
representing a measurement over the largest volume of the Universe
ever sampled at \hbox{25 $h^{-3}$ Gpc$^{3}$} (comoving) assuming the
current $\Lambda$CDM cosmology. We find that:

\begin{itemize}
\item{The two-point redshift-space correlation function is
    adequately described by a 
    single power-law of the form $\xi = (s/s_{0})^{-\gamma}$ where
    $s_{0}=5.95\pm0.45\hmpc$ and $\gamma_{s}=1.16^{+0.11}_{-0.16}$ 
    over $1 \leq s \leq 25 \hmpc.$}
\item{We see no evidence for significant clustering ($\xis>0$) 
    at scales of $s>100 \hmpc$.}
\item{There are strong redshift-space distortions present in the
    2-D $\xi(r_{p},\pi)$ measurement, with ``Fingers of God'' seen 
    at small scales. However, these are most likely primarily dominated 
    by redshift errors. }
\item{We find no significant evolution of clustering amplitude
    of the SDSS quasars to $z\sim2.5$, though we note that the luminosity
    threshold of the sample also increases steadily with redshift and
    the clustering strength does increase at higher redshift. This 
    is investigated further in Shen et al. (2009).}
\item{Comparing our results with recent deep X-ray surveys, 
    our clustering measurements are in reasonable agreement 
    in some cases e.g. Gilli et al. (2005), Miyaji et al. (2007) 
    and XMM-COSMOS \citet{Gilli08} but
    significantly lower correlation lengths in others.
    However, there is still much scatter in the deep X-ray data, 
    potentially due to cosmic variance and the small samples used for these
    analyses.}
\item{The linear bias for SDSS quasars over the
    redshift range of $0.3 \leq z \leq 2.2$ is $b(z=1.27)=2.06\pm0.03$. 
    Using this bias measurement and assuming $\Omm(z=1.27)=0.81$, 
    but not taking into account the effects of small-scale 
    redshift-space distortions, we find $\beta(z=1.27)=0.43$. 
    Both these values are consistent with measurements from
    previous surveys, i.e. the 2QZ.} 
\item{Using models which relate dark halo mass to clustering
    strength \citep[e.g. ][]{Sheth01}, we find that
    the dark halo mass at which a `typical SDSS quasar' resides 
    remains roughly constant with redshift at 
    $M_{\rm DMH}\sim 2 \times10^{12} h^{-1} M_{\odot}$. 
    This non-evolution of quasar host halo mass agrees
    very well with previous studies by e.g. \citet[][]{Croom05} and
    \citet{daAngela08}.
    Therefore, as dark halo masses grow with time, 
    the ratio of the typical halo mass for a quasar to 
    other haloes at the same epoch drops with redshift.}
\item{Using current clustering data, we are unable to 
    discriminate between the ``Inefficient Feedback'', ``Efficient Feedback''
    and ``Maximal Growth'' models proposed by \citet{Hopkins07} at $z<2$.
    The measured evolution of the clustering amplitude is in reasonable
    agreement with recent theoretical models, although measurements to
    fainter limits will be needed to distinguish different scenarios for
    quasar feeding and black hole growth.}
\end{itemize}

Shen et al. (2009) study the clustering properties of DR5 quasars as a
function of luminosity, virial mass, colour and radio loudness.

The SDSS is now complete and the final quasar catalogue from Data
Release 7 \citep{Abazajian08} is being prepared. This catalogue should
be a $\sim60\%$ increment over DR5, containing about \hbox{130,000}
quasars with spectroscopic observations, and will almost double the
number of quasars in the UNIFORM sub-sample.  DR7 will not change the
luminosity dynamic range of the SDSS quasar survey but with final
analysis of data from, e.g., the 2SLAQ QSO Survey \citep{Croom08}, and
extension of the deep X-ray surveys \citep[e.g. Extended CDF-S,
][]{Lehmer05}, connections between the ``luminous'' and ``average''
AGN luminosity regimes  should begin to converge. 

Looking further ahead, even with the dramatic increase in data that
surveys such as the 2QZ and SDSS have provided, the desire to increase
dynamic range continues. For instance, due to the steepness of the
faint end of the quasar luminosity function \citep{HRH07}, low
luminosity quasars should be relatively plentiful, as long as one can
identify these objects. This will be a strong challenge for the next
generation of quasar redshift surveys e.g. the Baryon Oscillation
Spectroscopic Survey \citep[BOSS, ][]{Schlegel07} but one that will
lead to another significant increase in our understanding of quasars,
supermassive black holes, galaxy formation and evolution and the
properties of the Universe.

\section*{Acknowledgments}
This work was partially supported by National Science Foundation
grants AST-0607634 (N.P.R. and D.P.S.) and AST-0707266 (Y.S. and
M.A.S.). We warmly thank S.M. Croom for providing the 2QZ data points,
S. Basilakos, A. Lidz and P. Hopkins for providing their model data
shown in Section 5 and R. Gilli for allowing us to report the most
recent XMM-Newton COSMOS results prior to publication.  P. Allen,
W.N. Brandt, A.D. Myers and R. Nemmen provided very useful
discussion. The JavaScript Cosmology Calculator was used whilst
preparing this paper \citep{Wright06}. This research made use of the
NASA Astrophysics Data System. The data and code used will become
publicly available at {\tt http://www.astro.psu.edu/users/npr/DR5/}
upon publication of this paper. We thank the anonymous referee for
comments that improved this work.

Funding for the SDSS and SDSS-II has been provided by the Alfred
P. Sloan Foundation, the Participating Institutions, the National
Science Foundation, the U.S. Department of Energy, the National
Aeronautics and Space Administration, the Japanese Monbukagakusho, the
Max Planck Society, and the Higher Education Funding Council for
England. The SDSS Web Site is http://www.sdss.org/.

The SDSS is managed by the Astrophysical Research Consortium for the
Participating Institutions. The Participating Institutions are the
American Museum of Natural History, Astrophysical Institute Potsdam,
University of Basel, University of Cambridge, Case Western Reserve
University, University of Chicago, Drexel University, Fermilab, the
Institute for Advanced Study, the Japan Participation Group, Johns
Hopkins University, the Joint Institute for Nuclear Astrophysics, the
Kavli Institute for Particle Astrophysics and Cosmology, the Korean
Scientist Group, the Chinese Academy of Sciences (LAMOST), Los Alamos
National Laboratory, the Max-Planck-Institute for Astronomy (MPIA),
the Max-Planck-Institute for Astrophysics (MPA), New Mexico State
University, Ohio State University, University of Pittsburgh,
University of Portsmouth, Princeton University, the United States
Naval Observatory, and the University of Washington.

\appendix
\section{A. SDSS Technical  details}\label{sec:Appendix_SDSS}

    \subsection{A.1 The Catalog Archive Server}
    The SDSS database can be interrogated through the Catalog Archive
    Server\footnote{http://cas.sdss.org} (CAS) using standard Structured
    Query Language (SQL) queries. When querying the CAS, one has a choice
    to query either the \BEST or \TARGET database for a given Data Release
    (in our case, DR5). The former database contains information on all
    the photometric and spectroscopic objects obtained using the latest
    (and thus the ``best'') versions of the data reduction and analysis
    pipelines \citep[Section 3,][]{Abazajian04}. The \TARGET database
    however, contains the information on objects available at the time that
    the targeting algorithm pipelines were run. An object's
    magnitude or colour can be subtly
    different between target allocation and the most recent data
    processing, and some objects change their target selection status
    between the two. More details regarding the CAS, \BEST and \TARGET are
    given in the relevant SDSS Data Release papers \citep{Stoughton02,
      Abazajian04, Adelman-McCarthy07}.
    
    In order to create a statistical data sample, or to mimic it for a
    comparative `random' sample, we need to know the properties of our
    chosen objects {\it at the time of targetting}, i.e. which objects
    were selected as quasar candidates. Thus here, we only use information
    from \TARGET.
    
    \begin{figure}
      \includegraphics[height=8.0cm,width=9.0cm]
      {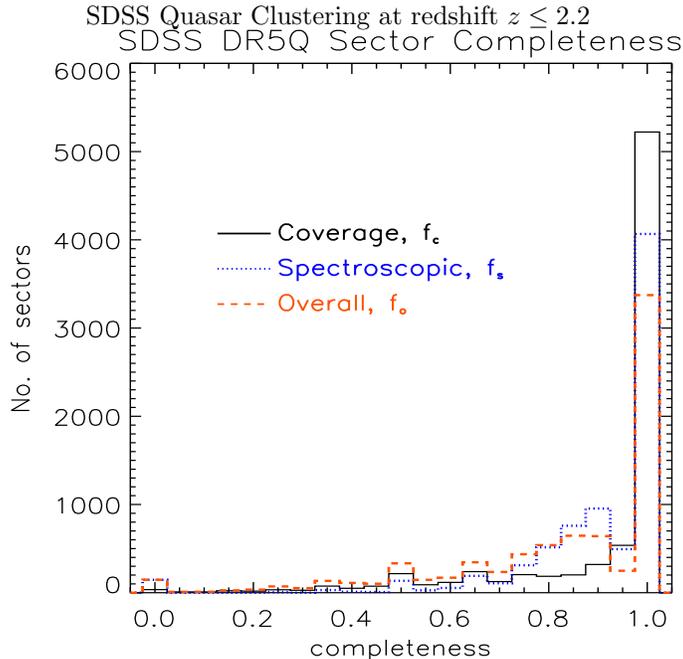}
      \centering
      \caption[Histogram showing the completeness of the DR5 Quasar survey
      by sector. ]
      {Histogram showing the completeness of the DR5 Quasar survey by sector.
        The solid (orange) line shows completeness distribution for
        all 7814 sectors, while the dotted (blue) line shows the 
        completeness distribution for the 5831 sectors which have
        one or more POAs in them. The summed area of sectors with
      given completenesses is shown by the dashed line.}
      \label{fig:sector_completeness}
    \end{figure}  
    \subsection{A.2 SDSS Survey Geometry} 
    As mentioned in Section 3.1 and 3.2, in order to  calculate the
    2PCF, one needs to assemble a ``random''  sample which reproduces the
    angular selection function (``the mask'')  and the radial distribution
    of the quasar data. In this section we describe the steps required to
    define the mask over which our sample was defined. 
    
    The first SQL query we run simply asks the CAS to return all the
    objects in the Photometric database that were targetted as
    ``primary'' candidate quasars. When run on DR5, this returns \hbox{203
      185} objects from the PhotoObjAll table.
    
    We next calculate which primary `PhotoObjAll' objects (POAs) fall
    within the spectroscopic survey plate boundaries. We do not use any of
    the ``Extra'', ``Special'', or ``ExtraSpecial'' plates for our
    analysis as these plates were not targetted with the normal quasar
    algorithm, or are duplicates \citep{Adelman-McCarthy06}. There are
    \hbox{145 524} POA objects that fall within 1.49 degrees of a given
    DR5 plate centre, noting that since plates overlap due to the tiling
    scheme, an object can be in more than one plate. 
    
    Of these \hbox{145,524} objects, \hbox{11,336} are duplicate
    objects, defined as being within $1''$ of another object in the
    catalogue. Of these \hbox{134,188} unique objects, we would next like
    to know how many were {\it (a)} designated as spectroscopic
    (``tilable'') targets by the process of `Tiling' and  {\it (b)}
    allocated fibres. A tile is a 1.49 degree radius circle on the sky
    which contains the locations of up to 592 tilable targets and other
    science targets (the other 48 fibres are assigned to calibration
    targets and blank sky). For each tile a physical aluminum plate is
    created. The plates will have holes drilled in them for fibres to be
    plugged, in order to observe the tiled targets. The goal of the tiling
    procedure, described in detail by \citet[][]{Blanton03}\footnote{see
      also http://www.sdss.org/dr6/algorithms/tiling.html}, is to maximise
    the total number of targets assigned fibres. Due to the large-scale
    structure in the quasar/galaxy distribution the procedure overlaps
    tiles with one other. 
    
    As described in \citet{Blanton03}, \citet{Tegmark04},
    \citet{Blanton05} and \citet{Percival07b}, a ``sector'' is defined as
    a set of tile overlap regions (spherical polygons) observed by a
    unique combination of tiles and survey ``chunks''. A `chunk' is a unit
    of SDSS imaging data and is a part of an SDSS `stripe', which is a
    2.5$^{\circ}$ wide cylindrical segment aligned along a great circle
    between the survey poles. These sectors are the natural areas on
    which to define the completeness of our sample. There are \hbox{7 814}
    sectors for DR5, \hbox{5 831} of which have one or more POA objects in
    them. Using the RegionID field in the \TARGET table (which gives the
    sector identification number if set, zero otherwise) we match the
    positions (R.A.'s and Decs) of objects in \TARGET to those that are in
    PhotoObjAll and the DR5Q. 
    
    The efficiency of the quasar targetting algorithm is $\sim95\%$
    \citep{Vanden_Berk05}. We can define two functions
    for the primary sample which have dependence on angular position in
    the sky only in order to calculate the completeness of the survey:
    \begin{itemize}
    \item{Coverage Completeness, $f_Q$. 
        The coverage completeness is the ratio of the number of
        quasar targets that are assigned a spectroscopic fibre to the total
        number of quasar candidates in a given sector. Fibre collisions will
        be one contributing factor in the coverage completeness. 
      }
    \item{Spectroscopic Completeness, $f_{s}(\theta)$. 
        This is the ratio of the number of high-quality spectra obtained
        in a sector to the number spectroscopically observed. 
        Due to the nature of the SDSS quasar survey, this ratio tends 
        to be very high.}
    \end{itemize}
    The `overall completeness', $f_{\rm O}$, is defined as $f_{\rm
      O}=f_{q} \times f_{s}$ and the distribution of this overall 
    sector completeness is shown in Fig.~\ref{fig:sector_completeness}.

\section{B. Jackknife Errors}\label{sec:jackknife_appendix}
\begin{figure}
  \includegraphics[height=7.0cm,width=12.0cm]
  {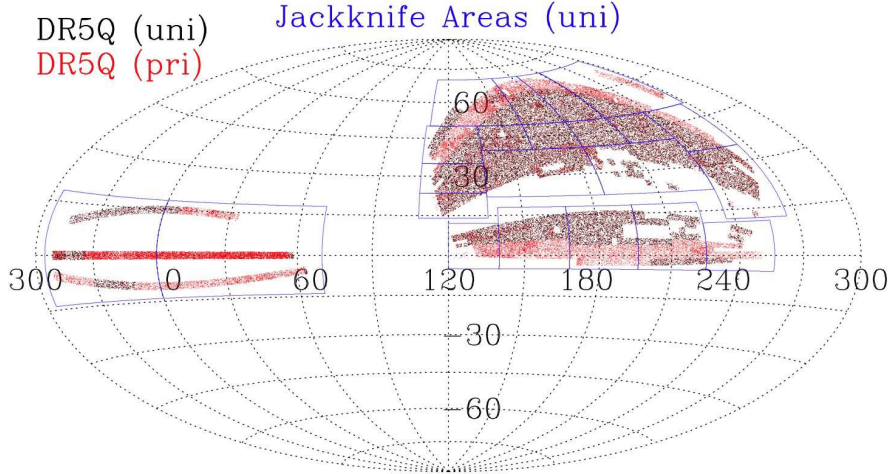}
  \centering
  \caption[Geometry of the SDSS DR5Q Jackknife areas, 2.]
  {Geometry of the SDSS DR5Q Jackknife areas,
    showing the location of the DR5Q PRIMARY (orange/grey dots)
    and the UNIFORM (black dots) samples. 
    The jackknife areas were chosen to follow the overall
    geometry of the SDSS Quasar survey. 
    The number of quasars in each area is approximately equal.
    Note the sparse coverage of the UNIFORM sample in the
    Southern Stripes.}
  \label{fig:jackknife_areas_data}
\end{figure}  
Here we follow \citet[][\S\S 3.4.5, 11.3 and their
Eq. 10]{Scranton02}, \citet[][\S 3.4 and equation 7]{Zehavi02} and
\citet[][Appendix A]{Myers07a} to calculate the jackknife error
estimates on our quasar clustering data. 

\citet{Myers07a} estimate errors using an ``inverse variance''
weighted jackknife technique. This method divides the data into $N$
sub-samples and then recalculates the given statistic (e.g. $\xis$)
using the Landy-Szalay estimator (equation~\ref{eqn:lseq}), {\it leaving out}
one sub-sample area at one time. Following their convention we denote
subsamples by the subscript $L$ and recalculate $\xis_{L}$ in each
jackknife realization via equation~\ref{eqn:lseq}. The
inverse-variance-weighted covariance matrix, $C(s_i,s_j) = C_{ij}$, is 
\begin{equation}
  C_{ij} = \sum_{L=1}^{N}\sqrt{\frac{RR_{L}(s_i)}{RR(s_i)}}
          \left[\xi_{L}(s_i)- \xi(s_i)\right] 
          \sqrt{\frac{RR_{L}(s_j)}{RR(s_j)}}
          \left[\xi_{L}(s_j)- \xi(s_j)\right]
 \label{eqn:CM}
\end{equation}
where $\xi$ denotes the correlation function for all data.
Jackknife errors $\sigma_{i}$ are obtained from the diagonal
elements ($\sigma_{i}^{2} = C_{ii}$), and the normalized covariance
matrix, also known as the regression matrix, is
\begin{equation}
  |C| = \frac{C_{ij}}{\sigma_i\sigma_j}
  \label{eqn:NCM}
\end{equation}

\begin{deluxetable*}{lrrrrrr}
  \tablecolumns{7}
  \tablewidth{10cm}
  \tablecaption{}
  \tablehead{Region & RA min & RA max & Dec min & Dec max & No. of  & No. of  \\
                    &        &        &         &         & Quasars & Randoms}
  \startdata
  N01    & 120.   & 140.   &	-5.    & 12.     & 29 445  & 870 558 \\
  N02    & 140.   & 168.   &	-5.    & 18.     & 28 456  & 841 541 \\
  N03    & 168.   & 196.   &	-5.    & 18.     & 27 904  & 825 442 \\
  N04    & 196.   & 225.   &	-5.    & 18.     & 28 717  & 846 926 \\
  N05    & 225.   & 256.   &	-5.    & 11.     & 29 891  & 879 837 \\
  \hline
  N06    & 108.   & 136.   &	14.    & 23.5    & 29 614 & 873 778 \\ 
  N07    & 108.   & 136.   &	23.5   & 35.     & 28 646 & 845 871 \\
  N08    & 136.   & 186.   &	22.    & 40.     & 26 942 & 798 307 \\	
  N09    & 186.   & 236.   &	22.    & 40.     & 27 957 & 820 491 \\	
  N10    & 236.   & 265.   &	12.    & 35.     & 28 253 & 831 920 \\	
  \hline
  N11    & 108.   & 136.   &	35.    & 50.     & 29 003 & 856 576 \\	
  N12    & 136.   & 161.   &	40.    & 50.     & 28 875 & 855 021 \\	
  N13    & 161.   & 186.   &	40.    & 50.     & 28 857 & 853 908 \\	
  N14    & 186.   & 211.   &	40.    & 50.     & 28 917 & 854 055 \\	
  N15    & 211.   & 236.   &	40.    & 50.     & 28 924 & 854 070 \\	
  N16    & 236.   & 265.   &	35.    & 50.     & 29 246 & 863 221 \\	
  \hline
  N17    & 110.   & 161.   &	50.    & 70.     & 29 253 & 863 420 \\	
  N18    & 161.   & 186.   &	50.    & 70.     & 28 899 & 853 792 \\	
  N19    & 186.   & 211.   &	50.    & 70.     & 28 911 & 853 175 \\	
  N20    & 211.   & 268.   &	50.    & 70.     & 29 404 & 868 561 \\
  \hline
  S      & 0$\vee$305  & 70$\vee$360  & -14  & 18 & 28 675 & 842 497 \\		
  \enddata
  \tablecomments{Details of the regions used for the Jackknife subsamples. The ``No. of quasars''
  column gives the number of quasars {\it left} in the remaining regions when the given region is cut out.}
  \label{tab:jackknife_areas}
\end{deluxetable*}
  
We divide the sample into $21$ sub-samples. The number of subdivisions
is chosen such that each represents a cosmologically significant
volume, while retaining sufficient numbers of objects that shot noise
will not dominate any subsequent analysis.  The detailed boundaries of
the sub-samples are given in Table~\ref{tab:jackknife_areas}
and described by Fig.~\ref{fig:jackknife_areas_data}.
\begin{figure}
  \includegraphics[height=7.0cm,width=8.0cm]
  {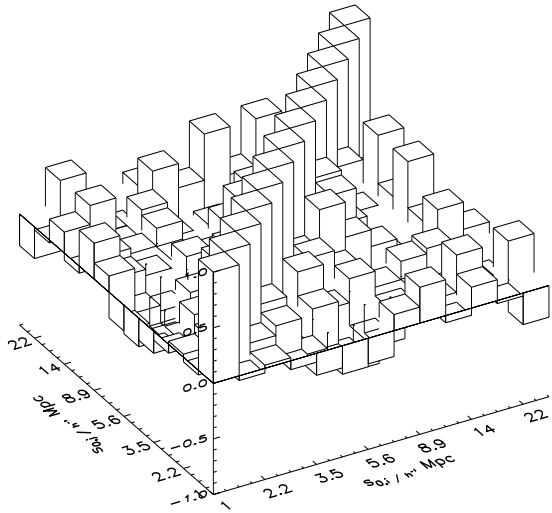}
  \centering
  \caption[The normalised covariance matrix, (the regression matrix) for $\xis$ 
  from jackknife error analysis on 21 sub-samples of the UNIFORM DR5Q]
  {The normalised covariance matrix, (the regression matrix) for $\xis$ 
    from jackknife error analysis on 21 sub-samples of the UNIFORM DR5Q, 
    for scales $1 \hmpc < s < 25 \hmpc$.} 
  \label{fig:regression_matrix}
\end{figure}     

\begin{figure}
  \includegraphics[height=7.0cm,width=8.0cm]
  {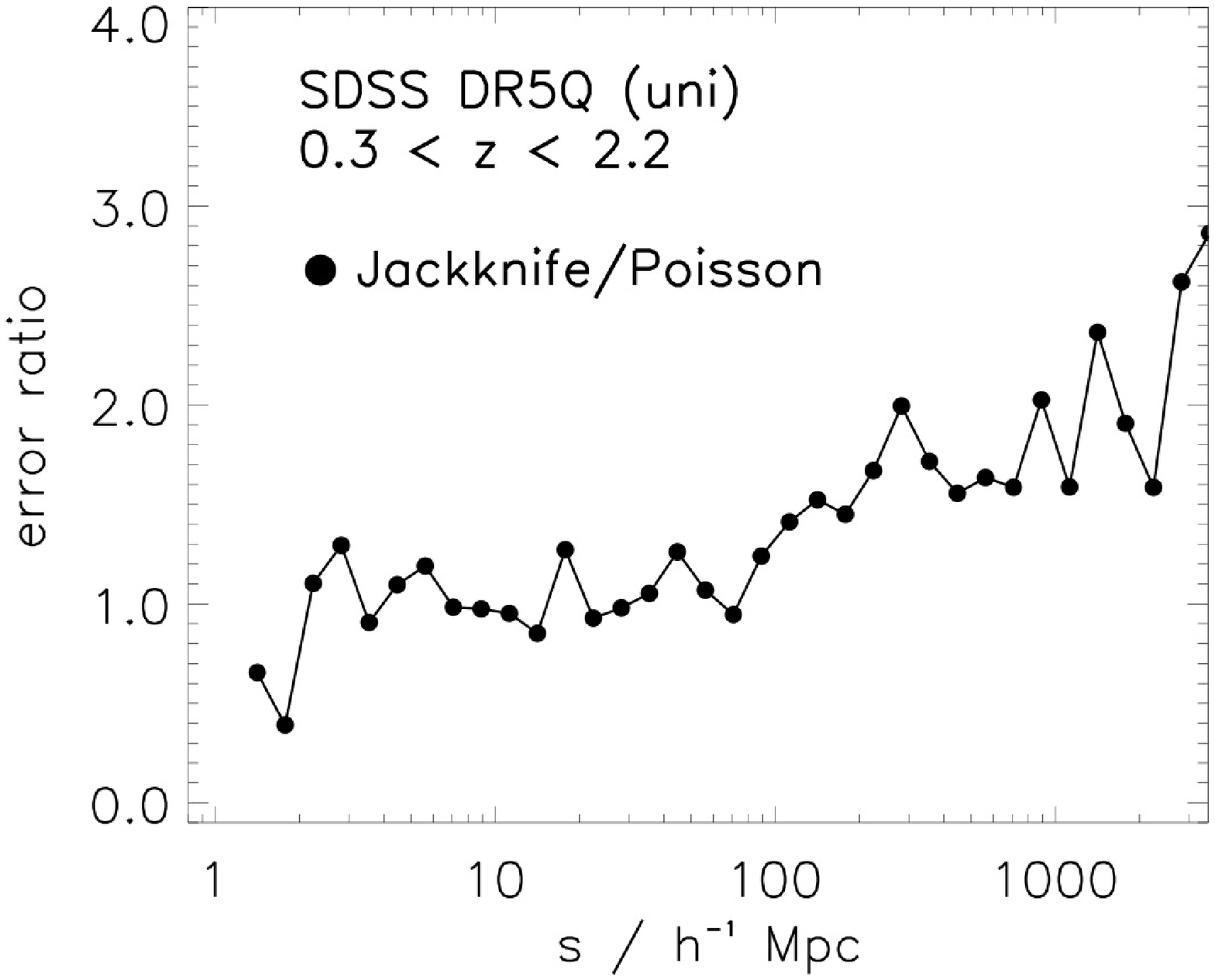}
  \centering
  \caption[Comparing Poisson and Jackknife errors.]
  {Comparison of Poisson and Jackknife errors for the UNIFORM 
    DR5 Quasar sample. The ratio between the Poisson
    and Jackknife errors (from the diagonal elements of the covariance 
    matrix only)  is very close to one at $s \lesssim 70\hmpc$, 
    while at $s \gtrsim 70\hmpc$, the Poisson errors are
    $\sim$~double that of the Jackknifes. 
  }
  \label{fig:error_ratio}
\end{figure}     
We find, as in previous quasar clustering work
\citep[e.g. ][]{Shanks94, Croom96}, that Poisson 
errors are a good description on scales where 
$DD_{q} \lesssim N_{q}$, where $N_{q}$ is the number
of quasars in a given sample and $DD_{q}$ is 
the number of quasar pairs in a given bin. On larger scales, 
the Poisson error tends to 
underestimate the Jackknife error, see Fig.~\ref{fig:error_ratio}.
The scale where $N_{q} \approx ~ DD_{q}$ is 
$\sim 70 \hmpc $ for the SDSS UNIFORM Quasar sample. 

Given the smallness of the off-diagonal elements of the covariance
matrix, we measure errors using the diagonal elements only. But here
we carry out a check using the full covariance matrix. 
We fit the observed $\xis$ to the power law model 
using the full covariance matrix. We calculate $\chi^{2}$ as
\begin{equation}
  \chi^{2} = \sum_{ij}
  [\xi(s_{i}) - \xi_{\rm mod}(s_{i})]
  C^{-1}_{ij} 
  [\xi(s_{j}) - \xi_{\rm mod}(s_{j})]
\end{equation}
where $C_{ij}^{-1}$ is the inverse covariance matrix, and $\xi_{\rm
mod}(s) = (s/s_{0})^{-\gamma_{s}}$ is our model, where we vary $s_{0}$
over the range $s_{0}=0.0-15.0 \hmpc$ in steps of $0.05\hmpc$ and
$\gamma_{s}$ over the range $\gamma_{s}=0.00-3.00$ in steps of 0.01,
fitting $\xis$ on scales from  $1 \hmpc < s < 25.0 \hmpc$ scales. 

Our estimates of the redshift-space correlation length and power-law slope
are now $s_{0}=6.35^{+0.40}_{-0.35} \hmpc$ and
$\gamma_{s}=1.11^{+0.11}_{-0.08}$ respectively
(we found $s_{0}=5.95\pm0.45\hmpc$ and $\gamma_{s}=1.16^{+0.11}_{-0.08}$
using the diagonal elements only). However, fitting over
$1.0\hmpc<s<100.0\hmpc$ scales, we find there is some tension between
the best-fit values given in Section 4.1 of $s_{0}=5.90\pm0.30\hmpc$
and $\gamma_{s}=1.57^{+0.04}_{-0.05}$ and the best-fit values using
the covariance matrix of $s_{0}=6.95^{+0.45}_{-0.55}\hmpc$ and
$\gamma_{s}=1.53\pm0.09$. We believe this is due to the noisy matrix
inversion, where small values at large scales in the covariance matrix
will  dominate the signal in the inverse matrix. However, we are
confident that using the diagonal elements of the covariance matrix
only for our model fits does not change the interpretation of our
results.

\section{C. Systematics in the SDSS Quasar 2PCF.}\label{sec:Appendix_Systematics}
Here we explore the effects of how the various systematic effects in our data,
and our methodology affect our correlation results. 
We shall determine the effects of different quasar samples
(Sec.~\ref{sec:samples}), changing the high-redshift cut
(Sec.~\ref{sec:hiz}), the fields which had poor imaging
(Sec.~\ref{sec:bad_fields_appendix}), Galactic reddening 
(Sec.~\ref{sec:reddening}) and fibre collisions 
(Sec.~\ref{sec:fibre_collisions}). We shall
report on $\xis$ and $\wp/r_{p}$ and find that when using the UNIFORM
sample, our overall results  (and subsequent interpretations) 
are robust.

    \begin{figure}
      \includegraphics[height=8.0cm,width=8.0cm]
      {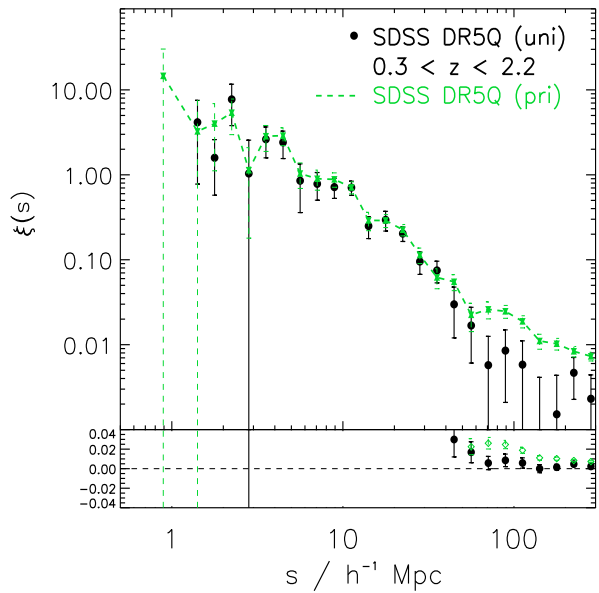}
      \centering
      \caption[The SDSS DR5 Quasar $\xis$ for the PRIMARY and UNIFORM samples.]
      {The SDSS DR5 Quasar $\xis$ for the PRIMARY and UNIFORM samples. 
        The lower panel shows the behaviour of  
        $\xis$ near zero on a linear scale.
        There is excellent agreement between the two samples
        at $s\leq 20\hmpc$, but the PRIMARY sample exhibits a
        higher clustering strength at large scales, $s \geq 40 \hmpc$. 
       This result provides our main motivation for using the UNIFORM sample
       exclusively in sections 3 and 4.}
      \label{fig:xis_DR5_UNI22_PRI22}
    \end{figure}
    \subsection{C.1 Effects of Different Samples on $\xis$}\label{sec:samples}
    Figure~\ref{fig:xis_DR5_UNI22_PRI22} shows the difference  in the
    redshift-space correlation function, $\xis$, for the PRIMARY
    sample and of the UNIFORM sample using the LS estimator. 
    The two samples are in excellent agreement at small
    scales, $s\leq 20 \hmpc$, but the PRIMARY sample exhibits a
    higher clustering strength at large scales, $s \geq 40 \hmpc$. One
    possible explanation for this discrepancy is due to the differing
    radial distributions in PRIMARY and UNIFORM resulting from the
    different target selection used before DR2. This
    result provides our main motivation for using the UNIFORM sample
    exclusively in sections 3 and 4.

     \begin{figure}
      \includegraphics[height=7.0cm,width=8.0cm]
      {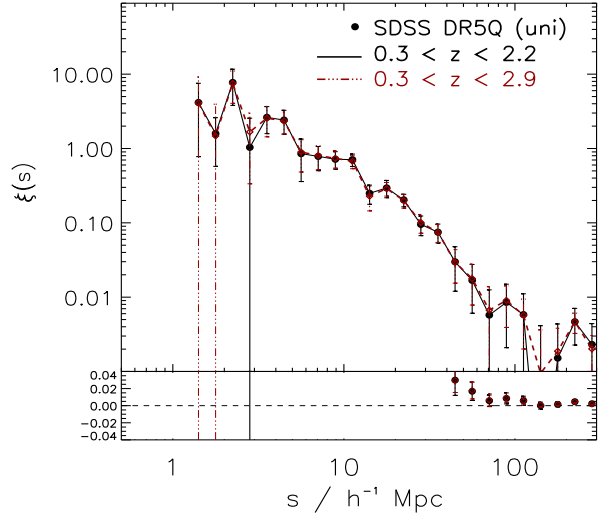}
      \centering
      \caption[The SDSS DR5 Quasar $\xis$ the UNIFORM sample with 
      $z<2.2$ and $z<2.9$.]
      {The SDSS DR5 Quasar UNIFORM $\xis$ with upper
        redshift cut-offs of $z \leq 2.2$ and $z \leq 2.9$. 
        The lower panel shows the behaviour of 
        $\xis$ near zero on a linear scale.
      The inclusion of data at $2.2 \leq z \leq 2.9$ barely
      changes the measured $\xis$.}
      \label{fig:xis_DR5_UNI22_UNI29}
    \end{figure}
    \subsection{C.2 High redshift cut-off}\label{sec:hiz}
    Figure~\ref{fig:xis_DR5_UNI22_UNI29} shows the 
    redshift-space 2-point correlation function
    $\xis$ for the UNIFORM sample with the high-redshift
    cut-off being changed from $z\leq2.2$ to $z\leq2.9$. 
    It is reassuring that the change between 
    $\xis$ is minimal, though this is somewhat unexpected
    since our the optical selection for the quasar
    sample is known to be affected between 
    $z=2.2$ and $z=2.9$ \citep{Richards06}.

    \begin{figure}
          \includegraphics[height=7.0cm,width=8.0cm]
          {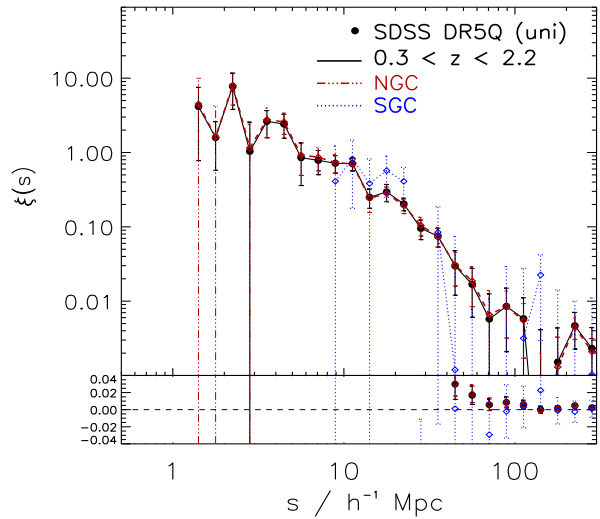}
          \centering
          \caption[The SDSS DR5 Quasar $\xis$ for the UNIFORM sample with 
                   the sample split into the NCG and SCG. ]
                   {The SDSS DR5 Quasar $\xis$ for the UNIFORM sample with 
                   the sample split into the NCG and SCG.
                   The two measurements are in good agreement.}
          \label{fig:xis_DR5_quasars_UNI22_NS}
    \end{figure}
    \subsection{C.3 The NGC vs. the SGC}
    Figure~\ref{fig:xis_DR5_quasars_UNI22_NS} shows the 
    redshift-space 2-point correlation function
    $\xis$ for the UNIFORM sample, split
    into quasars from the North Galactic Cap (NGC) and
    the South Galactic Cap (SGC). Note the data is
    heavily dominated by the NGC in the UNIFORM sample. 
    There is no detectable signal in the SGC $\xis$ below
    $s\approx 10 \hmpc$ and the two measurements are in
    good agreement.

    \begin{figure}
      \includegraphics[height=7.0cm,width=8.0cm]
      {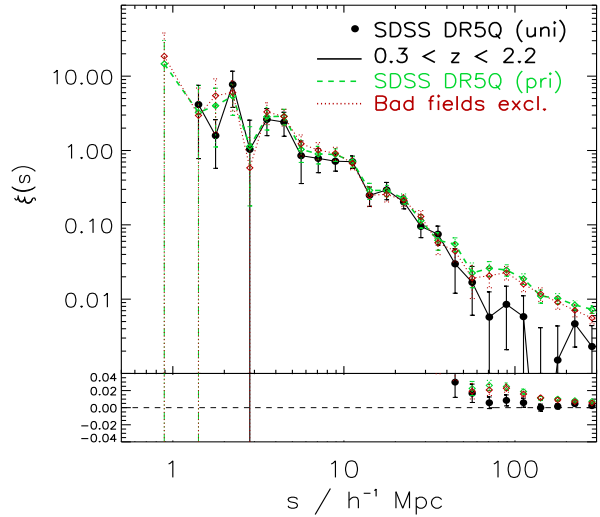}
      \centering
      \caption[The SDSS DR5 Quasar $\xis$ the UNIFORM sample with 
      the sample split into the NCG and SCG. ]
      {The SDSS DR5 Quasar $\xis$ for the 
        UNIFORM sample is given by the solid (black) circles. 
        $\xis$ for the PRIMARY sample 
        is given with including, green (solid) line and
        excluding, dashed (red) line bad imaging fields.}
      \label{fig:xis_DR5_quasars_UNI22_PRI22_BadFields}
    \end{figure}
    \subsection{C.4 Bad Fields}\label{sec:bad_fields_appendix}
    In the SDSS, a ``field'' is an image in all five bands, with
    approximate dimensions of 13' $\times$ 10'.  Since the quasar target
    selection algorithm searches for outliers from the stellar locus in
    colour space it is very sensitive to data with large photometric
    errors due to problems in photometric calibration or in point-spread
    function (PSF) determination \citep{Richards06}. Thus, using the
    definitions of ``bad fields'' given by \citet{Richards06} and
    \citet{Shen07}, based on the position of the stars in colour-colour
    space \citep[][]{Ivezic04}, we calculate the correlation function both 
    including and excluding data from these areas. 

    Figure~\ref{fig:xis_DR5_quasars_UNI22_PRI22_BadFields} shows the
    redshift-space 2-point correlation function $\xis$ for the UNIFORM
    sample (solid black circles).  Also shown is $\xis$ for the PRIMARY
    sample including,  (solid green) and excluding, (dashed red) lines,
    the ``Bad Fields'' as defined by \citet{Shen07}. Here we can
    see that there is minimal difference (for the PRIMARY DR5Q sample)
    between the $\xis$ estimates when including and excluding
    the bad fields. This results is reassuring but generally expected since
    at $z<2.2$ quasar selection using the UV excess technique is relatively
    insensitive to ``bad fields''. However, at higher redshift, \citet{Shen07}
    found this to be a major issue, where the selection is more sensitive.

    \begin{figure}
      \includegraphics[height=7.0cm,width=8.0cm]
      {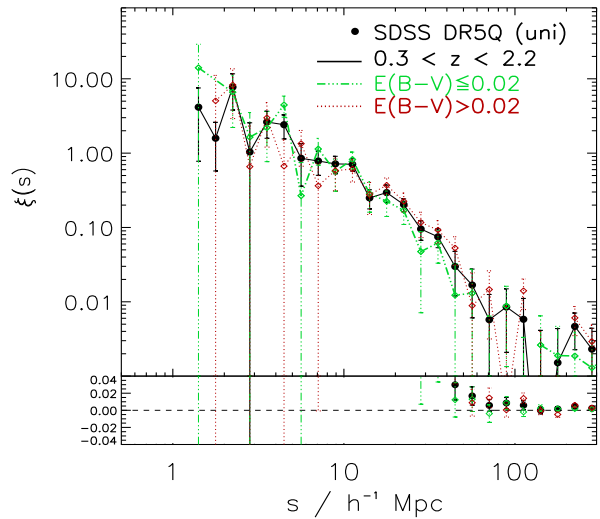}
      \centering
      \caption[The SDSS DR5 Quasar $\xis$ for the UNIFORM sample with 
      the sample split by $E(B-V)$. ]
      {The SDSS DR5 Quasar $\xis$ for the UNIFORM sample with 
        the sample split by $E(B-V)$. We see no
        systematic difference in the clustering signal between
        the two $\xis$ measurements that might have been 
        caused due to errors in the reddening correction model.}
      \label{fig:xis_DR5_quasars_UNI22_EBV0pnt0217_20080822}
    \end{figure}
    \subsection{C.5 Reddening}\label{sec:reddening}
    While all selection for the quasar sample is undertaken using
    dereddened colors \citep[][]{Richards02} following the Galactic
    extinction model of \citet[][]{Schlegel98}, any remaining
    systematic errors in the reddening model can induce excess power
    into the clustering in a number of different ways. The most obvious
    possibility comes from a modulation in the angular density of quasars
    as a function of position on the sky. In addition the color
    dependence of the reddening correction may preferentially exclude
    quasars at specific redshifts. As we currently assume a common $N(z)$
    for all quasars in the UNIFORM sample, an $N(z)$ that is reddening-dependent
    can also induce excess clustering. For this analysis we will
    assume that any artificial signal that might be induced by the
    reddening correction will scale with the magnitude of the reddening
    correction itself. We therefore subdivide the UNIFORM quasar sample
    into two subsets, of approximately equal number, a low reddening
    sample, with $0.0028 < E(B-V) \leq 0.0217$, and a high reddening
    sample $0.0217 < E(B-V) \leq 0.2603$. The reddening estimates are
    derived from the maps of \citet{Schlegel98}.
    
    Figure~\ref{fig:xis_DR5_quasars_UNI22_EBV0pnt0217_20080822} shows
    the redshift-space 2-point correlation function $\xis$ for the full
    UNIFORM sample, with the reddening split sub-samples. The low
    reddening component, dot-dashed (green) line and the high reddening
    sample, dotted (red) line are consistent within the errors for all
    scales out to $\sim 250 \hmpc$. There is no evidence for a systematic
    difference in the clustering signal on large scales that might be
    induced by any modulation due to errors in the reddening correction.

    \begin{figure}
      \includegraphics[height=8.0cm,width=7.0cm]
      {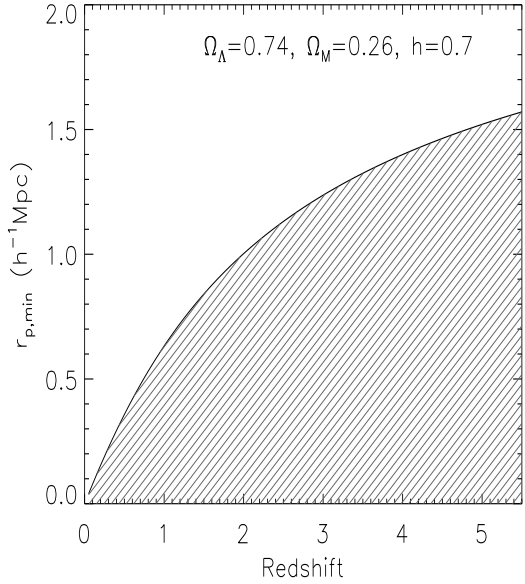}
      \centering
      \caption[The transverse comoving separation corresponding to 
                55'' as a function of redshift.]
              {The transverse comoving separation corresponding to 
                55'' as a function of redshift. This is the minimal projected
                comoving separation that can be probed with the SDSS 
                spectroscopic quasar sample as a function of redshift, 
                due to the fibre collision limit of the SDSS spectroscopy.}
      \label{fig:fibre_collisions}
    \end{figure}
    \begin{figure}
      \includegraphics[height=7.0cm,width=8.0cm]
      {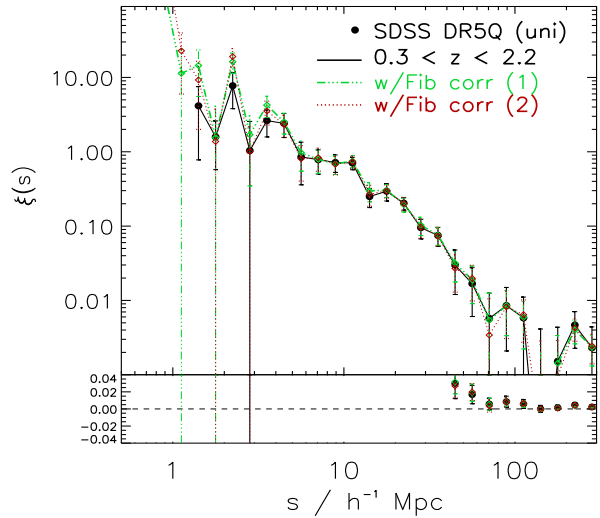}
      \centering
      \caption[The SDSS DR5 Quasar $\xis$ for the UNIFORM sample taking
      into account fibre collisions.]
      {The SDSS DR5 Quasar $\xis$ for the UNIFORM sample: 
        with no fibre collision correction (filled circles); 
        using photometric redshifts for quasar candidates 
        that we not observed due to fibre collisions (green, dot-dashed line)
        and using the redshifts from the nearest observed quasar
        for quasar candidates that we not observed due to fibre collisions
        (red, dotted line). We see very little difference on 
        scales $s>5\hmpc$ but do measure increased values of
        $\xis$ at $\sim 1-5 \hmpc$.} 
      \label{fig:xis_DR5_quasars_UNI22_wFibre}
    \end{figure}
    \begin{figure}
      \includegraphics[height=7.0cm,width=8.0cm]
      {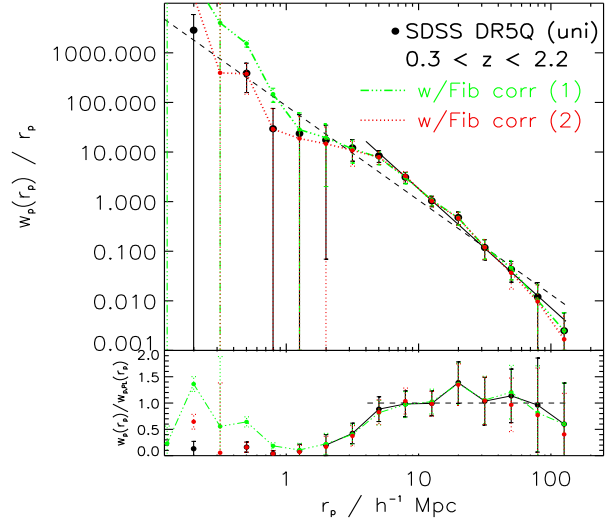}
      \centering
      \caption[The SDSS DR5 Quasar $\wp/r_{p}$ for the UNIFORM sample taking
      into account fibre collisions.]
      {The projected correlation function $\wp/r_{p}$ for the SDSS DR5Q 
        UNIFORM sample with $0.30 <z<2.2$ with no fibre collision correction 
        (filled circles); 
        using photometric redshifts for quasar candidates 
        that we not observed due to fibre collisions (green, dot-dashed line)
        and using the redshifts from the nearest observed quasar
        for quasar candidates that we not observed due to fibre collisions
        (red, dotted line). We see very little difference on 
        scales $r_{p}>1\hmpc$ but do measure increased values of
        $\wp/r_{p}$ at $r_{p}<1 \hmpc$.} 
      \label{fig:wp_div_sigma_DR5_quasars_UNI22_wFibre}
    \end{figure}
    \subsection{C.6 Fibre Collisions}\label{sec:fibre_collisions}

    Due to the design of the SDSS fibres and plates, no two
    spectroscopic fibres can be separated by less than $55''$ (Section
    2.1).  The corresponding minimum physical separation in $r_{p}$  that
    we can sample is shown by Fig.~\ref{fig:fibre_collisions}.
    
    To investigate this effect on our correlation function estimates,
    we find which of the \hbox{145 524} POA objects  were not observed due
    to fibre collisions. There are 431 objects that were within 55'' of a
    UNIFORM quasar that were not observed. We assign the ``collided''
    quasar candidates a redshift using two methods. First, (model (1) in
    Fig. \ref{fig:xis_DR5_quasars_UNI22_wFibre} and
    Fig.~\ref{fig:wp_div_sigma_DR5_quasars_UNI22_wFibre} ), using the new
    version of the SDSS Quasar photometric catalogue, \citep{Richards09},
    we assign the redshift of the nearest photometric quasar to the
    collided objects. We assume that all the collided objects are in fact
    quasars, though in reality this is not the case.  Second, (model (2)
    in Fig. \ref{fig:xis_DR5_quasars_UNI22_wFibre} and
    Fig.~\ref{fig:wp_div_sigma_DR5_quasars_UNI22_wFibre}) we assign the
    collided quasar candidate the redshift from the quasar that ``knocked
    it out''.  We then recalculate the 2PCF with these additional
    objects. 
    
    As we can see from Fig.~\ref{fig:xis_DR5_quasars_UNI22_wFibre},
    the inclusion of these collided objects makes very little difference
    to our measurement of $\xis$ at scales $\gtrsim 5 \hmpc$. However, we
    do measure increased values of $\xis$ at $s=1-5 \hmpc$. Therefore, we
    again fit a single power-law to the data which has been corrected for
    fibre collisions using the photometric quasar redshifts,  over the
    scales $1<s<25 \hmpc$ and find $s_{0}=6.70^{+0.45}_{-0.30}\hmpc$ and
    $\gamma_{s}=1.29^{+0.12}_{-0.10}$ (cf. $s_{0}=5.95\pm0.45\hmpc$ and
    $\gamma_{s}=1.16^{+0.11}_{-0.08}$ found in Section 4.1). With the
    inclusion of more data at small separations, the fibre-corrected
    $\xis$ has a higher $s_{0}$ value and steeper slope, but we find these
    results are consistent with our measurement of $\xis$ without the
    fibre collision corrections, given the errors. 
    
    By examining Fig.~\ref{fig:wp_div_sigma_DR5_quasars_UNI22_wFibre},
    we see that fibre collisions do {\it not} account for the  possible
    break in the slope of $\wp/r_{p}$ that was discussed in  Section
    4.3. We are thus satisfied that fibre collisions do not impact the
    results presented herein and refer the reader to \citet{Hennawi06} and
    \citet{Myers08} for more detailed investigations of quasar clustering
    and quasar binaries on these very small scales.

    \begin{figure}
      \includegraphics[height=7.0cm,width=8.0cm]
      {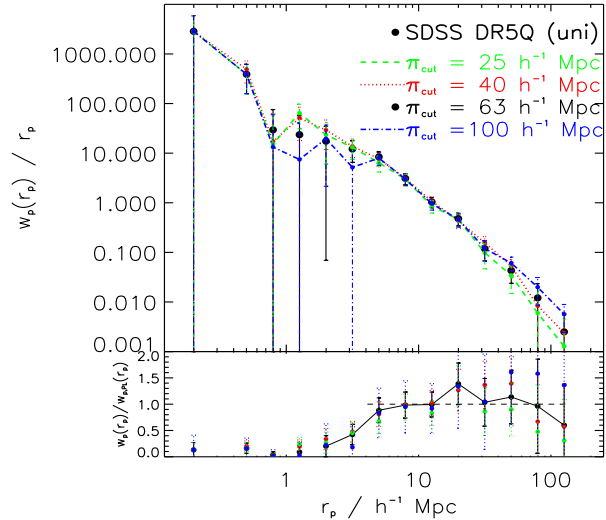}
      \centering
      \caption[The projected correlation function
      $\wp/r_{p}$ for the SDSS DR5Q UNIFORM sample with $0.30 <z<2.2$, 
      varying $\pi_{\rm max}$ from equation~\ref{eq:wp_sigma_def}.]
      {The projected correlation function
        $\wp/r_{p}$ for the SDSS DR5Q UNIFORM sample with $0.30 <z<2.2$, 
        varying $\pi_{\rm max}$ from equation~\ref{eq:wp_with_pi_max}.
        The lower panel has the data divided by the best fit power 
        law from Section 4, with $r_{0}=8.75\hmpc$ and $\gamma=2.40$.}
      \label{fig:wp_sigma_DR5_quasars_UNI22_picutComparisons_20080909}
    \end{figure}
    \subsection{C. 7 Varying $\pi_{\rm max}$ limits for $\wp$ }\label{sec:pi_max}

    Figure~\ref{fig:wp_sigma_DR5_quasars_UNI22_picutComparisons_20080909} 
    shows the projected correlation function
    $\wp/r_{p}$ for the SDSS DR5Q UNIFORM sample with $0.30 <z<2.2$, 
    varying $\pi_{\rm max}$ from equation~\ref{eq:wp_with_pi_max}.
    We vary $\pi_{\rm max}$ in intervals of $10^{0.2}$ over the range
    $\pi_{\rm max}=10^{1.4 - 2.0} = 25.1 - 100.0 \hmpc$.
    Although changing the $\pi_{\rm max}$ cut
    does produce a noticeable effect in estimates
    of $\wp/r_{p}$, when fitting our single power law over
    the scales $4.0 < r_{p} <130.0 \hmpc$, we do not see a significant 
    change in the best-fit $r_{0}$ or power-law slope values, 
    with the former constant at $r_{0}\approx8.3 \hmpc$ and the 
    latter constant at $\gamma\approx2.3$. We are therefore confident that the integration 
    limit of  
    $\pi_{\rm max}=63.1 \hmpc$ provides a good balance between
    larger $\pi$ values which would add noise to our $\wp/r_{p}$
    estimate, and lower $\pi$ values which might not recover
    the full signal at the largest separations.


\bibliographystyle{mn2e}
\bibliography{Ross_etal_2009}

\begin{thebibliography}{}

\bibitem[\protect\citeauthoryear{{Abazajian} et~al.,}{{Abazajian}
  et~al.}{2003}]{Abazajian03}
{Abazajian} K.,  et~al., 2003, \aj, 126, 2081

\bibitem[\protect\citeauthoryear{{Abazajian} et~al.,}{{Abazajian}
  et~al.}{2004}]{Abazajian04}
{Abazajian} K.,  et~al., 2004, \aj, 128, 502

\bibitem[\protect\citeauthoryear{{Abazajian} et~al.,}{{Abazajian}
  et~al.}{2008}]{Abazajian08}
{Abazajian} K.,  et~al., 2008, ArXiv:0812.0649v1

\bibitem[\protect\citeauthoryear{{Adelberger} \& {Steidel}}{{Adelberger} \&
  {Steidel}}{2005a}]{Adelberger05b}
{Adelberger} K.~L.,  {Steidel} C.~C.,  2005a, \apj, 630, 50

\bibitem[\protect\citeauthoryear{{Adelberger} \& {Steidel}}{{Adelberger} \&
  {Steidel}}{2005b}]{Adelberger05}
{Adelberger} K.~L.,  {Steidel} C.~C.,  2005b, \apjl, 627, L1

\bibitem[\protect\citeauthoryear{{Adelberger}, {Steidel}, {Pettini}, {Shapley},
  {Reddy} \& {Erb}}{{Adelberger} et~al.}{2005}]{Adelberger05c}
{Adelberger} K.~L.,  {Steidel} C.~C.,  {Pettini} M.,  {Shapley} A.~E.,  {Reddy}
  N.~A.,    {Erb} D.~K.,  2005, \apj, 619, 697

\bibitem[\protect\citeauthoryear{{Adelman-McCarthy} et~al.,}{{Adelman-McCarthy}
   et~al.}{2006}]{Adelman-McCarthy06}
{Adelman-McCarthy} J.~K.,  et~al., 2006, \apjs, 162, 38

\bibitem[\protect\citeauthoryear{{Adelman-McCarthy} et~al.,}{{Adelman-McCarthy}
   et~al.}{2007}]{Adelman-McCarthy07}
{Adelman-McCarthy} J.~K.,  et~al., 2007, \apjs, 172, 634

\bibitem[\protect\citeauthoryear{{Alexander} et~al.,}{{Alexander}
  et~al.}{2003}]{Alexander03}
{Alexander} D.~M.,  et~al., 2003, \aj, 126, 539

\bibitem[\protect\citeauthoryear{{Arp}}{{Arp}}{1970}]{Arp70}
{Arp} H.,  1970, \aj, 75, 1

\bibitem[\protect\citeauthoryear{{Baes}, {Buyle}, {Hau} \& {Dejonghe}}{{Baes}
  et~al.}{2003}]{Baes03}
{Baes} M.,  {Buyle} P.,  {Hau} G.~K.~T.,    {Dejonghe} H.,  2003, \mnras, 341,
  L44

\bibitem[\protect\citeauthoryear{{Baldwin}, {Phillips} \&
  {Terlevich}}{{Baldwin} et~al.}{1981}]{BPT}
{Baldwin} J.~A.,  {Phillips} M.~M.,    {Terlevich} R.,  1981, \pasp, 93, 5

\bibitem[\protect\citeauthoryear{{Barger} et~al.,}{{Barger}
  et~al.}{2003}]{Barger03}
{Barger} A.~J.,  et~al., 2003, \aj, 126, 632

\bibitem[\protect\citeauthoryear{{Basilakos}, {Georgakakis}, {Plionis} \&
  {Georgantopoulos}}{{Basilakos} et~al.}{2004}]{Basilakos04}
{Basilakos} S.,  {Georgakakis} A.,  {Plionis} M.,    {Georgantopoulos} I.,
  2004, \apjl, 607, L79

\bibitem[\protect\citeauthoryear{{Basilakos}, {Plionis} \&
  {Ragone-Figueroa}}{{Basilakos} et~al.}{2008}]{Basilakos08}
{Basilakos} S.,  {Plionis} M.,    {Ragone-Figueroa} C.,  2008, \apj, 678, 627

\bibitem[\protect\citeauthoryear{{Becker}, {White} \& {Helfand}}{{Becker}
  et~al.}{1995}]{Becker95}
{Becker} R.~H.,  {White} R.~L.,    {Helfand} D.~J.,  1995, \apj, 450, 559

\bibitem[\protect\citeauthoryear{{Blanton}, {Cen}, {Ostriker} \&
  {Strauss}}{{Blanton} et~al.}{1999}]{Blanton99}
{Blanton} M.,  {Cen} R.,  {Ostriker} J.~P.,    {Strauss} M.~A.,  1999, \apj,
  522, 590

\bibitem[\protect\citeauthoryear{{Blanton}, {Eisenstein}, {Hogg} \&
  {Zehavi}}{{Blanton} et~al.}{2006}]{Blanton06}
{Blanton} M.~R.,  {Eisenstein} D.,  {Hogg} D.~W.,    {Zehavi} I.,  2006, \apj,
  645, 977

\bibitem[\protect\citeauthoryear{{Blanton} et~al.,}{{Blanton}
  et~al.}{2005}]{Blanton05}
{Blanton} M.~R.,  et~al., 2005, \aj, 129, 2562

\bibitem[\protect\citeauthoryear{{Blanton}, {Lin}, {Lupton}, {Maley}, {Young},
  {Zehavi} \& {Loveday}}{{Blanton} et~al.}{2003}]{Blanton03}
{Blanton} M.~R.,  {Lin} H.,  {Lupton} R.~H.,  {Maley} F.~M.,  {Young} N.,
  {Zehavi} I.,    {Loveday} J.,  2003, \aj, 125, 2276

\bibitem[\protect\citeauthoryear{{Bonoli}, {Marulli}, {Springel}, {White},
  {Branchini} \& {Moscardini}}{{Bonoli} et~al.}{2008}]{Bonoli08}
{Bonoli} S.,  {Marulli} F.,  {Springel} V.,  {White} S.~D.~M.,  {Branchini} E.,
     {Moscardini} L.,  2008, ArXiv:0812.0003v1

\bibitem[\protect\citeauthoryear{{Boyle}, {Georgantopoulos}, {Blair},
  {Stewart}, {Griffiths}, {Shanks}, {Gunn} \& {Almaini}}{{Boyle}
  et~al.}{1998}]{Boyle98}
{Boyle} B.~J.,  {Georgantopoulos} I.,  {Blair} A.~J.,  {Stewart} G.~C.,
  {Griffiths} R.~E.,  {Shanks} T.,  {Gunn} K.~F.,    {Almaini} O.,  1998,
  \mnras, 296, 1

\bibitem[\protect\citeauthoryear{{Boyle}, {Shanks}, {Croom}, {Smith}, {Miller},
  {Loaring} \& {Heymans}}{{Boyle} et~al.}{2000}]{Boyle00}
{Boyle} B.~J.,  {Shanks} T.,  {Croom} S.~M.,  {Smith} R.~J.,  {Miller} L.,
  {Loaring} N.,    {Heymans} C.,  2000, \mnras, 317, 1014

\bibitem[\protect\citeauthoryear{{Brandt} \& {Hasinger}}{{Brandt} \&
  {Hasinger}}{2005}]{Brandt05}
{Brandt} W.~N.,  {Hasinger} G.,  2005, \araa, 43, 827

\bibitem[\protect\citeauthoryear{{Coil}, {Hennawi}, {Newman}, {Cooper} \&
  {Davis}}{{Coil} et~al.}{2007}]{Coil07}
{Coil} A.~L.,  {Hennawi} J.~F.,  {Newman} J.~A.,  {Cooper} M.~C.,    {Davis}
  M.,  2007, \apj, 654, 115

\bibitem[\protect\citeauthoryear{{Coles} \& {Erdogdu}}{{Coles} \&
  {Erdogdu}}{2007}]{Coles07}
{Coles} P.,  {Erdogdu} P.,  2007, Journal of Cosmology and Astro-Particle
  Physics, 10, 7

\bibitem[\protect\citeauthoryear{{Coles} \& {Lucchin}}{{Coles} \&
  {Lucchin}}{2002}]{Coles02}
{Coles} P.,  {Lucchin} F.,  2002, {Cosmology: The Origin and Evolution of
  Cosmic Structure, Second Edition}.
Wiley-VCH.

\bibitem[\protect\citeauthoryear{{Constantin} \& {Vogeley}}{{Constantin} \&
  {Vogeley}}{2006}]{Constantin06}
{Constantin} A.,  {Vogeley} M.~S.,  2006, \apj, 650, 727

\bibitem[\protect\citeauthoryear{{Croom} et~al.,}{{Croom}
  et~al.}{2005}]{Croom05}
{Croom} S.~M.,  et~al., 2005, \mnras, 356, 415

\bibitem[\protect\citeauthoryear{{Croom} et~al.,}{{Croom}
  et~al.}{2008}]{Croom08}
{Croom} S.~M.,  et~al., 2008, \mnras, in press.

\bibitem[\protect\citeauthoryear{{Croom} \& {Shanks}}{{Croom} \&
  {Shanks}}{1996}]{Croom96}
{Croom} S.~M.,  {Shanks} T.,  1996, \mnras, 281, 893

\bibitem[\protect\citeauthoryear{{Croom}, {Shanks}, {Boyle}, {Smith}, {Miller},
  {Loaring} \& {Hoyle}}{{Croom} et~al.}{2001}]{Croom01}
{Croom} S.~M.,  {Shanks} T.,  {Boyle} B.~J.,  {Smith} R.~J.,  {Miller} L.,
  {Loaring} N.~S.,    {Hoyle} F.,  2001, \mnras, 325, 483

\bibitem[\protect\citeauthoryear{{Croom}, {Smith}, {Boyle}, {Shanks}, {Miller},
  {Outram} \& {Loaring}}{{Croom} et~al.}{2004}]{Croom04}
{Croom} S.~M.,  {Smith} R.~J.,  {Boyle} B.~J.,  {Shanks} T.,  {Miller} L.,
  {Outram} P.~J.,    {Loaring} N.~S.,  2004, \mnras, 349, 1397

\bibitem[\protect\citeauthoryear{{da {\^A}ngela} et~al.,}{{da {\^A}ngela}
  et~al.}{2008}]{daAngela08}
{da {\^A}ngela} J.,  et~al., 2008, \mnras, 383, 565

\bibitem[\protect\citeauthoryear{{da {\^A}ngela}, {Outram}, {Shanks}, {Boyle},
  {Croom}, {Loaring}, {Miller} \& {Smith}}{{da {\^A}ngela}
  et~al.}{2005}]{daAngela05}
{da {\^A}ngela} J.,  {Outram} P.~J.,  {Shanks} T.,  {Boyle} B.~J.,  {Croom}
  S.~M.,  {Loaring} N.~S.,  {Miller} L.,    {Smith} R.~J.,  2005, \mnras, 360,
  1040

\bibitem[\protect\citeauthoryear{{Davis} et~al.,}{{Davis}
  et~al.}{2003}]{Davis03}
{Davis} M.,  et~al., 2003, in {Guhathakurta} P.,  ed., Discoveries and Research
  Prospects from 6- to 10-Meter-Class Telescopes II. Proceedings of the SPIE,
  p.161

\bibitem[\protect\citeauthoryear{{Davis}, {Newman}, {Faber} \&
  {Phillips}}{{Davis} et~al.}{2001}]{Davis01}
{Davis} M.,  {Newman} J.~A.,  {Faber} S.~M.,    {Phillips} A.~C.,  2001, in
  {Cristiani} S.,  {Renzini} A.,   {Williams} R.~E.,  eds, Deep Fields:
  Springer-Verlag, p. 241

\bibitem[\protect\citeauthoryear{{Davis} \& {Peebles}}{{Davis} \&
  {Peebles}}{1983}]{Davis83}
{Davis} M.,  {Peebles} P.~J.~E.,  1983, \apj, 267, 465

\bibitem[\protect\citeauthoryear{{Dekel} \& {Lahav}}{{Dekel} \&
  {Lahav}}{1999}]{Dekel99}
{Dekel} A.,  {Lahav} O.,  1999, \apj, 520, 24

\bibitem[\protect\citeauthoryear{{Eisenstein} et~al.,}{{Eisenstein}
  et~al.}{2001}]{Eisenstein01}
{Eisenstein} D.~J.,  et~al., 2001, \aj, 122, 2267

\bibitem[\protect\citeauthoryear{{Fan}}{{Fan}}{1999}]{Fan99}
{Fan} X.,  1999, \aj, 117, 2528

\bibitem[\protect\citeauthoryear{{Fine} et~al.,}{{Fine}  et~al.}{2006}]{Fine06}
{Fine} S.,  et~al., 2006, \mnras, 373, 613

\bibitem[\protect\citeauthoryear{{Fisher}, {Davis}, {Strauss}, {Yahil} \&
  {Huchra}}{{Fisher} et~al.}{1994}]{Fisher94}
{Fisher} K.~B.,  {Davis} M.,  {Strauss} M.~A.,  {Yahil} A.,    {Huchra} J.~P.,
  1994, \mnras, 267, 927

\bibitem[\protect\citeauthoryear{{Fukugita}, {Ichikawa}, {Gunn}, {Doi},
  {Shimasaku} \& {Schneider}}{{Fukugita} et~al.}{1996}]{Fukugita96}
{Fukugita} M.,  {Ichikawa} T.,  {Gunn} J.~E.,  {Doi} M.,  {Shimasaku} K.,
  {Schneider} D.~P.,  1996, \aj, 111, 1748

\bibitem[\protect\citeauthoryear{{Gilli} et~al.,}{{Gilli}
  et~al.}{2005}]{Gilli05}
{Gilli} R.,  et~al., 2005, \aap, 430, 811

\bibitem[\protect\citeauthoryear{{Gilli} et~al.,}{{Gilli}
  et~al.}{2008}]{Gilli08}
{Gilli} R.,  et~al., 2008, ArXiv:0810.4769v2

\bibitem[\protect\citeauthoryear{{Gunn} et~al.,}{{Gunn}  et~al.}{1998}]{Gunn98}
{Gunn} J.~E.,  et~al., 1998, \aj, 116, 3040

\bibitem[\protect\citeauthoryear{{Gunn} et~al.,}{{Gunn}  et~al.}{2006}]{Gunn06}
{Gunn} J.~E.,  et~al., 2006, \aj, 131, 2332

\bibitem[\protect\citeauthoryear{{Guzzo} et~al.,}{{Guzzo}
  et~al.}{2008}]{Guzzo08}
{Guzzo} L.,  et~al., 2008, \nat, 451, 541

\bibitem[\protect\citeauthoryear{{Haiman} \& {Hui}}{{Haiman} \&
  {Hui}}{2001}]{Haiman01}
{Haiman} Z.,  {Hui} L.,  2001, \apj, 547, 27

\bibitem[\protect\citeauthoryear{{Hamilton}}{{Hamilton}}{1992}]{Hamilton92}
{Hamilton} A.~J.~S.,  1992, \apjl, 385, L5

\bibitem[\protect\citeauthoryear{{Hawkins} et~al.,}{{Hawkins}
  et~al.}{2003}]{Hawkins03}
{Hawkins} E.,  et~al., 2003, \mnras, 346, 78

\bibitem[\protect\citeauthoryear{{Hawkins} \& {Reddish}}{{Hawkins} \&
  {Reddish}}{1975}]{Hawkins75}
{Hawkins} M.~R.~S.,  {Reddish} V.~C.,  1975, \nat, 257, 772

\bibitem[\protect\citeauthoryear{{Hennawi} et~al.,}{{Hennawi}
  et~al.}{2006}]{Hennawi06}
{Hennawi} J.~F.,  et~al., 2006, \aj, 131, 1

\bibitem[\protect\citeauthoryear{{Hogg}, {Finkbeiner}, {Schlegel} \&
  {Gunn}}{{Hogg} et~al.}{2001}]{Hogg01}
{Hogg} D.~W.,  {Finkbeiner} D.~P.,  {Schlegel} D.~J.,    {Gunn} J.~E.,  2001,
  \aj, 122, 2129

\bibitem[\protect\citeauthoryear{{Hopkins}, {Hernquist}, {Cox} \& {Kere{\v
  s}}}{{Hopkins} et~al.}{2008}]{Hopkins08}
{Hopkins} P.~F.,  {Hernquist} L.,  {Cox} T.~J.,    {Kere{\v s}} D.,  2008,
  \apjs, 175, 356

\bibitem[\protect\citeauthoryear{{Hopkins}, {Lidz}, {Hernquist}, {Coil},
  {Myers}, {Cox} \& {Spergel}}{{Hopkins} et~al.}{2007}]{Hopkins07}
{Hopkins} P.~F.,  {Lidz} A.,  {Hernquist} L.,  {Coil} A.~L.,  {Myers} A.~D.,
  {Cox} T.~J.,    {Spergel} D.~N.,  2007, \apj, 662, 110

\bibitem[\protect\citeauthoryear{{Hopkins}, {Richards} \&
  {Hernquist}}{{Hopkins} et~al.}{2007}]{HRH07}
{Hopkins} P.~F.,  {Richards} G.~T.,    {Hernquist} L.,  2007, \apj, 654, 731

\bibitem[\protect\citeauthoryear{{Hoyle}, {Outram}, {Shanks}, {Boyle}, {Croom}
  \& {Smith}}{{Hoyle} et~al.}{2002}]{Hoyle02}
{Hoyle} F.,  {Outram} P.~J.,  {Shanks} T.,  {Boyle} B.~J.,  {Croom} S.~M.,
  {Smith} R.~J.,  2002, \mnras, 332, 311

\bibitem[\protect\citeauthoryear{{Iovino} \& {Shaver}}{{Iovino} \&
  {Shaver}}{1988}]{Iovino88}
{Iovino} A.,  {Shaver} P.~A.,  1988, \apjl, 330, L13

\bibitem[\protect\citeauthoryear{{Ivezi{\'c}} et~al.,}{{Ivezi{\'c}}
  et~al.}{2004}]{Ivezic04}
{Ivezi{\'c}} {\v Z}.,  et~al., 2004, Astronomische Nachrichten, 325, 583

\bibitem[\protect\citeauthoryear{{Jackson}}{{Jackson}}{1972}]{Jackson72}
{Jackson} J.~C.,  1972, \mnras, 156, 1P

\bibitem[\protect\citeauthoryear{{Jing}}{{Jing}}{1998}]{Jing98}
{Jing} Y.~P.,  1998, \apjl, 503, L9

\bibitem[\protect\citeauthoryear{{Kaiser}}{{Kaiser}}{1987}]{Kaiser87}
{Kaiser} N.,  1987, \mnras, 227, 1

\bibitem[\protect\citeauthoryear{{Kauffmann} et~al.,}{{Kauffmann}
  et~al.}{2003}]{Kauffmann03}
{Kauffmann} G.,  et~al., 2003, \mnras, 346, 1055

\bibitem[\protect\citeauthoryear{{Kerscher}, {Szapudi} \& {Szalay}}{{Kerscher}
  et~al.}{2000}]{Kerscher00}
{Kerscher} M.,  {Szapudi} I.,    {Szalay} A.~S.,  2000, \apjl, 535, L13

\bibitem[\protect\citeauthoryear{{Kewley}, {Dopita}, {Sutherland}, {Heisler} \&
  {Trevena}}{{Kewley} et~al.}{2001}]{Kewley01}
{Kewley} L.~J.,  {Dopita} M.~A.,  {Sutherland} R.~S.,  {Heisler} C.~A.,
  {Trevena} J.,  2001, \apj, 556, 121

\bibitem[\protect\citeauthoryear{{Kollmeier} et~al.,}{{Kollmeier}
  et~al.}{2006}]{Kollmeier06}
{Kollmeier} J.~A.,  et~al., 2006, \apj, 648, 128

\bibitem[\protect\citeauthoryear{{Kundic}}{{Kundic}}{1997}]{Kundic97}
{Kundic} T.,  1997, \apj, 482, 631

\bibitem[\protect\citeauthoryear{{La Franca}, {Andreani} \& {Cristiani}}{{La
  Franca} et~al.}{1998}]{LaFranca98}
{La Franca} F.,  {Andreani} P.,    {Cristiani} S.,  1998, \apj, 497, 529

\bibitem[\protect\citeauthoryear{{Landy} \& {Szalay}}{{Landy} \&
  {Szalay}}{1993}]{LS93}
{Landy} S.~D.,  {Szalay} A.~S.,  1993, \apj, 412, 64

\bibitem[\protect\citeauthoryear{{Lehmer} et~al.,}{{Lehmer}
  et~al.}{2005}]{Lehmer05}
{Lehmer} B.~D.,  et~al., 2005, \apjs, 161, 21

\bibitem[\protect\citeauthoryear{{Lidz}, {Hopkins}, {Cox}, {Hernquist} \&
  {Robertson}}{{Lidz} et~al.}{2006}]{Lidz06}
{Lidz} A.,  {Hopkins} P.~F.,  {Cox} T.~J.,  {Hernquist} L.,    {Robertson} B.,
  2006, \apj, 641, 41

\bibitem[\protect\citeauthoryear{{Linder}}{{Linder}}{2005}]{Linder05}
{Linder} E.~V.,  2005, \prd, 72, 043529

\bibitem[\protect\citeauthoryear{{Lupton}, {Gunn}, {Ivezi{\'c}}, {Knapp} \&
  {Kent}}{{Lupton} et~al.}{2001}]{Lupton01}
{Lupton} R.,  {Gunn} J.~E.,  {Ivezi{\'c}} Z.,  {Knapp} G.~R.,    {Kent} S.,
  2001, in {Harnden} Jr. F.~R.,  {Primini} F.~A.,   {Payne} H.~E.,  eds,
  Astronomical Data Analysis Software and Systems X, Astronomical Society of
  the Pacific Conference Series, 238, 269

\bibitem[\protect\citeauthoryear{{Lynden-Bell}}{{Lynden-Bell}}{1969}]{LyndenBe%
ll69}
{Lynden-Bell} D.,  1969, \nat, 223, 690

\bibitem[\protect\citeauthoryear{{Mart{\'{\i}}nez} \& {Saar}}{{Mart{\'{\i}}nez}
  \& {Saar}}{2002}]{Martinez02book}
{Mart{\'{\i}}nez} V.~J.,  {Saar} E.,  2002, Statistics of the Galaxy
  Distribution.
Chapman \& Hall/CRC

\bibitem[\protect\citeauthoryear{{Martini} \& {Weinberg}}{{Martini} \&
  {Weinberg}}{2001}]{Martini01}
{Martini} P.,  {Weinberg} D.~H.,  2001, \apj, 547, 12

\bibitem[\protect\citeauthoryear{{Miller}, {Nichol}, {G{\'o}mez}, {Hopkins} \&
  {Bernardi}}{{Miller} et~al.}{2003}]{Miller03}
{Miller} C.~J.,  {Nichol} R.~C.,  {G{\'o}mez} P.~L.,  {Hopkins} A.~M.,
  {Bernardi} M.,  2003, \apj, 597, 142

\bibitem[\protect\citeauthoryear{{Miyaji} et~al.,}{{Miyaji}
  et~al.}{2007}]{Miyaji07}
{Miyaji} T.,  et~al., 2007, \apjs, 172, 396

\bibitem[\protect\citeauthoryear{{Mountrichas} \& {Shanks}}{{Mountrichas} \&
  {Shanks}}{2007}]{Mountrichas07}
{Mountrichas} G.,  {Shanks} T.,  2007, \mnras, 380, 113

\bibitem[\protect\citeauthoryear{{Myers}, {Brunner}, {Nichol}, {Richards},
  {Schneider} \& {Bahcall}}{{Myers} et~al.}{2007}]{Myers07a}
{Myers} A.~D.,  {Brunner} R.~J.,  {Nichol} R.~C.,  {Richards} G.~T.,
  {Schneider} D.~P.,    {Bahcall} N.~A.,  2007, \apj, 658, 85

\bibitem[\protect\citeauthoryear{{Myers}, {Brunner}, {Richards}, {Nichol},
  {Schneider} \& {Bahcall}}{{Myers} et~al.}{2007}]{Myers07b}
{Myers} A.~D.,  {Brunner} R.~J.,  {Richards} G.~T.,  {Nichol} R.~C.,
  {Schneider} D.~P.,    {Bahcall} N.~A.,  2007, \apj, 658, 99

\bibitem[\protect\citeauthoryear{{Myers} et~al.,}{{Myers}
  et~al.}{2006}]{Myers06}
{Myers} A.~D.,  et~al., 2006, \apj, 638, 622

\bibitem[\protect\citeauthoryear{{Myers}, {Outram}, {Shanks}, {Boyle}, {Croom},
  {Loaring}, {Miller} \& {Smith}}{{Myers} et~al.}{2005}]{Myers05}
{Myers} A.~D.,  {Outram} P.~J.,  {Shanks} T.,  {Boyle} B.~J.,  {Croom} S.~M.,
  {Loaring} N.~S.,  {Miller} L.,    {Smith} R.~J.,  2005, \mnras, 359, 741

\bibitem[\protect\citeauthoryear{{Myers}, {Richards}, {Brunner}, {Schneider},
  {Strand}, {Hall}, {Blomquist} \& {York}}{{Myers} et~al.}{2008}]{Myers08}
{Myers} A.~D.,  {Richards} G.~T.,  {Brunner} R.~J.,  {Schneider} D.~P.,
  {Strand} N.~E.,  {Hall} P.~B.,  {Blomquist} J.~A.,    {York} D.~G.,  2008,
  \apj, 678, 635

\bibitem[\protect\citeauthoryear{{Norberg}, {Baugh}, {Gaztanaga} \&
  {Croton}}{{Norberg} et~al.}{2008}]{Norberg08}
{Norberg} P.,  {Baugh} C.~M.,  {Gaztanaga} E.,    {Croton} D.~J.,  2008,
  ArXiv:0810.1885v1

\bibitem[\protect\citeauthoryear{{Oke} \& {Gunn}}{{Oke} \&
  {Gunn}}{1983}]{Oke83}
{Oke} J.~B.,  {Gunn} J.~E.,  1983, \apj, 266, 713

\bibitem[\protect\citeauthoryear{{Osmer}}{{Osmer}}{1981}]{Osmer81}
{Osmer} P.~S.,  1981, \apj, 247, 762

\bibitem[\protect\citeauthoryear{{Outram}, {Shanks}, {Boyle}, {Croom}, {Hoyle},
  {Loaring}, {Miller} \& {Smith}}{{Outram} et~al.}{2004}]{Outram04}
{Outram} P.~J.,  {Shanks} T.,  {Boyle} B.~J.,  {Croom} S.~M.,  {Hoyle} F.,
  {Loaring} N.~S.,  {Miller} L.,    {Smith} R.~J.,  2004, \mnras, 348, 745

\bibitem[\protect\citeauthoryear{{Padmanabhan} et~al.,}{{Padmanabhan}
  et~al.}{2008}]{Padmanabhan08a}
{Padmanabhan} N.,  et~al., 2008, \apj, 674, 1217

\bibitem[\protect\citeauthoryear{{Padmanabhan}, {White}, {Norberg} \&
  {Porciani}}{{Padmanabhan} et~al.}{2008}]{Padmanabhan08}
{Padmanabhan} N.,  {White} M.,  {Norberg} P.,    {Porciani} C.,  2008,
  ArXiv:0802.2105v2

\bibitem[\protect\citeauthoryear{{Peacock}}{{Peacock}}{1999}]{Peacock99}
{Peacock} J.~A.,  1999, {Cosmological Physics}.
Cambridge University Press

\bibitem[\protect\citeauthoryear{{Peacock} et~al.,}{{Peacock}
  et~al.}{2001}]{Peacock01}
{Peacock} J.~A.,  et~al., 2001, \nat, 410, 169

\bibitem[\protect\citeauthoryear{{Peebles}}{{Peebles}}{1973}]{Peebles73}
{Peebles} P.~J.~E.,  1973, \apj, 185, 413

\bibitem[\protect\citeauthoryear{{Peebles}}{{Peebles}}{1980}]{Peebles80}
{Peebles} P.~J.~E.,  1980, {The Large-Scale Structure of the Universe}.
Princeton University Press.

\bibitem[\protect\citeauthoryear{{Peebles}}{{Peebles}}{1993}]{Peebles93book}
{Peebles} P.~J.~E.,  1993, {Principles of Physical Cosmology}.
Princeton, NJ: Princeton University Press

\bibitem[\protect\citeauthoryear{{Percival} et~al.,}{{Percival}
  et~al.}{2007}]{Percival07b}
{Percival} W.~J.,  et~al., 2007, \apj, 657, 645

\bibitem[\protect\citeauthoryear{{Pier}, {Munn}, {Hindsley}, {Hennessy},
  {Kent}, {Lupton} \& {Ivezi{\'c}}}{{Pier} et~al.}{2003}]{Pier03}
{Pier} J.~R.,  {Munn} J.~A.,  {Hindsley} R.~B.,  {Hennessy} G.~S.,  {Kent}
  S.~M.,  {Lupton} R.~H.,    {Ivezi{\'c}} {\v Z}.,  2003, \aj, 125, 1559

\bibitem[\protect\citeauthoryear{{Porciani}, {Magliocchetti} \&
  {Norberg}}{{Porciani} et~al.}{2004}]{PMN04}
{Porciani} C.,  {Magliocchetti} M.,    {Norberg} P.,  2004, \mnras, 355, 1010

\bibitem[\protect\citeauthoryear{{Rees}}{{Rees}}{1984}]{Rees84}
{Rees} M.~J.,  1984, \araa, 22, 471

\bibitem[\protect\citeauthoryear{{Richards} et~al.,}{{Richards}
  et~al.}{2001}]{Richards01}
{Richards} G.~T.,  et~al., 2001, \aj, 122, 1151

\bibitem[\protect\citeauthoryear{{Richards} et~al.,}{{Richards}
  et~al.}{2002}]{Richards02}
{Richards} G.~T.,  et~al., 2002, \aj, 123, 2945

\bibitem[\protect\citeauthoryear{{Richards} et~al.,}{{Richards}
  et~al.}{2004}]{Richards04}
{Richards} G.~T.,  et~al., 2004, \apjs, 155, 257

\bibitem[\protect\citeauthoryear{{Richards} et~al.,}{{Richards}
  et~al.}{2006}]{Richards06}
{Richards} G.~T.,  et~al., 2006, \aj, 131, 2766

\bibitem[\protect\citeauthoryear{{Richards} et~al.,}{{Richards}
  et~al.}{2009}]{Richards09}
{Richards} G.~T.,  et~al., 2009, \apjs, 180, 67

\bibitem[\protect\citeauthoryear{{Rosati}, {Tozzi}, {Giacconi}, {Gilli},
  {Hasinger}, {Kewley}, {Mainieri}, {Nonino}, {Norman}, {Szokoly}, {Wang},
  {Zirm}, {Bergeron}, {Borgani}, {Gilmozzi}, {Grogin}, {Koekemoer}, {Schreier}
  \& {Zheng}}{{Rosati} et~al.}{2002}]{Rosati02}
{Rosati} P.,  {Tozzi} P.,  {Giacconi} R.,  {Gilli} R.,  {Hasinger} G.,
  {Kewley} L.,  {Mainieri} V.,  {Nonino} M.,  {Norman} C.,  {Szokoly} G.,
  {Wang} J.~X.,  {Zirm} A.,  {Bergeron} J.,  {Borgani} S.,  {Gilmozzi} R.,
  {Grogin} N.,  {Koekemoer} A.,  {Schreier} E.,    {Zheng} W.,  2002, \apj,
  566, 667

\bibitem[\protect\citeauthoryear{{Ross} et~al.,}{{Ross}  et~al.}{2007}]{Ross07}
{Ross} N.~P.,  et~al., 2007, \mnras, 381, 573

\bibitem[\protect\citeauthoryear{{Salpeter}}{{Salpeter}}{1964}]{Salpeter64}
{Salpeter} E.~E.,  1964, \apj, 140, 796

\bibitem[\protect\citeauthoryear{{S{\'a}nchez}, {Baugh}, {Percival}, {Peacock},
  {Padilla}, {Cole}, {Frenk} \& {Norberg}}{{S{\'a}nchez}
  et~al.}{2006}]{Sanchez06}
{S{\'a}nchez} A.~G.,  {Baugh} C.~M.,  {Percival} W.~J.,  {Peacock} J.~A.,
  {Padilla} N.~D.,  {Cole} S.,  {Frenk} C.~S.,    {Norberg} P.,  2006, \mnras,
  366, 189

\bibitem[\protect\citeauthoryear{{Scherrer} \& {Weinberg}}{{Scherrer} \&
  {Weinberg}}{1998}]{Scherrer98}
{Scherrer} R.~J.,  {Weinberg} D.~H.,  1998, \apj, 504, 607

\bibitem[\protect\citeauthoryear{{Schlegel} et~al.,}{{Schlegel}
  et~al.}{2007}]{Schlegel07}
{Schlegel} D.~J.,  et~al., 2007, in BAAS Vol.~38, {SDSS-III: The Baryon
  Oscillation Spectroscopic Survey (BOSS)}.
p.~966

\bibitem[\protect\citeauthoryear{{Schlegel}, {Finkbeiner} \&
  {Davis}}{{Schlegel} et~al.}{1998}]{Schlegel98}
{Schlegel} D.~J.,  {Finkbeiner} D.~P.,    {Davis} M.,  1998, \apj, 500, 525

\bibitem[\protect\citeauthoryear{{Schneider} et~al.,}{{Schneider}
  et~al.}{2007}]{Schneider07}
{Schneider} D.~P.,  et~al., 2007, \aj, 134, 102

\bibitem[\protect\citeauthoryear{{Schulz} \& {White}}{{Schulz} \&
  {White}}{2006}]{Schulz06}
{Schulz} A.~E.,  {White} M.,  2006, Astroparticle Physics, 25, 172

\bibitem[\protect\citeauthoryear{{Scoville} et~al.,}{{Scoville}
  et~al.}{2007}]{Scoville07}
{Scoville} N.,  et~al., 2007, \apjs, 172, 1

\bibitem[\protect\citeauthoryear{{Scranton} et~al.,}{{Scranton}
  et~al.}{2002}]{Scranton02}
{Scranton} R.,  et~al., 2002, \apj, 579, 48

\bibitem[\protect\citeauthoryear{{Scranton} et~al.,}{{Scranton}
  et~al.}{2005}]{Scranton05}
{Scranton} R.,  et~al., 2005, \apj, 633, 589

\bibitem[\protect\citeauthoryear{{Serber}, {Bahcall}, {M{\'e}nard} \&
  {Richards}}{{Serber} et~al.}{2006}]{Serber06}
{Serber} W.,  {Bahcall} N.,  {M{\'e}nard} B.,    {Richards} G.,  2006, \apj,
  643, 68

\bibitem[\protect\citeauthoryear{{Shankar}, {Weinberg} \&
  {Miralda-Escud\'e}}{{Shankar} et~al.}{2007}]{Shankar07}
{Shankar} F.,  {Weinberg} D.~H.,    {Miralda-Escud\'e} J.,  2007,
  ArXiv:0710.4488v2

\bibitem[\protect\citeauthoryear{{Shanks} \& {Boyle}}{{Shanks} \&
  {Boyle}}{1994}]{Shanks94}
{Shanks} T.,  {Boyle} B.~J.,  1994, \mnras, 271, 753

\bibitem[\protect\citeauthoryear{{Shanks}, {Fong}, {Boyle} \&
  {Peterson}}{{Shanks} et~al.}{1987}]{Shanks87}
{Shanks} T.,  {Fong} R.,  {Boyle} B.~J.,    {Peterson} B.~A.,  1987, \mnras,
  227, 739

\bibitem[\protect\citeauthoryear{{Shanks}, {Fong}, {Green}, {Clowes} \&
  {Savage}}{{Shanks} et~al.}{1983}]{Shanks83c}
{Shanks} T.,  {Fong} R.,  {Green} M.~R.,  {Clowes} R.~G.,    {Savage} A.,
  1983, \mnras, 203, 181

\bibitem[\protect\citeauthoryear{{Shen} et~al.,}{{Shen}  et~al.}{2007}]{Shen07}
{Shen} Y.,  et~al., 2007, \aj, 133, 2222

\bibitem[\protect\citeauthoryear{{Shen} et~al.,}{{Shen}  et~al.}{2009}]{Shen09}
{Shen} Y.,  et~al., 2009, \apj, in press, ArXiv:0810.4144v1

\bibitem[\protect\citeauthoryear{{Shen}, {Greene}, {Strauss}, {Richards} \&
  {Schneider}}{{Shen} et~al.}{2008}]{Shen08b}
{Shen} Y.,  {Greene} J.~E.,  {Strauss} M.~A.,  {Richards} G.~T.,    {Schneider}
  D.~P.,  2008, \apj, 680, 169

\bibitem[\protect\citeauthoryear{{Sheth}, {Mo} \& {Tormen}}{{Sheth}
  et~al.}{2001}]{Sheth01}
{Sheth} R.~K.,  {Mo} H.~J.,    {Tormen} G.,  2001, \mnras, 323, 1

\bibitem[\protect\citeauthoryear{{Sheth} \& {Tormen}}{{Sheth} \&
  {Tormen}}{1999}]{Sheth99}
{Sheth} R.~K.,  {Tormen} G.,  1999, \mnras, 308, 119

\bibitem[\protect\citeauthoryear{{Smith} et~al.,}{{Smith}
  et~al.}{2002}]{Smith02}
{Smith} J.~A.,  et~al., 2002, \aj, 123, 2121

\bibitem[\protect\citeauthoryear{{Smith} et~al.,}{{Smith}
  et~al.}{2003}]{Smith03}
{Smith} R.~E.,  et~al., 2003, \mnras, 341, 1311

\bibitem[\protect\citeauthoryear{{Smith}, {Scoccimarro} \& {Sheth}}{{Smith}
  et~al.}{2007}]{Smith07}
{Smith} R.~E.,  {Scoccimarro} R.,    {Sheth} R.~K.,  2007, \prd, 75, 063512

\bibitem[\protect\citeauthoryear{{Somerville}, {Lee}, {Ferguson}, {Gardner},
  {Moustakas} \& {Giavalisco}}{{Somerville} et~al.}{2004}]{Somerville04}
{Somerville} R.~S.,  {Lee} K.,  {Ferguson} H.~C.,  {Gardner} J.~P.,
  {Moustakas} L.~A.,    {Giavalisco} M.,  2004, \apjl, 600, L171

\bibitem[\protect\citeauthoryear{{Spergel} et~al.,}{{Spergel}
  et~al.}{2007}]{Spergel07}
{Spergel} D.~N.,  et~al., 2007, \apjs, 170, 377

\bibitem[\protect\citeauthoryear{{Springel} et~al.,}{{Springel}
  et~al.}{2005}]{Springel05}
{Springel} V.,  et~al., 2005, \nat, 435, 629

\bibitem[\protect\citeauthoryear{{Steidel}, {Shapley}, {Pettini}, {Adelberger},
  {Erb}, {Reddy} \& {Hunt}}{{Steidel} et~al.}{2004}]{Steidel04}
{Steidel} C.~C.,  {Shapley} A.~E.,  {Pettini} M.,  {Adelberger} K.~L.,  {Erb}
  D.~K.,  {Reddy} N.~A.,    {Hunt} M.~P.,  2004, \apj, 604, 534

\bibitem[\protect\citeauthoryear{{Stoughton} et~al.,}{{Stoughton}
  et~al.}{2002}]{Stoughton02}
{Stoughton} C.,  et~al., 2002, \aj, 123, 485

\bibitem[\protect\citeauthoryear{{Strand}, {Brunner} \& {Myers}}{{Strand}
  et~al.}{2008}]{Strand08}
{Strand} N.~E.,  {Brunner} R.~J.,    {Myers} A.~D.,  2008, \apj, 688, 180

\bibitem[\protect\citeauthoryear{{Strauss} et~al.,}{{Strauss}
  et~al.}{2002}]{Strauss02}
{Strauss} M.~A.,  et~al., 2002, \aj, 124, 1810

\bibitem[\protect\citeauthoryear{{Swanson}, {Tegmark}, {Blanton} \&
  {Zehavi}}{{Swanson} et~al.}{2008}]{Swanson08}
{Swanson} M.~E.~C.,  {Tegmark} M.,  {Blanton} M.,    {Zehavi} I.,  2008,
  \mnras, 385, 1635

\bibitem[\protect\citeauthoryear{{Tegmark} et~al.,}{{Tegmark}
  et~al.}{2004}]{Tegmark04}
{Tegmark} M.,  et~al., 2004, \apj, 606, 702

\bibitem[\protect\citeauthoryear{{Tucker} et~al.,}{{Tucker}
  et~al.}{2006}]{Tucker06}
{Tucker} D.~L.,  et~al., 2006, Astronomische Nachrichten, 327, 821

\bibitem[\protect\citeauthoryear{{Ueda}, {Akiyama}, {Ohta} \& {Miyaji}}{{Ueda}
  et~al.}{2003}]{Ueda03}
{Ueda} Y.,  {Akiyama} M.,  {Ohta} K.,    {Miyaji} T.,  2003, \apj, 598, 886

\bibitem[\protect\citeauthoryear{{Vanden Berk} et~al.,}{{Vanden Berk}
  et~al.}{2005}]{Vanden_Berk05}
{Vanden Berk} D.~E.,  et~al., 2005, \aj, 129, 2047

\bibitem[\protect\citeauthoryear{{Wake} et~al.,}{{Wake}  et~al.}{2004}]{Wake04}
{Wake} D.~A.,  et~al., 2004, \apjl, 610, L85

\bibitem[\protect\citeauthoryear{{Wright}}{{Wright}}{2006}]{Wright06}
{Wright} E.~L.,  2006, \pasp, 118, 1711

\bibitem[\protect\citeauthoryear{{Wyithe} \& {Loeb}}{{Wyithe} \&
  {Loeb}}{2005}]{WyitheLoeb05a}
{Wyithe} J.~S.~B.,  {Loeb} A.,  2005, \apj, 621, 95

\bibitem[\protect\citeauthoryear{{Wyithe} \& {Padmanabhan}}{{Wyithe} \&
  {Padmanabhan}}{2006}]{Wyithe06}
{Wyithe} J.~S.~B.,  {Padmanabhan} T.,  2006, \mnras, 366, 1029

\bibitem[\protect\citeauthoryear{{York} et~al.,}{{York}  et~al.}{2000}]{York00}
{York} D.~G.,  et~al., 2000, \aj, 120, 1579

\bibitem[\protect\citeauthoryear{{Zehavi}, {Blanton}, {Frieman}, {Weinberg},
  {Waddell}, {Yanny} \& {York}}{{Zehavi} et~al.}{2002}]{Zehavi02}
{Zehavi} I.,  {Blanton} M.~R.,  {Frieman} J.~A.,  {Weinberg} D.~H.,  {Waddell}
  P.,  {Yanny} B.,    {York} D.~G.,  2002, \apj, 571, 172

\bibitem[\protect\citeauthoryear{{Zehavi} et~al.,}{{Zehavi}
  et~al.}{2005}]{Zehavi05b}
{Zehavi} I.,  et~al., 2005, \apj, 630, 1

\end{thebibliography}

\end{document}